%% file: non-degenerate_v1.tex
\documentclass[12pt,a4paper,oneside]{article}

\usepackage[margin=2cm]{geometry}

\usepackage[english]{babel}
\usepackage{amsmath,amssymb}
\usepackage[dvips]{graphicx}
\usepackage{footnote}
\usepackage{cite}

\newcommand{\Sla}[1]{/\!\!\!\!#1}
\newcommand {\beq} {\begin{equation}}
\newcommand {\eeq} {\end{equation}}
\newcommand {\bea} {\begin{eqnarray}}
\newcommand {\eea} {\end{eqnarray}}

\begin{document}

\title{ {\tiny \vspace*{-2.cm}\hspace*{14cm} CERN-PH-TH/2013-259}\\
{\tiny \vspace*{-.5cm} \hspace*{14cm} LAPTh-064/13} \vspace*{.9cm}\\ 
\bf \Large Light Non-degenerate Composite Partners at the LHC}

\author{\fontsize{12}{16}\selectfont C\'edric Delaunay$^{\, a,b}$, Thomas Flacke$^{\, c}$, J.\ Gonzalez--Fraile$^{\, d}$,\\ 
\fontsize{12}{16}\selectfont Seung J. Lee$^{\, c,e}$, Giuliano Panico$^{\, b}$ and Gilad Perez$^{\, b,f}$
\vspace{6pt}\\
\fontsize{11}{16}\selectfont\textit{$^a$ LAPTH, Universit\'e de Savoie, CNRS, B.P.110, F-74941 Annecy-le-Vieux, France}\\
\fontsize{11}{16}\selectfont\textit{$^b$ CERN Physics Department, Theory Division, CH-1211 Geneva 23, Switzerland}\\
\fontsize{11}{16}\selectfont\textit{$^c$ Department of Physics, Korea Advanced Institute of Science and Technology,}\\  
\fontsize{11}{16}\selectfont\textit{335 Gwahak-ro, Yuseong-gu, Daejeon 305-701, Korea}\\
\fontsize{11}{16}\selectfont\textit{$^d$ Departament d'Estructura i Constituents de la Mat\`eria and ICC-UB,}\\
\fontsize{11}{16}\selectfont\textit{Universitat de Barcelona, 647 Diagonal, E-08028 Barcelona, Spain}\\
\fontsize{11}{16}\selectfont\textit{$^e$ School of Physics, Korea Institute for Advanced Study, Seoul 130-722, Korea}\\
\fontsize{11}{16}\selectfont\textit{$^f$ Department of Particle Physics and Astrophysics, Weizmann Institute of Science,}\\
\fontsize{11}{16}\selectfont\textit{Rehovot 76100, Israel}}
\date{}
\maketitle

\begin{abstract}
We study the phenomenological implications of a large degree of compositeness
for the light generation quarks in composite pseudo-Nambu-Goldstone-boson Higgs models. We focus in particular on
phenomenologically viable scenarios where the right-handed up-type quarks have a sizable mixing with the
strong dynamics. For concreteness we assume the latter to be characterized by an SO(5)$/$SO(4) symmetry with fermionic resonances in the SO(4) singlet and fourplet representations. 
Singlet partners dominantly decay to a Higgs boson and jets. Since  no dedicated searches are currently looking for these final states, singlet partners can still be rather light. 
Conversely, some fourplet partner components dominantly decay to an electroweak gauge boson and a jet, a type of signature which has been analysed at the LHC.
We have reinterpreted various ATLAS and CMS analyses in order to constrain the parameter space of this class of models. 
In the limit of first two generation degeneracy, as in minimal flavor violation or U(2)-symmetric flavor models, fourplet partners need to be relatively heavy, with masses above $1.8\,$TeV, or the level of compositeness needs to be rather small.
The situation is significantly different in models which deviate from the first two generation degeneracy paradigm, as charm quark parton distribution functions are suppressed relative to the up quark ones.
We find that the right-handed charm quark component
can be mostly composite together with their partners being as light as $600\,$GeV, while the right-handed up quark needs either to be mostly elementary or to have partners as heavy as $2\,$TeV.
Models where right-handed up-type quarks are fully composite fermions are also analysed and yield qualitatively similar conclusions. 
Finally, we consider the case where both the fourplet and the singlet states are present. We demonstrate that in this case the fourplet bounds could be significantly weaken due to a combination of smaller production rates and the opening of new channels including cascade processes. 
\end{abstract}

\newpage
\section{Introduction}\label{sec:intro}
\input{intro.tex}
\section{Modeling the composite light quark flavors}\label{sec:mod}
\input{model.tex}\section{Hadron collider signatures of light composite partners}\label{sec:signal}
\input{signatures.tex}
\section{Existing direct searches and indirect constraints}\label{sec:searches}
\input{searches.tex}

\section{Bounds on non-degenerate composite light partners}\label{sec:results}
\input{results.tex}
\section{Conclusions}\label{sec:conclusions}
\input{conclusions.tex}

\bigskip
\emph{Acknowledgements:} \\
 The work of CD is supported by the ANR project ENIGMASS.
 TF and SL are supported by the National Research Foundation of Korea (NRF) grant funded by the Korea government (MEST) N01120547. 
J. G--F is supported by MICINN FPA2010-20807, CSD-2008-0037, EU grant FP7 ITN INVISIBLES (Marie Curie Actions
PITN-GA-29011-2989442) and by Spanish ME FPU grant AP2009-2546.
The work of GP is supported by grants from GIF, IRG, ISF, Minerva and the Gruber award.
TF and SL would like to thank the CERN theory group and the Weizmann Institute of Science for their
hospitality when part of this work was done.\\
J. G--F would also like to thank CERN theory group and ITP Heidelberg for their
hospitality during part of this work. 
\bigskip

\appendix
\section{$SO(5)/SO(4)$ Essentials}\label{sec:app1}
\input{appendix1.tex}

\section{Couplings derivation in partially composite models}\label{sec:app3}
\input{appendix3.tex}

\section{Predicted cross sections and exclusion limits}\label{sec:app4}
\input{appendix4.tex}

\section{Leptoquark search recast}\label{sec:app2}
\input{appendix2.tex}
\section{Pair production from cut-off physics}\label{UVpairprod}
\input{appendix5.tex}

\end{document}

%% file: intro.tex
The discovery of a Higgs-like boson at the LHC~\cite{Aad:2012tfa,Chatrchyan:2012ufa} is a great victory for the Standard Model (SM) of particle physics. With its minimal scalar sector of electroweak (EW) symmetry breaking,
the SM is a complete weakly coupled theory up to short distances far below the proton radius. Although the SM dynamics cannot explain several experimental evidences such as neutrino masses, the baryon asymmetry of the universe and the origin of dark matter, one cannot infer with any certainty an energy scale at which the SM would have to be extended, besides the well-known exceptions of the Planck scale related to gravity and the fantastically high scale of the Landau pole associated with the U(1) hyper-charge interaction.
Nevertheless, the fact that the Higgs mass is subject to additive renormalization implies that the EW scale is technically unnatural~\cite{thooft}. Any solution of this UV sensitivity (or fine-tuning) problem of the Higgs mass requires new dynamics beyond the SM (BSM) characterized by an energy scale close to the EW one\footnote{Giving up on this solution typically implies accepting a ``desert-like'' paradigm, in which the Higgs boson and the top quark, which couples rather strongly to the Higgs field, do not significantly couple to any form of new dynamics down to microscopic scales many of orders of magnitude below those currently probed at high-energy colliders. (See {\it e.g.} Refs.~\cite{Farina:2013mla,Giudice:2013yca} for recent discussions.) This  approach somehow resembles the overall state of mind of the physics community towards the end of the nineteenth century, when it was commonly believed that, apart from some small puzzles, the understanding of the basic laws of nature was almost complete. Another alternative approach to fine-tuning problem argues that, in analogy to the present explanation of the smallness of the cosmological constant~\cite{Weinberg:1987dv}, the EW scale is set by an environmental selection principle~\cite{Agrawal:1997gf}. However, this explanation seem to be less robust as life-permitting universes quite similar to ours may arise without weak interactions~\cite{Harnik:2006vj,Gedalia:2010iy}.}. 
From a low-energy perspective, the most severe UV sensitivity problem arises from quantum processes which involve a Higgs boson  splitting into a top-anti-top quark pair with arbitrarily large virtuality which gets absorbed back into the Higgs field pushing its mass towards the UV boundary of the theory. A simple way to stabilize the EW scale in a controlled manner is to postulate the existence of new particles carrying the same gauge quantum numbers as the top quarks. The UV insensitivity of the Higgs mass is obtained in practice from the virtual contributions of the new particles which exactly cancel those coming from the SM tops as dictated by some underlying symmetry. New physics states displaying this property are collectively denoted as top partners. In known BSM examples the partners might be scalar quarks, as in the celebrated case of supersymmetry, or vector-like fermions as in composite Higgs models (CHMs). In these two distinct realizations of the naturalness paradigm, the rest of the flavor sector, beyond the top partners, and its coupling are left unspecified. 

Top partners are defined according to their coupling to the Higgs field which is set in order to satisfy their role of EW-scale stabilizers.
Therefore, one might naively conclude that flavor physics is completely decoupled from naturalness considerations. 
However, even within a minimal sector, the flavor structure of the top partners could still be non-trivial, as
top partners need not be mass eigenstate fields in order to yield a sufficient cancellation in the Higgs mass. This feature was recently explored in low-energy supersymmetry in Ref.~\cite{Blanke:2013uia}, where it was demonstrated that the top squark flavor eigenstate can consist of an admixture of would be stop-like and scharm-like mass eigenstates. In such cases, flavor and CP violation effects may even arise from a minimally extended top sector.

The possibility of  top partners being admixtures of several mass eigenstates raises the important question of what could be robustly assumed regarding the flavor structure of the partners. Usually, this question is overlooked due to theoretical prejudices, as well as a possibly too naive interpretation of the bounds coming form low-energy flavor-changing neutral current processes (FCNCs). Most studies of naturalness assumed either flavor universality among the partners or an approximate U(2) symmetry which acts on the partner of the first two generations. However, a recent analysis of flavor constraints coming from $D-\bar D$ and $K-\bar K$ mixing observables showed that partners need not be degenerate within models of flavor alignment in the down-type quark sector~\cite{Gedalia-Kamenik-Ligeti-Perez}.
Models in which the new physics couplings are diagonal in the mass basis were considered both in the context of supersymmetry~\cite{Nir-Seiberg, Galon-Perez-Shadmi} and within the framework of composite Higgs~\cite{Fitzpatrick-Perez-Randall, Csaki-Perez-Surujon-Weiler}. 
The non-degeneracy of partners becomes even more interesting thanks to the following two facts which were shown to hold in supersymmetric models:
\begin{itemize}
\item[(i)] Direct experimental bounds on the second generation squarks are rather weak, of  $\mathcal{O}(400-500)\,$GeV,  since the associated searches are mainly sensitive to ``valence''
squark masses (masses of the first generation squarks) and are optimized for heavy squarks~\cite{Mahbubani:2012qq}.
\item[(ii)] If the top partners are not pure mass eigenstates but rather form an admixture of {\it e.g.} top-like and charm-like squarks, the direct search strategies need to be modified, as the relevant final states would not only involve top  pairs (and eventually missing energy) but also charm pairs and top-charm final states resulting in a weaker bound on the top partner mass and potentially improving on the EW scale fine-tuning~\cite{Blanke:2013uia}.
\end{itemize}
Combining (i) and (ii) leads to a supersymmetric ``flavorful'' naturalness scenario where the non-trivial flavor structure of the top sector yields a level of fine-tuning similar or, in some cases, even improved compared to the more conventional pure stop mass eigenstate scenario. In this paper we transpose the logic of Ref.~\cite{Mahbubani:2012qq} and item (i) above in the context of minimal CHMs where the Higgs doublet is realized as a pseudo-Nambu--Goldstone boson (pNGB)~\cite{HPGB,Agashe:2004rs} (see also~\cite{DP}). 
In particular we address the question of how light can the first and second generation quark partners be without assuming degenerate compositeness parameters.\\

The collider phenomenology of the quark partners within the compositeness framework is controlled by two important flavor parameters, namely the mass of the partners and the degree of compositeness of the SM quarks. Note that this is qualitatively different than in supersymmetric models where the only relevant parameter is the squark mass. 
The possibility of non-degenerate composite quarks is subject to a set of potentially strong indirect constraints  
arising  from two classes of precision tests which are coming from flavor physics and EW precision observables. However, as already argued above, flavor constraints can be sufficiently ameliorated thanks to flavor alignment. Furthermore,
it was shown in Refs.~\cite{flavor-triviality1,Redi-Weiler} that while the degree of compositeness of the SM quark doublets is severely constrained by EW precision tests, bounds on the degree of compositeness of the SM quark singlets can be rather weak thanks to an approximate custodial parity~\cite{Zbb}.
This observation may seem insignificant as in most minimal CHMs the spectrum consists of a single multiplet of top partners, and so any discussion related to flavor structure  of the partners is absent.
Note that this minimal approach fits very well with the flavor anarchic paradigm of CHMs, as in this case only the third generation quarks are sizably composite and the relevant phenomenology is well described by mass eigenstate top partners\footnote{This is similar in spirit to  supersymmetric models where the first two generation squarks are much heavier than the top squarks and thus less relevant to the current and near future collider program.}.
However,  the assumption that the top partner is not only a flavor but also a mass eigenstate is not required by naturalness arguments and the flavor-depending part of the collider phenomenology is not necessarily orthogonal.
In this work we relax the flavor anarchy assumption and focus on the implications of non-degenerate first two generation composite partners for LHC phenomenology. This possibility leads to a series of interesting experimental consequences which have been partially discussed in Ref.~\cite{flavor-triviality1,Redi-Weiler,flavor-triviality2,Da-Rold-Delaunay-Grojean-Perez,Delaunay-Grojean-Perez,Redi:2013eaa}. We focus here on analysing signals which could be already probed by direct searches at the LHC.\\

The remainder of the paper is organized as follows. In Section~\ref{sec:mod} we layout the modeling of composite right-handed quarks of the first two generations in the framework of minimal CHMs, and we outline the most important direct signatures at hadron colliders in Section~\ref{sec:signal}. Existing direct searches sensitive to the composite light quark signals are summarized in Section~\ref{sec:searches} and we derive the corresponding bounds on non-degenerate composite light quark scenarios in Section~\ref{sec:results}. We present our conclusions in Section~\ref{sec:conclusions}.

%% file: model.tex
We use a general low-energy parametrization of the strong sector dynamics which
only includes the lightest fermionic degrees of freedom connected to the up-type quarks. Possible vector resonances are  ``integrated out'' and do not appear directly in
the effective description.
This approach is motivated by EW precision bounds which tend to push the mass scale
of the vector resonances towards the multi-TeV range, while sub-TeV fermionic resonances
are typically present in realistic CHMs~\cite{Matsedonskyi:2012ym,Redi:2012ha,Marzocca:2012zn,Pomarol:2012qf,Panico:2012uw}.
Motivated by minimal composite Higgs realizations~\cite{Agashe:2004rs,Contino:2006qr}, we focus on implementations where the strong dynamics has a global ${\rm SO}(5)$ symmetry broken at the scale $f\lesssim \mathcal{O}(1\,$TeV$)$ down to its ${\rm SO}(4)$ subgroup. The Higgs field is identified with the NGB spanning the ${\rm SO}(5)/{\rm SO}(4)$ coset.
The symmetry structure of the strong dynamics does not fix the embedding of the fermionic resonances. For simplicity we assume that the up-type partners live in the fundamental representation, $\bf 5$, of ${\rm SO}(5)$.
We also neglect all flavor violation effects and focus on fermionic partners of the up and charm quarks, with the same coupling structure in both cases.

We adopt the Callan--Coleman--Wess--Zumino notation (CCWZ)~\cite{ccwz} in order to write down the effective Lagrangian in a non-linearly invariant way under ${\rm SO}(5)$.  (See {\it e.g.} Refs~\cite{DeSimone:2012fs,Grojean:2013qca} for a detailed presentation in CHM.)
In CCWZ the strong sector resonances are classified
in terms of irreducible representations of the unbroken global ${\rm SO}(4)$. In particular
the lightest composite fermions contained in the $\bf 5$ of ${\rm SO}(5)$
decompose as a fourplet, $Q$, and a singlet, $\tilde U$, under ${\rm SO}(4)$.
As well known, an extra global ${\rm U}(1)_X$
symmetry must be added to the strong dynamics in order to accommodate the correct fermion hypercharges~\cite{Zbb}. The ${\rm U}(1)_Y$ generator is then
identified with the combination $Y = T^3_R + X$, where $T^3_R$ is the diagonal generator of
the ${\rm SU}(2)_R$ subgroup of ${\rm SO}(4) \simeq {\rm SU}(2)_L \times {\rm SU}(2)_R$. Both composite fermions $Q$ and $\tilde U$ have charge $+2/3$ under ${\rm U}(1)_X$.

In terms of ${\rm SU}(2)_L \times {\rm U}(1)_Y$ representations, the fourplet $Q$ gives rise to two doublets.
One doublet with quantum numbers ${\bf 2}_{1/6}$, as the SM left-handed doublets, contains a charge $2/3$
state, $U$, and a charge $-1/3$ state, $D$. The second doublet of quantum numbers
${\bf 2}_{7/6}$ contains an exotic state with charge $5/3$, $X_{5/3}$, and a charge $2/3$ state, $X_{2/3}$. The composite states are embedded in a fundamental
${\rm SO}(5)$ representation $\psi$ as\footnote{{\it c.f.}~Appendix~\ref{sec:app1} for details on the conventions  used in the paper in regard to ${\rm SO}(5)$ representations.}
\beq
\psi=\left(\begin{array}{c}
                Q\\
		  \tilde{U}
               \end{array}\right)
               =\frac{1}{\sqrt{2}}\left(\begin{array}{c}
                iD-iX_{5/3}\\
		 D+X_{5/3}\\
		 iU+iX_{2/3}\\
		 -U+X_{2/3}\\
		 \sqrt{2}\tilde{U}
               \end{array}\right)\,.
\label{defpsi}
\eeq

The left-handed elementary quark doublets $q_L=\left(u_L, d_L\right)^T$ are incorporated as incomplete embeddings in the $\bf 5$ of ${\rm SO}(5)$ as
\beq\label{elfermL}
q_L^5\equiv \frac{1}{\sqrt{2}}\left( i d_L\,, d_L\,, i u_L\,, -u_L \,,  0\right)^T\,.
\eeq
$q_L$ then mixes with states of the composite sector through Yukawa interactions, leading to partially composite SM quark doublets~\cite{Kaplan:1991dc}.

The SM right-handed quark singlets could be realized as partially composite fermions as well by introducing elementary singlets $u_R$ embedded in incomplete $\bf 5$ of ${\rm SO}(5)$ as
\beq
u_R^5\equiv \left( 0\,,  0\,,  0\,,  0\,, u_R\right)^T\,.\label{elfermR}
\eeq
Since a large degree of compositeness will be considered for the SM singlets, an alternative possibility consists in directly identifying the latter with chiral ${\rm SO}(5)$ singlet states of the composite sector. This approach leads to  fully composite right-handed SM  quarks, similarly to the construction proposed in Ref.~\cite{DeSimone:2012fs} for
the right-handed top quark.

In all cases the total effective Lagrangian, $\mathcal{L}$, consists of two parts
\begin{equation}
{\cal L} = {\cal L}_{\rm comp} + {\cal L}_{\rm elem}.
\label{PCLag}
\end{equation}
${\cal L}_{\rm comp}$ describes the dynamics of the composite sector resonances, while ${\cal L}_{\rm elem}$ contains the kinetic terms of the elementary fermions as well as their mixing with the composite resonances.
We consider both scenarios where the right-handed singlets are either partially and fully composite states and we describe in the following subsections the details of their respective realizations.

\subsection{Models with partially composite right-handed up-type quarks}\label{sec:pcmodel}

We consider here a class of models based on the standard
partial compositeness construction~\cite{Kaplan:1991dc} in which both the SM doublets and singlets
have an elementary counterpart.
In CCWZ the Lagrangian for the composite fermionic sector reads
 \begin{eqnarray}
 \mathcal{L}_{\rm comp}=i\ \bar{Q}(D_\mu +i e_\mu) \gamma^\mu Q + i \bar{\tilde U}\Sla{D}\tilde U
 -M_4\bar{Q}Q
 -M_1\bar{\tilde{U}}\tilde{U}+\left(i c\, \bar{Q}^i \gamma^\mu d^i_\mu \tilde{U}+\mbox{h.c.}\right),
 \label{eq:pcL1}
\end{eqnarray}
where here and below $D_\mu$ contains the QCD gauge interaction and the $B_\mu$ coupling coming from the ${\rm U}(1)_X$ symmetry,
the $e_\mu$ and $d_\mu$ symbols are needed to reconstruct the
CCWZ ``covariant derivative'' and to restore the full non-linearly realized ${\rm SO}(5)$ invariance
({\it c.f.}~Appendix~\ref{sec:app1}).
The Lagrangian for the elementary fermions contains the usual kinetic terms,
including interactions with the SM gauge fields, and a set of linear mass mixings with the composite fermions
\begin{eqnarray}
 \mathcal{L}_{\rm elem}=i\ \bar{q}_L\,\Sla{D}q_L+i\ \bar{u}_R\,\Sla{D}u_R-y_L f\bar{q}^5_L U_{gs}\psi_R
 -y_R f\bar{u}^5_R U_{gs}\psi_L+\mbox{h.c.},
 \label{eq:pcL2}
\end{eqnarray}
where $q_L^5$ and $u_R^5$ are incomplete embeddings of the elementary fermions in the fundamental representation of ${\rm SO}(5)$ as given in Eqs.~\eqref{elfermL},\eqref{elfermR}. $U_{gs}$ is the Goldstone matrix containing the Higgs doublet components, which reads in unitary gauge
\beq
U_{gs}=\left(\begin{array}{ccccc}
                1 & 0 & 0 & 0 & 0 \\
		 0 & 1 & 0 & 0 & 0 \\
		 0 & 0 & 1 & 0 & 0 \\
		 0 & 0 & 0 & \cos{\bar{h}/f} & \sin{\bar{h}/f} \\
		 0 & 0 & 0 & -\sin{\bar{h}/f} & \cos{\bar{h}/f}
               \end{array}\right)\,.
               \label{gmatrU}
\eeq
 $\bar{h}\equiv v+h$ denotes the Higgs field
with the EWSB vacuum expectation value (VEV) $v$, which is related to the Fermi constant $G_F$ through
\beq
v=f\sin^{-1}\left(\frac{(\sqrt{2}G_F)^{-1/2}}{f}\right)\,,
\eeq
 and the physical Higgs boson $h$.
Notice that we work in an ${\rm SO}(5)$ basis where the elementary fermions $q_L$ and $u_R$ couple to the composite states $\psi$
only through the Goldstone matrix $U_{gs}$~\cite{DeSimone:2012fs,Grojean:2013qca}.

For simplicity, we assumed that the mixings  in Eq.~\eqref{eq:pcL2} respect an ${\rm SO}(5)$
structure, {i.e.} the mixing parameters of the elementary quarks with the fourplet
and the singlet are the same. In more general parametrizations two independent mixings
can be introduced, one for each ${\rm SO}(4)$ multiplet in $\psi$~\cite{Grojean:2013qca}.
The ${\rm SO}(5)$ mixing structure we consider is actually naturally predicted in explicit models with
a calculable Higgs potential, as the $2$-site model of Refs.~\cite{Panico:2011pw,Matsedonskyi:2012ym} whose effective description coincides with Eqs.~\eqref{eq:pcL1},\eqref{eq:pcL2} for $c=0$.
Moreover, the partial compositeness construction implies that the two mixing
parameters should be of comparable size as each elementary state mixes with only one operator from the strong dynamics~\cite{DeSimone:2012fs}.
The effect of this assumption on our analysis is marginal.
In particular our results are not modified in the limiting cases where only one ${\rm SO}(4)$ multiplet is light and present in the effective description.\\

We now discuss the mass spectrum of the model outlined above.
First of all, the exotic state $X_{5/3}$ does not mix with any other
states since electric charge is conserved, so its mass is simply $M_4$.
Conversely, the other composite fermions  mix with the elementary states. The complete
mass Lagrangian for the up- and down-type fermions is
\beq
 \mathcal{L}_{\rm mass}=-\left(\bar{u}\ \bar{U}\ \bar{X}_{2/3}\ \bar{\tilde{U}}\ \right)_L \mathcal{M}_u
 \left(\begin{array}{c}
           u\\
           U\\
           X_{2/3}\\
           \tilde{U}
       \end{array}\right)_R
       -\left(\bar{d}\ \bar{D}\right)_L \mathcal{M}_d
 \left(\begin{array}{c}
           d\\
           D
       \end{array}\right)_R
       +\mbox{h.c.}\,,
 \eeq
where
 \beq
 \mathcal{M}_u=\left(\begin{array}{cccc}
                0 & y_L f\cos^2\frac{\epsilon}{2} & y_L f\sin^2\frac{\epsilon}{2} & -\frac{y_L f}{\sqrt{2}}\sin\epsilon \\
		 \frac{y_R f}{\sqrt{2}}\sin\epsilon & M_4 & 0 & 0 \\
		 -\frac{y_R f}{\sqrt{2}}\sin\epsilon & 0 & M_4 & 0 \\
		 y_R f\cos\epsilon & 0 & 0 & M_1
               \end{array}\right)\,,\quad\quad \epsilon \equiv \frac{v}{f}\,,
\label{uMassmat}
\eeq
with $\mathcal{M}_u$ being mass matrix of the charge $2/3$ states, and
\beq
\mathcal{M}_d=\left(\begin{array}{cc}
                0 & y_L f\\
		 0 & M_4
               \end{array}\right),
\label{dMassmat}
\eeq
 the mass matrix for the charge $-1/3$ states.
The mass of the lightest charge $2/3$ quarks, which are
identified with the SM up-type quarks, is
\beq\label{eq:mass_up}
m_u\simeq \frac{v}{\sqrt{2}f}\times \big|M_1-M_4\big |\times \frac{y_Lf}{\sqrt{(M_4^2+y_L^2f^2)}} \times \frac{y_Rf}{\sqrt{(M_1^2+y_R^2f^2)}}\,,
\eeq
to leading order in the $\epsilon$.
We focus here on significantly composite right-handed up-type quarks. These states are associated with order one eigenvalues of $y_R$.
Then, the small mass of the light generation SM quarks implies very small values for the mixing parameters of the left-handed
elementary states, $y_L \ll 1$ (suppressing the flavor indices), unless the composite multiplets are nearly degenerate, $|M_1-M_4|\ll M_{1,4}$. However, the fourplet/singlet splitting is dominantly induced by the ${\rm SO}(5)$ breaking of the strong dynamics and is therefore expected to be large.
We thus assume $|M_1-M_4|\sim \mathcal{O}(M_{1,4})$, so that setting $y_L\ll 1$ is always a good approximation. We work in the $y_L=0$ limit in the remainder of the analysis.
To understand why $m_u\to 0$ in the limit $M_1 =M_4$ notice that in this case the free Lagrangian (setting the Higgs to its VEV) is having an enhanced {\rm SO}(5) symmetry. It can be used to bring $U_{gs}$ to trivial form by redefining the field $\psi$.
This implies that electroweak symmetry is not ``felt" by the elementary fermions. Thus, one expects to have two chiral massless states. Another more explicit way to see it is to notice that in this limit we can define two new linear combinations of $u_R$ and $\tilde U_R$ and similar for the left-handed fields that do not appear in any of the mass terms. These would correspond to the two zero modes. This enhanced chiral symmetry is broken at the quantum level due to the interaction terms. 

Notice that in the $y_L=0$ limit the Lagrangian
for the composite states and the elementary right-handed up quarks is exactly invariant
under the custodial ${\rm SO}(3)_c$ subgroup of ${\rm SO}(4)$. In fact, the $y_L$ mixing in Eq.~\eqref{eq:pcL2} is the only term which breaks the custodial
invariance, besides the usual ${\rm U}(1)_Y$ gauging of the SM. The $y_R$ mixing
preserves the custodial symmetry since the elementary $u_R$ is embedded as an ${\rm SO}(4)$ singlet.
We will show that this custodial invariance determines the structure of mixings and
couplings of the model. It is thus convenient to classify the
states in terms of ${\rm SO}(3)_c$ representations. $u_R$ and $\tilde U$
are ${\rm SO}(3)_c$ singlets, while the fourplet $Q$ splits into a singlet with charge $2/3$, which we denote
by $U_m$, and a triplet made of $D$, $X_{5/3}$ and a charge $2/3$ state, $U_p$.
In terms of the original fields the $U_{p,m}$ states are given by the combinations
\beq
U_{p,m}\equiv \frac{1}{\sqrt{2}}\left(U\pm X_{2/3}\right).
\label{defUpm}
\eeq
The Higgs field $\bar h$ is a singlet of custodial symmetry, while the
EW Goldstones form a triplet. Therefore the triplet states $D, U_p, X_{5/3}$
are mass eigenstates with mass $M_4$, and $u_R$ quarks can only mix
with $\tilde U$ and $U_m$. 
The mass Lagrangian for the custodial singlets is
\beq
\mathcal{L}_{\rm mass}^{\rm singlet} =  -\left(\bar{U}_m\,, \bar{\tilde U}\right)_L \mathcal{M}_u^{\rm singlet}
 \left(\begin{array}{c}
           u\\
           U_m\\
           \tilde U
       \end{array}\right)_R+\mbox{h.c.}\,,
\eeq
\beq
\mathcal{M}_u^{\rm singlet}=\left(\begin{array}{ccc}
		 y_R f \sin\epsilon & M_4 & 0 \\
		 y_R f \cos\epsilon &  0 & M_1
               \end{array}\right)\,,
\label{uMassmatapp}
\eeq
which yields the following masses for the heavy eigenstates $U_{l,h}$
\bea\label{U12mass}
m^2_{U_{l,h}}=\frac{1}{2}\left[M_1^2+M_4^2+y_R^2f^2 \mp \sqrt{\left(M_1^2-M_4^2+y_R^2f^2\right)^2-4 \sin^2 \epsilon\left(M_1^2-M_4^2\right)y_R^2f^2}\right]\,.
\eea
For $\sqrt{M_4^2+(y_R f \sin\epsilon)^2} \ll \sqrt{M_1^2+(y_R f \cos\epsilon)^2}$, the lighter eigenstate $U_l$ is dominantly the fourplet state $U_m$ mixed with the elementary quark, while $U_h$ is dominantly  $\tilde U$ mixed with the elementary fermion, while in the opposite limit, the r\^ole of $U_l$ and $U_h$ is exchanged.\\

We summarize below the structure of the couplings between the
elementary $u_R$ and the composite resonances which are relevant for both production
and decay of the composite resonances at the LHC. The relevant couplings are defined through the interaction Lagrangian
\bea
\mathcal{L}_{\rm int} &=& -\lambda_{huU_l} h\bar u^{\rm SM}_R U_{l L}-\lambda_{huU_h} h\bar u^{\rm SM}_R U_{h L}+{\rm h.c.}\nonumber\\
&&+g_{WuD}\bar D\, \Sla{W^-}u^{\rm SM}+g_{WuX}\bar X_{5/3}\,\Sla{W^+}u^{\rm SM}+g_{ZuU_p}\bar U_p\, \Sla{Z} u^{\rm SM}+{\rm h.c.}\,.\label{Lint}
\eea
We first consider two simplified limits where only one of the composite multiplets, either $Q$ or $\tilde U$,
is present in the low energy effective description, and then move to the generic case where both multiplets
are light.

\begin{figure}[tb!]
\begin{center}
\includegraphics[width=0.55\textwidth]{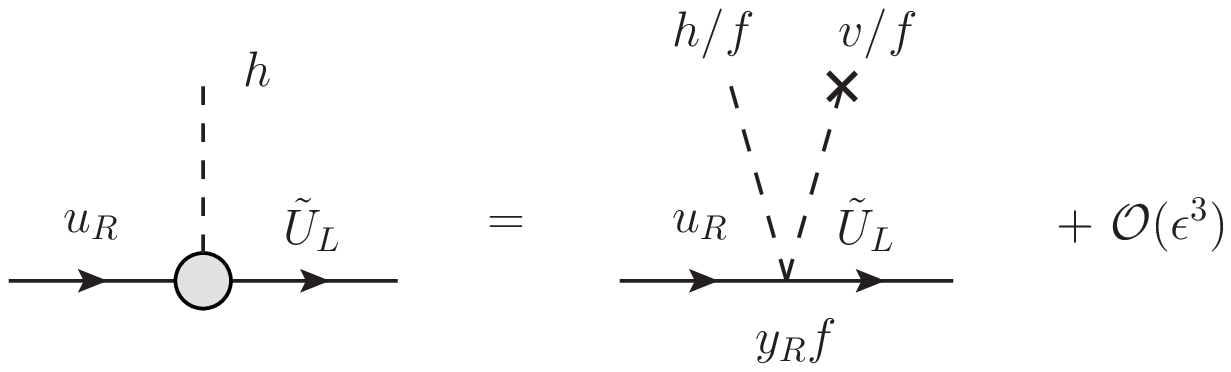}\\
(a)\\
\ \\
\includegraphics[width=0.83\textwidth]{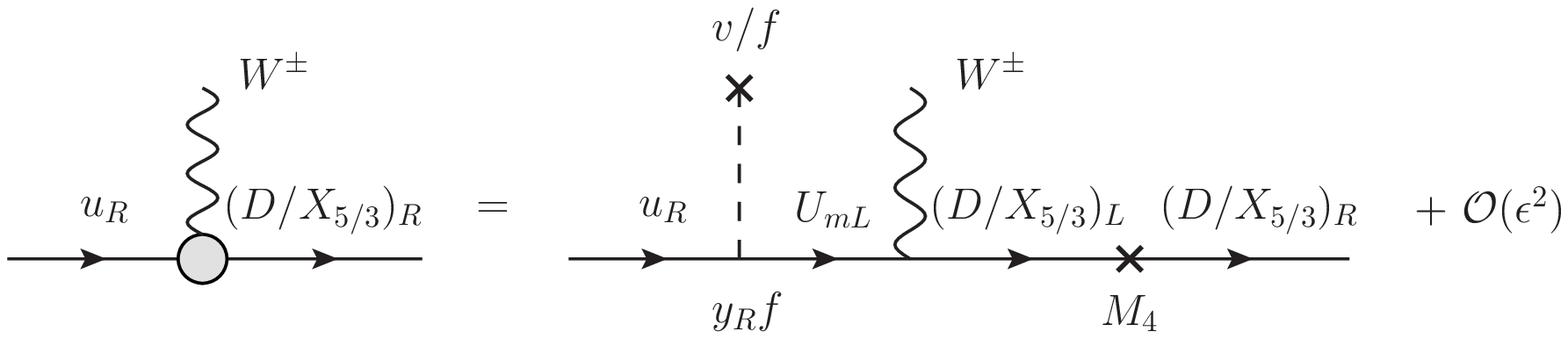}\\
(b)\\
\ \\
\includegraphics[width=0.99\textwidth]{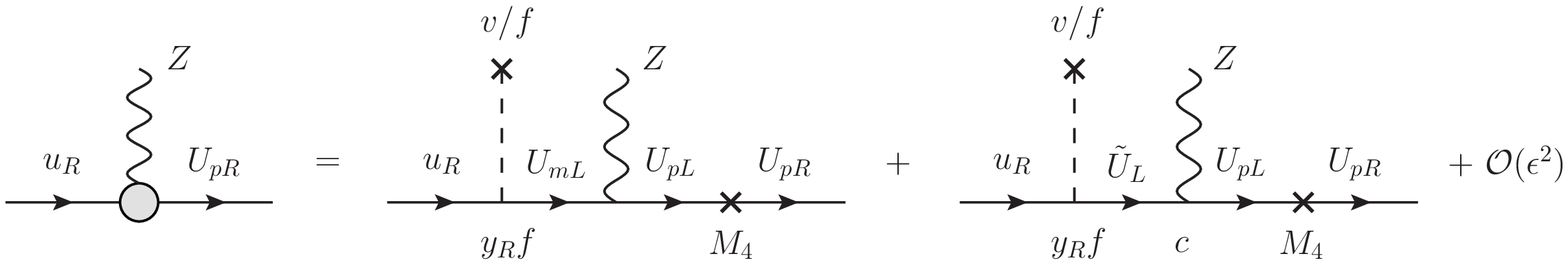}\\
(c)
\end{center}
\caption{Interaction vertices between the partially composite SM right-handed up-type quarks and their fermionic partners from the strong dynamics. All  vertices are drawn to leading order in both $\epsilon\simeq v/f$ and $y_R$, the elementary-composite mixing in the right-handed up sector. (a) Linear interaction between $u_R$, the custodial singlet resonances and the Higgs boson. (b), (c) Linear interaction between $u_R$, the custodial triplet resonances and the W and Z bosons. For the $Z$ vertex, the second diagram on the right hand side is absent when the singlet $\tilde U$ is decoupled.}
\label{vertices}
\end{figure}

\subsubsection{Light singlet partner interactions}\label{sec:PCsing}

We consider the case where the fourplet $Q$ is decoupled from the the low-energy theory, $M_4\to\infty$, and only a light singlet $\tilde U$
is present. In this limit the only light partner state is $U_l=U_{lL}+U_{lR}$, with $U_{lL}=\tilde U_L$ and
$U_{lR}=\sin\varphi_1 u_R+\cos\varphi_1 \tilde U_R$,
where $\varphi_1\equiv \tan^{-1}(y_Rf\cos\epsilon/M_1)$ is the elementary/composite mixing angle of the right-handed quarks.
The finite mass from Eq.~\eqref{U12mass} reduces to
\beq
m_{U_l}=\sqrt{ M_1^2+\left(y_R f \cos \epsilon\right)^2} =\frac{M_1}{\cos\varphi_1}\,,
\eeq
while the SM up quark $u^{\rm SM}=u^{\rm SM}_L+u^{\rm SM}_R$, with $u^{\rm SM}_L=u_L$ and $u^{\rm SM}_R=\cos\varphi_1 u_R-\sin\varphi_1\tilde U_R$, remains massless in the $y_L=0$ limit.

Custodial invariance implies that the only interaction of $\tilde U$ with the elementary quarks arises through the Higgs boson $h$.
Expanding the Goldstone matrix in Eq.~(\ref{gmatrU}) yield the following linear interaction with the Higgs
\beq\label{eq:HuUtilde}
\mathcal{L} \supset y_R \sin \epsilon\, \bar{u}_R\,  h\, \tilde{U}_L + \mbox{h.c.}\,.
\eeq
Notice the interaction in Eq.~\eqref{eq:HuUtilde} originates solely from the non-linear
Higgs dynamics, since $u$ and $\tilde{U}$, being both ${\rm SO}(4)$ singlets, can only couple
to an even number of Higgs doublets.
Diagrammatically the coupling can be understood as shown in Fig.~\ref{vertices}(a).
In the mass eigenstate basis, the $u^{\rm SM}_{R}-h-U_{lL}$ coupling becomes
\beq
\lambda_{huU_l}= -y_R\sin\epsilon\cos\varphi_1\,.
\label{couplingsinglet}
\eeq

\subsubsection{Light fourplet partners interactions}\label{sec:pc4}

We now consider the case where only a light fourplet $Q$ is present in the low-energy theory while the singlet $\tilde U$ is decoupled, $M_1\to\infty$. The  custodial triplet, made of $D$, $U_p$ and $X_{5/3}$, have mass $M_4$, while the custodial singlet $U_m$ state mixes with the elementary $u_R$ through EWSB. The other mass eigenstate is $U_l=U_{lL}+U_{lR}$, with $U_{lL}=U_{mL}$ and $U_{lR}=\sin\varphi_4 u_R+\cos\varphi_4 U_{mR}$, where $\varphi_4\equiv \tan^{-1}(y_Rf\sin\epsilon/M_4)$ is the elementary/composite mixing angle of the right-handed quarks. The finite mass from Eq.~\eqref{U12mass} reduces to
\beq
m_{U_l}=\sqrt{M_4^2+(y_R f \sin \epsilon)^2}=\frac{M_4}{\cos\varphi_4}\,,
\label{mtres}
\eeq
while the SM quark $u^{\rm SM}=u_L^{\rm SM}+u_R^{\rm SM}$, with $u_L^{\rm SM}=u_L$ and $u_R^{\rm SM}=\cos\varphi_4 u_R-\sin\varphi_4 U_{mR}$, remains massless in the $y_L=0$ limit. Notice that the $y_R$ contribution to the heavy
resonance mass is suppressed by a $v/f$ factor and thus it is only relevant
for large  $y_R$ values. For $y_R \lesssim 1$, this EWSB contribution turns out to  be typically negligible numerically, in which case
 all the fourplet states become nearly degenerate.

Custodial symmetry implies that $U_m$ only
interacts with $u_R$ through a vertex containing the Higgs boson. The linear interaction of $U_m$ with the Higgs is
\beq
\mathcal{L} \supset - y_R \cos \epsilon\, \bar{u}_R\, h\, U_{m,L}+\mbox{h.c.},
\eeq
This interaction
is understood diagrammatically the same way as in the previous case with a light $\tilde U$, up to the fact that here
the vertex is between an ${\rm SO}(4)$ singlet, $u_R$, and a fourplet component, $U_m$, which requires
 an odd number of Higgs insertions.
In the mass eigenbasis the $u_R^{\rm SM}-h-U_{lL}$ coupling becomes
\beq
\lambda_{huU_l}=y_R\cos \epsilon\cos \varphi_4\,.
\label{tres}
\eeq

The custodial triplet states interactions with the SM up quarks are also determined by custodial symmetry. The triplet states $D$, $U_p$ and $X_{5/3}$  only interact with the singlet $U_m$ through the triplet of EW gauge bosons (or equivalently through EW Goldstone bosons within the Higgs doublet). The interactions of the triplet states with $u^{\rm SM}$ then arise from their interactions with
$U_m$ through $y_R$ mixing. In  unitary gauge the relevant couplings in the original basis come from the fourplet kinetic term
in Eq.~\eqref{eq:pcL1}
\begin{equation}
{\cal L} \supset \frac{g}{2} \cos \epsilon \left(\bar D\, \Sla{W^-}
- \bar X_{5/3}\, \Sla{W^+} + \frac{1}{c_w} \bar U_p\, \Sla{Z}\right) U_m + {\rm h.c.}\,,
\end{equation}
where $g$ is the ${\rm SU}(2)_L$ gauge coupling and $c_w$ is the cosine of the weak mixing angle. The origin of these interactions is understood diagrammatically as shown in Fig.~\ref{vertices}(b)
 and Fig.~\ref{vertices}(c)
for the $W$ and $Z$ vertices, respectively. In the mass eigenbasis, the $u_R^{\rm SM}-Z-U_{pR}$, $u_R^{\rm SM}-W^+-D_R$ and $u_R^{\rm SM}-W^--X_{5/3R}$ couplings are then
\beq
g_{WuX} = - g_{WuD} = -c_w\, g_{ZuU_p} = \frac{g}{2} \cos \epsilon \sin\varphi_4.
\label{uno}
\eeq

\subsubsection{Generic partially composite case}\label{sec:pc5plet}

Finally, we consider here the more general situation where both the fourplet $Q$ and the
singlet $\tilde U$ composite states are light and below cut-off of the effective theory.
The structure of the SM up quark interactions with the custodial singlet and the triplet composite states is similar to the previous cases with only one multiplet in the effective theory. However, additional interactions between the singlet and the
fourplet arise from the $d_\mu$ term in Eq.~\eqref{eq:pcL1}. In particular the singlet $\tilde U$ interacts with the
custodial triplet via the $W$ and $Z$ bosons and with the custodial singlet $U_m$ via the physical Higgs boson. 
These additional interactions are relevant
for cascade decays like $U_h \rightarrow U_l h$ whenever $m_{U_h} > m_{U_l} + m_h$.

As in cases where only one multiplet is light, the $u_R$ quark interacts with the triplet states $D$, $U_p$ and $X_{5/3}$ only through EW gauge bosons, as dictated by custodial symmetry. In the original basis the couplings are diagrammatically understood from the same diagrams as in the light fourplet case, except for the $Z$ coupling which receives an additional contribution from the $d_\mu$ term, leading to the second diagram on the right-hand side of Fig.~\ref{vertices}(c).
The couplings take the form
\begin{equation}\label{gcpl_pc_generic}
g_{WuX} = - g_{WuD} = - c_w\, g_{ZuU_p} = \frac{g}{2}\cos \epsilon
\sin \varphi_4 \cos \tilde{\varphi}_1\,,
\end{equation}
with the effective mixing angle 
\begin{equation}
\tan \tilde \varphi_1 \equiv \frac{y_R f \cos \epsilon}{M_1}\frac{1}{\sqrt{1 + \left(y_R f \sin \epsilon\right)^2/M_4^2}}=\tan\varphi_1\cos\varphi_4\,.
\end{equation}

The right handed component of the up quark interacts with $U_m$ and $\tilde{U}$ only through the Higgs boson, thanks to custodial symmetry. The corresponding couplings in the mass eigenbasis can be calculated analytically, but the expressions are lengthy as they involve the diagonalization of the mass matrix in Eq.~\eqref{uMassmatapp}. The details on the calculation can be found in Appendix~\ref{sec:app3}. For $c=0$, approximate expressions can however be derived in the limit in which the fourplet is much lighter than the singlet; 
one finds
\begin{equation}
\lambda_{huU_l} \approx y_R \cos \epsilon \cos \varphi_4 \cos \tilde\varphi_1,
\qquad {\rm and} \qquad
\lambda_{huU_h} \approx - y_R \sin \epsilon \cos \varphi_4 \cos \tilde\varphi_1,
\end{equation}
where $U_{l,h}$ are the mass eigenstates with masses given by Eq.~\eqref{U12mass}.
Similar expressions are obtained in the opposite limit with a lighter singlet
through the replacement $U_l \leftrightarrow U_h$.

\subsection{Models with fully composite right-handed up-type quarks}\label{sec:fcmodel}

We follow here an alternative approach and identify directly the right-handed SM up quarks with chiral composite states of the strong dynamics. The right-handed up quarks are thus fully composite fermions in this scenario, without any elementary counterpart.
Moreover, the composite chiral fermions must be ${\rm SO}(5)$ singlets in order to avoid exotic massless quarks and reproduce the quantum numbers of the
right-handed SM up quarks.
The left-handed SM quark doublets are still realized as partially composite fermions whose mixing with the strong dynamics is small enough to account for the SM up and charm quark masses.

In CCWZ the composite Lagrangian becomes~\cite{DeSimone:2012fs,Grojean:2013qca}
\bea
\mathcal{L}_{\rm comp}&=&i\ \bar{\psi}(D_\mu+i e_\mu)\gamma^\mu\psi+i\
\bar{u}_R\, \Sla{D} u_R -   M_4\bar{Q}Q -M_1\bar{\tilde{U}}\tilde{U}\nonumber\\
&&
+\left(ic_L\, \bar{Q}_L^i d^i_\mu \gamma^\mu \tilde{U}_L+ i c_R\, \bar{Q}_R^i
d^i_\mu \gamma^\mu \tilde{U}_R+\mbox{h.c.}\right)
+\left(i c_1\, \bar{Q}_R^i d^i_\mu \gamma^\mu u_R+\mbox{h.c.}\right)\,,
\label{fcLag}
\eea
where $Q$ and $\tilde U$ are an ${\rm SO}(4)$ fourplet and singlet, respectively, embedded in a fundamental representation $\psi=(Q,\tilde U)^T$ of ${\rm SO}(5)$, as in Eq.~\eqref{defpsi} for the partially composite model. The chiral ${\rm SO}(5)$ singlet $u_R$ denotes the fully composite up quark. The Lagrangian describing the elementary fields $q_L$ and their mixings with the
composite states becomes
\bea
\mathcal{L}_{\rm elem}&=&
i\ \bar{q}_L\,\Sla{D}q_L -\left[y_L\, f\left(\bar{q}^5_L U_{gs}\right)_{i} {Q}_R^i + \mbox{h.c.}\right]\nonumber\\
&& - y_L\, c_2\, f\left(\bar{q}_L^5 U_{gs}\right)_5 u_R -  y_L\, c_3\, f\left(\bar{q}_L^5 U_{gs}\right)_5 \tilde{U}_R  + \mbox{h.c.}\,.
\label{eq:lagr_elem_compuR}
\eea
The partial compositeness assumption implies that $q_L$ only mixes
 with a single composite operator of the strong dynamics. Thus, we expect
all its mixings with the resonances to have comparable strengths.
We weighted the mass mixings in Eq.~\eqref{eq:lagr_elem_compuR} with an overall factor $y_L$ in order to account for this expectation. Possible deviations are parameterized by the $\mathcal{O}(1)$
 parameters $c_2$ and $c_3$. 

The spectrum of the model goes as follows. $X_{5/3}$ does not mix and has mass $M_4$. The mass matrix of the up-type sector in Eq.~\eqref{uMassmat} now reads 
\beq\label{upmatrixfc}
\mathcal{M}_u=\left(
\begin{array}{cccc}
-\frac{y_L c_2 f}{\sqrt{2}} \sin \epsilon  &y_L f \cos^2 \frac{\epsilon}{2} &  y_L f \sin^2 \frac{\epsilon}{2} &-\frac{y_L c_3 f}{\sqrt{2}} \sin \epsilon\\
0 & M_4 & 0 & 0\\
0 & 0 & M_4 & 0\\
0 & 0 & 0 & M_1
\end{array}
\right)\,,
\eeq
while the mass matrix in the down-type sector is the same as in Eq.~\eqref{dMassmat}.
The lightest up-type eigenvalue, which we identify with the mass of the SM up quark, is 
\beq
m_u \simeq c_2 y_L v \cos\varphi 
\approx c_2 y v \cos \varphi, 
\label{eq:mass_compuR}
\eeq
to leading order in $v/f$, where $\varphi \equiv \tan^{-1}(y_L f/M_4)$.
Therefore $y_L$ has to be small $\sim \mathcal{O}(m_u/v)$ in order to reproduce the light SM quark masses, and we set $y_L=0$  in the following.
In this limit $\mathcal{M}_u$ in Eq.~\eqref{upmatrixfc} is diagonal and the masses
of the up-type quark partners are simply
\beq
m_U=m_{X_{2/3}}=m_D=m_{X_{5/3}}= M_4\,, 
\qquad \mbox{and} \qquad m_{\tilde{U}}=M_1\,, 
\eeq
while $u_R^{\rm SM}=u_R$ remains massless.
As for the partially composite case with an elementary $u_R$, the only
terms which break the custodial ${\rm SO}(3)_c$ symmetry in the Lagrangian is the mixing of the
elementary doublet $q_L$. In the $y_L=0$ limit, the custodial invariance
is thus exact and dictates the structure of mixings and interactions among fermions.
It thus proves useful to classify the latter in terms of ${\rm SO}(3)_c$ representations. $u_R$, $\tilde U$ and $U_m$ fields are custodial singlets, while $X_{5/3}$,
$D$ and $U_p$ form a triplet, where $U_{p,m}$ are defined in terms of the original fields $U$ and $X_{2/3}$ as in Eq.~\eqref{defUpm}.

The other $d_\mu$ terms in Eq.~\eqref{fcLag} with coefficients
$c_L$ and $c_R$ also induce interactions between the fourplet $Q$ and the singlet $\tilde U$.\\

We now discuss the interactions of the fully composite $u_R$ with the composite resonances which are relevant for production and decay of the partners at the LHC. These interactions are characterized by the Lagrangian in Eq.~\eqref{Lint}.
We first consider the limiting cases with only one multiplet,  either the singlet $\tilde U$ or the fourplet $Q$, present in the low-energy spectrum. We close with the more general case where both multiplets are below the cut-off of the effective theory.\\

\subsubsection{Light singlet partner interactions}

When the fourplet is decoupled, $M_4\to \infty$, and  only $\tilde U$ is light, the effective Lagrangian significantly simplifies.
In particular the  SM up-type quark interactions with the heavy partners are
necessarily mediated by the $y_L c_2$ mixing and are thus extremely small. Heavy partners production at the LHC is therefore very suppressed which does not yield any interesting signal.

\subsubsection{Light fourplet partner interactions}

Although the mixing between the elementary states and the composite fermions disappears
completely in the $y_L=0$ limit, sizable interactions between the composite
states and $u_R$, coming from the $d_\mu$ term
 controlled by $c_1\sim \mathcal{O}(1)$ in Eq.~\eqref{fcLag},  are still present.
In the limit where only the fourplet is light and the singlet is decoupled ($M_1\to \infty$), $u_R$ interactions with the fourplet
states from the $d_\mu$ term in Eq.~\eqref{fcLag} are 
\beq
{\cal L} \supset -i \sqrt{2}\,\frac{c_1}{f}\bar{U}_{m,R} \gamma^\mu\left(\partial_\mu h \right)u_R - c_1 \frac{g}{\sqrt{2}} \sin \epsilon 
\left(\frac{1}{c_w}\bar{U}_{p,R}\,\Sla{Z} + \bar{D}\,\Sla{W}^-
-\bar{X}_{5/3}\,\Sla{W}^+\right)u_R +\mbox{h.c.}. \label{c1term}
\eeq
The EW gauge bosons mediate the interactions between the custodial triplet and $u_R$ with the following
couplings
\bea
g_{WuD}=-g_{WuX}=c_w\, g_{ZuU_p} = - c_1 \sin\epsilon \frac{g}{\sqrt{2}}.
\label{funo}
\eea
The linear Higgs term is a derivative interaction as expected from
the NGB nature of the Higgs.
Since we will only work at tree-level, we simply integrate the first term  by part in Eq.~\eqref{c1term}
and use the quark equations of motion in order to obtain the $u_R^{\rm SM}-h-U_{lL}$ coupling
\beq
\lambda_{huU_l}= - \sqrt{2} c_1 \frac{M_4}{f}\,,
\label{ftres}
\eeq
where $U_l=U_m$.

Note that the coupling structure of a fully composite $u_R$ is qualitatively similar to that of the partially composite case. In particular $U_p$ only couples to $u_R^{\rm SM}$ through the $Z$ boson, while $U_m$ does so only through the Higgs boson.

\subsubsection{Generic fully composite case}

In the generic case where both the fourplet $Q$ and the singlet $\tilde U$ are present in the
effective theory, the $d_\mu$ terms of coefficients $c_{L,R}$ yield
additional couplings between the fourplet states and $\tilde{U}$, which are defined through the interaction Lagrangian
\beq
\mathcal{L}_{\rm int}^{\rm heavy} = -\lambda_{h\tilde U U_l} h\bar{\tilde U}_L U_{l R}+g_{W\tilde U D} \bar D\,\Sla{W^-}\tilde U+g_{W\tilde U X}\bar X_{5/3}\,\Sla{W^+}\tilde U+g_{Z\tilde U U_p} \bar U_p\,\Sla{Z}\tilde U+{\rm h.c.}\,.
\eeq
$\tilde U$ interacts either through the EW gauge bosons with couplings 
\bea\label{funocp}
g_{W\tilde{U}D} &=& -g_{W\tilde{U}X}=c_w\, g^{L,R}_{Z\tilde{U}U_p}=- c_{L,R} \sin\epsilon\frac{g}{\sqrt{2}}
\eea
or through the Higgs boson with coupling
\bea
\lambda_{h\tilde{U}U_l} &=& \sqrt{2} c_{L,R} \frac{M_1-M_4}{f}\,,
\eea
with $U_l=U_m$.
While partner production proceeds as in the limiting cases where only $Q$ or $\tilde U$ is light, the decays are modified. For instance, if $M_4 > M_1+m_{W,Z,h}$,
the fourplet states can cascade decay through $\tilde U$, in addition to the direct decay into light quarks and $W^\pm,Z$ and $h$.

%% file: signatures.tex
We describe in this section the main phenomenological implications at hadron colliders of the existence of light up or charm quark composite partners. We present the dominant production and decay mechanisms of the partners, and then identify
the most promising channels for their discovery at the LHC. Also, the collider phenomenology of the up and charm quark partners differs significantly from that of top partners. (See {\it e.g.} Ref.~\cite{DeSimone:2012fs} for a recent discussion of top partner signatures at the LHC.) Hence, we also point out the main phenomenological differences between top and up/charm partners in regard to production mechanisms and final states from their decay. 

We base our discussion on the class of models described in Section~\ref{sec:mod}. We consider both scenarios where the right-handed up and charm quarks are partially or fully composite fermions, yet assuming a large degree of compositeness in the former case.
As we showed in the previous section, the structure of interactions is driven by an approximate custodial symmetry ${\rm SO}(3)_c$ in the limit where the left-handed SM quarks are mostly elementary fermions, and it is thus qualitatively similar in both partially and fully composite scenarios. In particular, the ${\rm SO}(4)$ singlet partner $\tilde U$ and the fourplet state $U_m$ are custodial singlets which couple to the SM quarks only through a Higgs interaction. Conversely, the remaining fourplet states $D,X_{5/3},U_p$ form a custodial triplet which therefore only couples to the SM quarks through EW gauge bosons.

\subsection{Production mechanisms}

\begin{figure}[tb!]
\begin{center}
\begin{tabular}{ccc}
\begin{tabular}{c}
\includegraphics[width=0.3\textwidth]{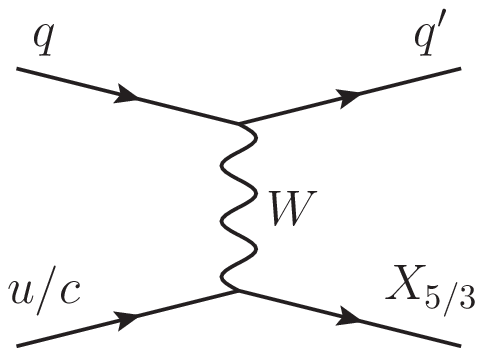}\\
\vspace{1cm}
\includegraphics[width=0.3\textwidth]{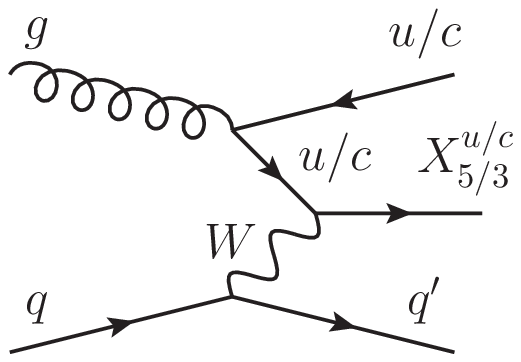}\\
\end{tabular}&
\includegraphics[width=0.3\textwidth]{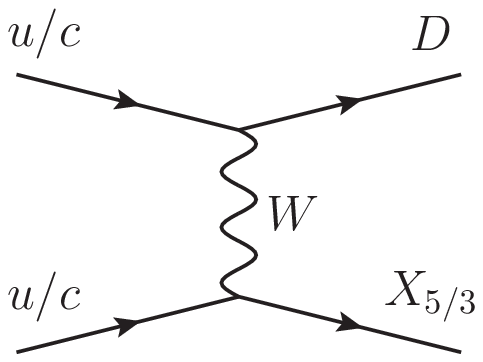}&
\begin{tabular}{c}
\includegraphics[width=0.3\textwidth]{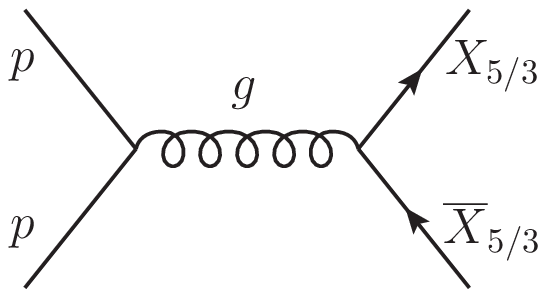}\\
\vspace{1cm}
\includegraphics[width=0.3\textwidth]{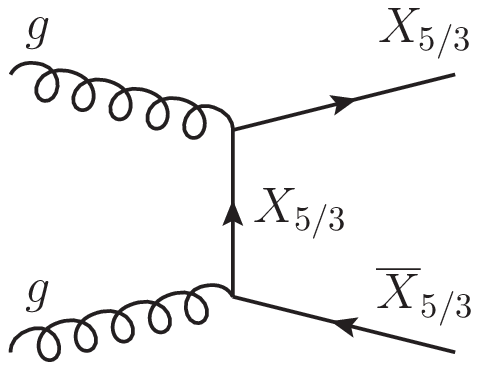}
\end{tabular}\\
(a) EW single production&
(b) EW pair production&
(c) QCD pair production\\
\end{tabular}
\end{center}
\caption{Dominant production channels of the up and charm quark composite partners. Similar diagrams with a neutral $Z$ exchange also exist. The $pp$ label in the top diagram in (c) collectively denotes the possible $q\bar q$ and $gg$ intial states.}
\label{fig:prodch}
\end{figure}

Since all the partners are colored, all of them can be produced in pairs at hadron colliders through universal QCD interactions as in Fig.~\ref{fig:prodch}c. QCD pair production is the same for all generations. It is furthermore completely model-independent, and its cross-section only depends on the partner mass. In particular, it does not depend on the degrees of compositeness of the associated SM quarks. We now consider in turns all other specific production mechanism of the singlet and fourplet partners. Note that the qualitative features of partner production do not depend on whether both or only one multiplet is present in the effective theory.\\

\begin{figure}[tb!]
\begin{center}
\begin{tabular}{ccc}
\includegraphics[width=0.3\textwidth]{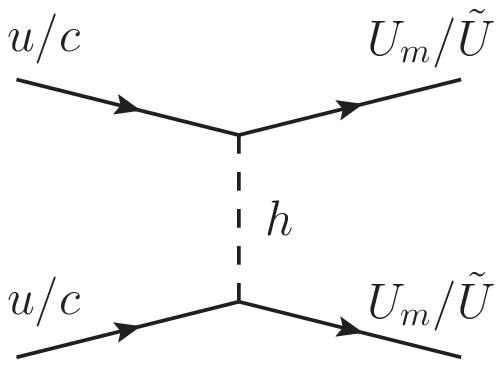}&
\includegraphics[width=0.3\textwidth]{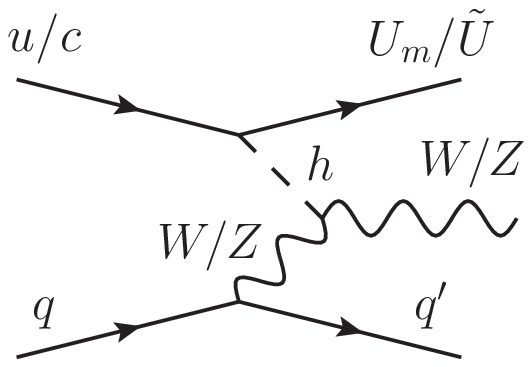}&
\includegraphics[width=0.3\textwidth]{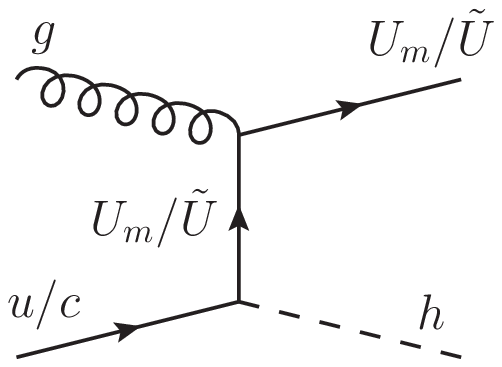}
\end{tabular}
\end{center}
\caption{Other relevant production channels which will be probed in a near future.}
\label{otherprod}
\end{figure}

We begin with production of the ${\rm SO}(4)$ singlet partners. Since the sole interaction of $\tilde U$ with the SM quark is through a Higgs boson, single production of the up and charm partner is suppressed by the square of the SM-like up and charm Yukawa coupling, respectively, and thus negligible\footnote{${\rm SO}(4)$ singlet partners can however be singly produced in the presence of color octet resonances from the strong dynamics~\cite{Carmona:2012jk,Redi:2013eaa}.}. 
This contrasts with the top partner case for which the large top mass makes single production one of the dominant mechanism, especially at large top partner mass~\cite{ContinoServant,MrazekWulzer}. 
However, as first pointed out in Ref.~\cite{Atre:2013ap}, single production in association with an EW gauge boson or a Higgs boson is possible and occurs through diagrams shown in Fig.~\ref{otherprod}. 
Finally, the first two generation $\tilde U$ partners can be produced in pairs, either through QCD interactions or through a $t$-channel Higgs exchange as shown in Fig.~\ref{fig:prodch}c and Fig.~\ref{otherprod}, respectively. Besides the partner mass dependence, QCD pair production is completely model-independent, while amplitudes involving a Higgs boson are also controlled by $\lambda_{huU_1}\propto v/f$. As a result cross-sections for Higgs-associated single production and Higgs-mediated double production are suppressed by a factor of $(v/f)^2$ and $(v/f)^4$, respectively.\\

The partner states within the ${\rm SO}(4)$ fourplet are produced in different ways depending on their respective custodial representation. On the one hand, the custodial singlet $U_m$ only couples to the Higgs. Thus it is produced either in pair or in association with a Higgs or an EW gauge boson, as $\tilde U$, albeit with a coupling $\lambda_{huU_m}$ which is not suppressed by EWSB.
On the other hand, the custodial triplet states can be singly produced through EW gauge boson exchange, as depicted in Fig.\ref{fig:prodch}a.
Besides QCD pair production, the triplet states are also pair produced through EW interactions as exemplified in Fig.\ref{fig:prodch}b. Both single and double production mechanisms of the triplet states are controlled by the model-dependent couplings $g_{WuX},g_{WuD},g_{ZuU_p}$.\\ 

All single production through the $qg$ initial state collisions (bottom diagram of Fig.~\ref{fig:prodch}a) occurs with the same luminosity for all generations. In contrast, single production through quark-quark initial states (top diagram of Fig.~\ref{fig:prodch}a) and EW pair-production have flavor dependent initial states. This  leads to significantly different production cross-sections at the LHC for up, charm and top partners due to the different PDFs of the initial quarks. For instance, we find that $uu$-mediated single and pair productions of up quark partners are completely dominated by the $t$-channel $W$ exchange. 
The situation differs qualitatively from that of top partners, as the large top mass implies that pair production is QCD dominated (top PDF vanishes at leading QCD order), while single production only occurs through $qg$ collisions~\cite{ContinoServant,MrazekWulzer}. Charm partner production sort of interpolates between the last two cases. Single production is dominated by $uc$ collisions (top diagram in Fig.~\ref{fig:prodch}a), while pair production is typically driven by QCD. EW pair production could however become more important than QCD production for large enough values of $y_R$ (in the partially composite $c_R$) or $c_1$ (in the fully composite $c_R$).  Note that the two diagrams of Fig.~\ref{fig:prodch}a contribute to different processes only in the kinematical region where the jet resulting from the gluon splitting is requested to have a large $p_T$. If the latter is either soft or collinear, the bottom diagram in Fig.~\ref{fig:prodch}a simply becomes part of the NLO correction to the process mediated at leading order by the top diagram in Fig.~\ref{fig:prodch}a.
 
\subsection{Decay channels and expected final states kinematics}

The decay of the partners typically goes through the vertex which dominate their production, with the exception of partners  produced in pair by QCD interactions.
For instance, the custodial singlet partners, $\tilde U$ and $U_m$, decay into SM quarks and a Higgs boson. For the first two generations these partners are produced either in pairs or in association with a Higgs, a $W$ or a $Z$ boson. Hence, the best channel to look for them involve $hhj$, $hWjj$ or $hZjj$, and $hhjj$ final states. 
Note that the $v/f$ suppression factor in the SO(4) singlet coupling to the Higgs boson (see Eq.\eqref{couplingsinglet}) can lead to a significant suppression of the singlet width in the limit of large compositeness scale $f$. In this case it is important to check whether decays through higher-order operators can become competitive. As pointed out in Ref.~\cite{Redi:2013eaa}, two higher-order effective operators can be relevant for the singlet decay. The first one is the loop-generated chromomagnetic operator, which leads to a decay into two jets ($\tilde U \rightarrow q g$). The second one is a four-fermion interaction mediated by an off-shell heavy gauge resonance, which leads to a decay into three jets ($\tilde U \rightarrow q q q$). The estimates for the partial widths of the singlet can be easily obtained from Ref.~\cite{Redi:2013eaa}. In the limit of a light singlet $m_{\tilde U} < m_\rho$, $m_\rho$ being the gauge resonance mass, and for couplings among the heavy states of order $m_\rho/f$, the decay channel $\tilde U \rightarrow hj$ is always dominant. Moreover, among the multi-jet channels, the $\tilde U \rightarrow q q q$ decay has typically a larger branching ratio than $\tilde U \rightarrow q g$.

The triplet states $D,X_{5/3}$ and $U_p$ decay into an EW gauge boson and a SM quark. For up and charm quark partners, the best search channels are thus
$Wjj$ and $Zjj$ for singly produced $D,X_{5/3}$ and $U_p$, respectively, and 
$WWjj$ and $ZZjj$ for pair production.

When both $Q$ and $\tilde U$ are present, composite partners can preferentially cascade decay into SM states through lighter partners, provided there is enough phase-space.
For instance, in partially composite $u_R$ scenarios, $D,X_{5/3}$ $U_p$ can first decay into an EW gauge boson and $\tilde U$, provided the latter is sufficiently light, which subsequently decays into a Higgs boson and a jet. 
In this case the signature is, respectively, one or two additional Higgs bosons in the final state, for singly or pair produced $D,X_{5/3}$ and $U_p$ partners. 

The  final states identified above  have rather peculiar kinematics which could be profitably used in better extracting NP signals from the SM background.
The heaviness of the produced partners typically implies high-$p_T$ jets and leptons in the final state and highly collimated  $W$ and $Z$ boson decay products.  The latter expectation usually allows usage of the kinematic variable $H_T$, defined as the sum of transverse momenta of the particles and missing momenta in the event, together with a fitted invariant mass method, in order to
increase signal to background ratio. In addition, at least one leptonically decaying EW gauge boson should be required in order to further reduce background.
Moreover, for pair production channels, at least two high-$p_T$ jets are expected in the final state, whereas for single production channels, the final state typically contains one hard jet from the heavy partner decay and one forward jet produced in association with the heavy partner.\\  

This contrast with top partners as they typically lead to taggable top or bottom quarks in the final states. Top partners are thus searched for in dedicated channels with much less background at ATLAS~\cite{ATLAS:topsearches} and CMS~\cite{CMS:topsearches}. One particular example is the $X_{5/3}$ top partner~\cite{ATLAS:53searches,CMS:53searches} which dominantly decays into $W^++t \rightarrow W^+ + W^+ b$. When both $W$ decay leptonically, the final state contains two same-sign charged leptons, a signature which was shown to have a significantly higher signal over background ratio~\cite{ContinoServant,MrazekWulzer}. This peculiar signal does not exist for the up and charm $X_{5/3}$ partners, since they promptly decay into a light quark jet instead of a heavy quark. 

%% file: searches.tex
We collect in this section the relevant collider searches performed at Tevatron and LHC experiments which we use  in order to constrain the existence of heavy quark partners of  sizably composite up and charm SM quarks. 
First of all, we review existing searches involving EW gauge bosons and hard jets as they are directly sensitive to the presence of up and charm partners with sizable couplings to right-handed first two generation quarks. Then, we discuss other searches which are also sensitive probes of up and charm partners, albeit to a lesser extend due to kinematical cuts tailored to search for different signals. In particular we argue that searches for leptoquarks at the LHC can be recast for our signal and hence be  relevant in constraining the existence of strong dynamics partners of composite up and charm SM quarks. 

Other generic collider signatures of heavy quark partners involve one or more Higgs bosons and high-$p_T$ jets. However, there are currently no available searches in Higgs bosons plus jets channels at the LHC due to small statistics. These channels could also be relevant in revealing (or further constraining) the existence of up and charm quark partners in the forthcoming 14$\,$TeV LHC run. 
Moreover they would be particularly important to study as they are the only ones sensitive to the presence of SO(4) singlet partners of composite right-handed up and charm quarks. As we focus here on the present experimental status of the composite up and charm quark partners, we do not consider these channels in the following and leave their analyses for future works\footnote{See Refs.~\cite{Carmona:2012jk} and~\cite{Atre:2013ap} for a study of some of these channels at the 14$\,$TeV LHC in a partial compositeness framework with a non-PGB Higgs dynamics.}.

\subsection{Relevant direct searches}\label{directsearches}

\subsubsection{Heavy quark searches in EW gauge bosons plus jets channels}
We review here existing experimental analysis seeking heavy fermionic partners which decay into light jets
and EW gauge bosons.
These include:

\begin{itemize}
\item CDF and D0 analyses based on, respectively, $Wjj$~\cite{CDF:wj} and $Wjj$ and $Zjj$~\cite{Abazov:2010ku} final states, and both using 5.4 fb$^{-1}$ of integrated luminosity.
These searches are sensitive to singly produced resonances within the custodial triplet.  They assume a leptonically decaying $W$ or $Z$ boson and further take advantage of special kinematics of the final
states arising from up and charm partner decays in order to suppress SM backgrounds. In particular,
cuts designed to single out a high-$p_T$ jet, together with a forward jet and one or more hard leptons from a highly boosted EW boson are imposed.
Moreover, for $Wjj$ final states, a high transverse missing energy collinear to the lepton is required. The $Wjj$ searches also benefit from the invariant mass reconstructed by the lepton, the hardest jet and the missing transverse
momenta in searching for fermionic resonances.

\item CDF and D0 analyses based on $WWjj$ and using, respectively, 5.6 fb$^{-1}$~\cite{Aaltonen:2011tq}
and 5.3 fb$^{-1}$~\cite{Abazov:2011vy} of integrated luminosity.
This channel is sensitive to pair produced $D$ and $X_{5/3}$ resonances within the custodial triplet.
The analyses focus on semi-leptonically decaying $W$ pairs. Thus they only select events with one hard isolated lepton, large
missing transverse energy and four isolated jets, one of which having large transverse momentum. Both analyses
use the $H_T$ variable together with a fitted mass method in order to derive exclusion bounds on pair production cross sections
of fermionic partners.

\item ATLAS analysis based on $Zjj$ and $Wjj$ final states, using 4.64 fb$^{-1}$ of integrated luminosity at the $7\,$TeV LHC run~\cite{ATLAS:pp}. Here the experiment searched for singly produced heavy quarks with large couplings to the SM up quark and ÊW gauge bosons by looking for final states with a jet with high
transverse momentum, a sub-leading jet in the forward direction and one or two isolated hard leptons originating from $W$ or $Z$ decay, respectively.
As for similar searches at the Tevatron, a large transverse missing energy is also required in $Wjj$ final states.
Advantage of the collimated decay products of $W$ and $Z$ bosons is also taken
by imposing rapidity and azimuthal angle cuts between the different reconstructed objects in the event. Cuts are further
optimized using multivariate analysis techniques. Finally, a fitted mass method is used in seeking resonances and placing limits
on their existence.

\item ATLAS analysis based on $WWjj$ final state, using 1.04 fb$^{-1}$ of integrated luminosity of the $7\,$TeV LHC run~\cite{Aad:2012bt}.
Both $W$ bosons are required to decay into leptons. Thus characteristic features of this search are at least
two jets, two opposite-sign leptons (out of the Z mass window) and missing transverse energy in the final state.
$H_T$ cut is also imposed. Finally, the heavy partner mass reconstruction benefits from the
large boost each W boson receives from the heavy quark decay, since each missing neutrino is nearly collinear with
its associated charged lepton.

\item CMS search for heavy resonances in the $W/Z$-tagged dijet mass spectrum, using 19.8 fb$^{-1}$ of integrated
luminosity at the $8\,$TeV LHC run~\cite{CMS:wzdijet}. The analysis looks for massive resonances which decay
into a light SM quark and a hadronically decaying weak boson. It takes advantage of the fact that for sufficiently heavy resonances decay products
of the $W$ or $Z$ boson merge into a single jet. This leads to an effective dijet signature in the event, where one jet is tagged as weak boson jet. Extra jets are  not vetoed.
The two hardest jets in each event are used to build a dijet spectrum.  Narrow resonances would reveal themselves as sharp peaks in the spectrum, in the absence of which bounds on the resonance masses are extracted.
This channel is sensitive to single production of heavy quark partners through EW interactions. In this case,
we checked that the jet from the prompt decay of the partner and the merged jet from W or Z decay 
are typically the hardest two in the events. More precisely we find that this is the case in more than 97$\%$ of the events in the mass range the analysis
is sensitive to. Therefore the presence of an extra forward jet coming from single production of the resonances does not significantly impact the efficiencies, so that
this analysis directly applies to composite up and charm SM quark partners.
\end{itemize}

We use direct searches reviewed above in order to bound the existence of the fermionic partners of up and charm SM quarks. 
For all analyses we implemented the models of Section~\ref{sec:mod} in FeynRules~\cite{Christensen:2008py}, interfaced 
with MadGraph 5~\cite{Alwall:2011uj}, we simulate our signals at the parton level. The exclusion limits from the above searches are then directly
applied to the models considered in this paper as they share the same kinematics than the theoretical setups assumed by the experimental collaborations.
We present our results for both partially and fully composite right-handed quark scenarios in Section~\ref{sec:results}.\\

\subsubsection{Recasting leptoquark searches}

Other experimental searches, designed to search for  different types of new physics particles, could also be used {\it a priori} to probe the presence of first two generation quark partners. They include for instance three-jet resonance searches~\cite{Chatrchyan:2012uxa}, originally designed to look for gluinos in $R$-parity-violating supersymmetric models, pair-produced top-like heavy quark searches~\cite{ATLAS:2012qe,Aad:2012xc,CMS:2012ab,Chatrchyan:2012vu}, bottom-like heavy quark searches~\cite{CMS:2012jwa} and pair-produced leptoquark searches~\cite{cms8lepto}. 
These searches are however much less efficient, relative to EW gauge bosons and jets channels, in looking for heavy quark partners of the first two generation SM quarks. The reduced efficiency mostly comes from specific requirements on the events, like the presence of $b$-tagged jets or different mass reconstruction assumptions, which are tailored to look for particles whose dynamics qualitatively differs from that of fermionic up and charm partners. 
Yet, among the above list, leptoquark searches are based on final states which are close enough to our signal to still yield relevant bounds on partner masses. In particular the pair-produced leptoquark search~\cite{cms8lepto} performed by the CMS collaboration is looking for a $\mu^+\mu^-jj$ final state which can be obtained from pair-produced $D$ and $X_{5/3}$ up and charm partners, each decaying into $Wj$ with a subsequent leptonic $W$ decay. We describe in the following how we recast the CMS leptoquark search of Ref.~\cite{cms8lepto} in order to derive bounds on these partners.\\

We present the qualitative features of the CMS search in Ref.~\cite{cms8lepto} and its recast, while further details in regard to how we performed the latter are collected in Appendix~\ref{sec:app2}.
The CMS analysis  is based on 19.6 fb$^{-1}$ of integrated luminosity at the $8\,$TeV LHC run with a mass reach extending to $1.2\,$TeV. 
We focus on the $\mu^+\mu^-+2$\,jets channel.
Preselection cuts are applied to isolate two hard muons and two hard jets. Further
cuts on $S_T\equiv p_T^{\mu_1}+p_T^{\mu_2}+p_T^{j_1}+p_T^{j_2}$, the dimuon invariant mass $M_{\mu\mu}$, and $M_{\mu j}^{\rm min}$ are then optimized for the leptoquark signal. $M_{\mu j}^{\rm min}$ is defined as the smallest of the two muon-jet invariant masses obtained for the muon-jet pairing which minimizes the difference between the two muon-jet invariant masses~\cite{cms8lepto}.

For recasting the results based on the above analysis, we use background estimations  and binned data reported by the CMS collaboration~\cite{cms8lepto}. 
We implemented the model of Section~\ref{sec:mod} using FeynRules~\cite{Christensen:2008py} and the corresponding up and charm quark partner signals were simulated with MadGraph 5~\cite{Alwall:2011uj} for event generation, interfaced with PYTHIA~\cite{Sjostrand:2006za} 
for parton shower and hadronization and with a PGS 4~\cite{pgs} detector simulation. We also simulate in the same way the leptoquark signal assumed in Ref.~\cite{cms8lepto}. This leptoquark simulation is then used 
to  further tune the heavy quark partner simulation in order to match CMS detection efficiencies quoted for leptoquark signals.
The CMS results are presented with
different invariant mass distributions and selection cut levels which we take advantage from in order to improve our recast
of the CMS analysis.
We then use the following statistical method in order to derive exclusion limits for the up and charm quark partners. First of all, we build a binned
log-likelihood function for each available distribution, 
where the number of observed events are assumed to follow a Poisson distribution. Then, these log-likelihood
functions are individually maximized (or equivalently the corresponding $\chi^2$'s are minimized) in order to derive partner mass values excluded at 95$\%$ confidence level (CL) for each distribution.
Finally, for each partner mass value, we quote as exclusion limit the strongest limit of those obtained out all available  kinematical distributions.

Bounds obtained from this leptoquark search  recast should however be taken with a grain of salt, when compared with that of
EW gauge boson plus jets reviewed in the previous subsection, as
additional assumptions were made in the determination of the former. First of all, while tuning our simulated efficiencies to match those of CMS for total event rates, we neglected a possible dependence of the
efficiencies on the energy and momentum of the particles in the events. Moreover, theoretical uncertainties
were included in a simplified way in the statistical analysis. Nevertheless,
we observe that the bounds are still statistically dominated, so that the exclusion limits that we derive from this recast are relatively accurate.

Besides deriving exclusion limits, another motivation for recasting the CMS leptoquark analysis is to encourage experimental collaborations to perform searches similar to that in Ref.~\cite{cms8lepto}, but with slightly different cuts optimized for first two generation quark partners.
Indeed, we show in the next section that exclusion limits based on this recast are, as expected, significantly diluted relative to the limits on leptoquark masses found in Ref.~\cite{cms8lepto}. We also show bounds from this recast are not considerably weaker than those from more straightforward EW gauge bosons and jets channels. Hence, we argue that a search similar to the leptoquark one in Ref.~\cite{cms8lepto} but with optimized cuts would potentially have a higher reach than EW gauge bosons plus jets channels,  given the much smaller luminosity of the latter.

\subsection{Indirect constraints from dijet production}\label{indirectsearches}

Strong dynamics near the TeV scale leads to significant new physics sources of dijet production at the LHC when the 
light SM quarks have a large degree of compositeness~\cite{Eichten}. Sizable dijet contributions arise in the presence of a light color octet vector resonance in the effective theory~\cite{flavor-triviality1,Redi-Weiler,Da-Rold-Delaunay-Grojean-Perez,Redi:2013eaa}. Even if such a state is absent, as assumed here, new physics dijet sources are generically induced by unknown physics at the cut-off scale $\Lambda\sim 4\pi f$. These effects are characterized below $\Lambda$ by four-fermion operators in the composite Lagrangian like\footnote{Other combinations of composite resonances, which are not captured in Eq.~\eqref{4F}, are allowed by the global symmetries of the strong dynamics. We do not aim here at a complete study of all four-fermion interactions, but we view  Eq.~\eqref{4F} as general enough to illustrate the typical level of dijet contributions  induced by the strong dynamics.}
\beq
\frac{1}{2f^2}\left[(\bar Q \gamma_\mu Q)^2+(\bar{\tilde U} \gamma_\mu \tilde U)^2+(\bar u_R\gamma_\mu u_R)^2\right]\,,\label{4F}
\eeq
where $\mathcal{O}(1)$ differences in their coefficients have been neglected and the last term is only present in fully composite scenarios. Note that the operators in Eq.~\eqref{4F} are not suppressed by the cut-off scale $\Lambda$, but rather by the compositeness scale $f$~\cite{Grojean:2013qca}. This is due to the fact that the UV physics is strongly coupled at $\Lambda$, so that generically the above operators are induced at that scale with coefficients of $\mathcal{O}(16\pi^2/\Lambda^2)$, according to naive dimensional analysis~\cite{NDA}. In the presence of mixings between the chiral quarks and the vector-like heavy resonances, operators like Eq.~\eqref{4F} yield four-fermion interactions  in terms of the SM quarks
\beq
{\cal L}_{\rm 4f}=\frac{c_{uu}}{2}\left(\bar u_R^{\rm SM} \gamma_\mu u_R^{\rm SM}\right)^2+\frac{c_{cc}}{2}\left(\bar c_R^{\rm SM} \gamma_\mu c_R^{\rm SM}\right)^2+c_{uc}\left(\bar u_R^{\rm SM} \gamma^\mu u_R^{\rm SM}\right)\left(\bar c_R^{\rm SM} \gamma_\mu c_R^{\rm SM}\right)\,,\label{4FSM}
\eeq 
where $c_{uu},\, c_{cc}$ and $c_{uc}$ have mass dimension $-2$.\\

Contact interactions like Eq.~\eqref{4FSM} have peculiar signatures in the angular distribution of dijet events at colliders. Indeed, despite their massive number in hadronic collisions, background dijet events from QCD are primarily produced in the forward direction, near the beam axis, due to a Rutherford-like scattering mediated by massless quarks and gluons in the $t$-channel. On the other hand, dijet events resulting from the contact interactions in Eq.~\eqref{4FSM} tend to be more isotropically distributed in the detector. This qualitative difference appears rather clearly in the event distribution in terms of the kinematical variable $\chi_j\equiv e^{2y_j}$, $y_j$ being the jet rapidity in the partonic center-of-mass frame, where QCD dijets are evenly distributed in $\chi_j$, while those originated from contact interactions are peaking at low $\chi_j$ values. The ATLAS and CMS~\cite{CMSdijet} collaborations searched for the presence of a new physics source in dijet production in the form of a representative contact interaction $c_{qq}/2\times(\bar q_L^{\rm SM}\gamma_\mu q_L^{\rm SM})^2$, involving the first generation left-handed SM quark doublet $q_L^{\rm SM}$. The consistency of the angular distribution of dijet events with QCD expectations leads to the following 95$\%$CL limits on the contact interaction above
\beq
|c_{qq}|^{-1/2}\gtrsim 2.2\,{\rm TeV\ \ \ \ \ for\ \ \ \ }\quad c_{qq}>0\,,
\eeq
from ATLAS~\cite{ATLASdijet}, and 
\beq
|c_{qq}|^{-1/2}\gtrsim 2.1\,(3.0)\,{\rm TeV\ \ \ \ \ \ for \ \ \ \ }\quad c_{qq}>0\,(c_{qq}<0)\,,
\eeq
from CMS~\cite{CMSdijet}. The bound is stronger for negative coefficient since the interference is constructive in this case. The sign of the Wilson coefficient in Eq.~\eqref{4F} is not resolved within the effective theory. Nonetheless, we assume constructive interference with QCD in order to remain on the conservative side when comparing with the data. Since neither collaboration analysed the set of operators in Eq.~\eqref{4FSM}, we follow the procedure of Ref.~\cite{Da-Rold-Delaunay-Grojean-Perez} and derive approximate lower bounds by demanding that the $\chi_j$ distributions for various dijet mass bins do not deviate from SM expectations more than in the presence of $c_{qq}/2\times(\bar q_L^{\rm SM}\gamma_\mu q_L^{\rm SM})^2$, with $|c_{qq}|^{-1/2}=3\,$TeV and $c_{qq}<0$. Assuming the presence of each operator in Eq.~\eqref{4FSM} at a time, we find\footnote{The bound on the first generation four-fermion operator is consistent with that obtained from the procedure used in Ref.~\cite{Pomarol-dijets}.}
\beq
|c_{uu}|^{-1/2}\gtrsim 2.8\,{\rm TeV}\,,\quad |c_{cc}|^{-1/2}\gtrsim 300\,{\rm GeV}\,,\quad |c_{uc}|^{-1/2}\gtrsim 800\,{\rm GeV}\,.\label{dijet_bounds}
\eeq
Notice that LHC experiments collected dijet events of invariant masses up to $\simeq4\,$TeV. The effective description breaks down at a scale of at most $\mathcal{O}(4\pi/\sqrt{c})$. We therefore expect $\mathcal{O}(1)$ modification in the $c_ {cc}$ bound due to the neglected radiative corrections.\\

In models where the right-handed up and charm quarks are fully composite fermions,  the four-fermion interactions in Eq.~\eqref{4FSM} arise at a scale $|c_{uu}|\sim |c_{cc}|\sim|c_{uc}|\simeq 1/f^2$, where $f\gtrsim 600\,$GeV in order not to introduce overly large tensions with EW precision tests~\cite{Grojean:2013qca}. By comparing with Eq.~\eqref{dijet_bounds}, we conclude that a fully composite $u_R^{\rm SM}$ is in  tension with dijet searches at the $7\,$TeV LHC, while the latter is not sensitive to a fully composite $c_R^{\rm SM}$. We nevertheless consider direct LHC signals of heavy partners of a fully composite right-handed up quark in order to illustrate the difference in sensitivity between the first two generation quarks.

For partially composite right-handed up and charm quarks, a smaller dijet contribution is expected, suppressed by the fourth power of the partial compositeness. Since a fully composite right-handed charm is not constrained by dijet data, no constraints are obtained on partially composite charms either. 
We thus consider only the first generation. Under the assumption that only the fourplet $Q^i$ or the singlet $\tilde U$ is present in the effective theory, Eq.~\eqref{4F} yields
\beq
c^4_{uu}=\frac{\sin^4\varphi_4}{f^2}\,,\quad c^1_{uu}=\frac{\sin^4\varphi_1}{f^2}\,,
\eeq
respectively, where $\varphi_{1,4}$ are the mixing angles for the first generation. Hence Eq.~\eqref{dijet_bounds} translates into an upper bound on the elementary/composite mixing angle of
\beq
\sin\phi_{1,4}^u\lesssim 0.5\times \left(\frac{f}{600\,{\rm GeV}}\right)^{1/2}\,.
\eeq
We conclude that a partially composite $c_R$ is not constrained by current dijet data, while the latter allows for a large elementary/composite mixing for $u_R$.

%% file: results.tex
We present in this section the LHC bounds on non-degenerate fermionic partners resulting from the analysis outlined in the previous sections. 
These are the main results of the paper.
We report two types of bounds:
\begin{itemize}
\item Bounds from QCD pair production, which are model independent and are the same for all generations;
\item Bounds from single production, which are model dependent and carry a very strong flavor dependence since the corresponding production mechanisms are based on either valence or sea quarks.
\end{itemize} 

We consider both scenarios where the right-handed up and charm quarks are either partially or fully composite fermions, as described in Section~\ref{sec:mod}. Since there is currently no search in the Higgs boson plus jets final state  probing the existence of SO(4) singlet $\tilde U$ partners, we only focus on bounding light SO(4) fourplet states.
For simplicity we thus assume a limit where the singlet states are decoupled from the low-energy effective theory, $M_1\to \infty$. Hence, only  the custodial triplet, made of $U_{p}$, $D$,
and $X_{5/3}$, and the custodial singlet $U_m$ partners are present for the first two generations. We discuss in Section~\ref{sec:res5plet} how additional light singlet resonances impact  bounds on the fourplet states. 

In the following we denote the first generation fourplet and singlet partners as $U_{p,m}$, $D$, $X_{5/3}^u$ and $\tilde U$, while we use $C_{p,m}$, $S$, $X_{5/3}^c$ and $\tilde C$ symbols for the corresponding second generation states. 
With only light fourplet resonances, the interaction structure of the models defined in Section~\ref{sec:mod}  considerably simplifies. This allows
for a complete survey of the parameter space, which consist of  the compositeness scale $f$, the mass scale of the
fourplet $M_4$, and the mixing between the would-be SM right-handed quarks with the composite dynamics. The mixing is characterized by the elementary/composite mixing parameter $y_R$ in Eq.~\eqref{eq:pcL2} for partially composite SM quarks, while it is
parameterized by the dimensionless couplings $c_1$ in Eq.~\eqref{fcLag} for fully composite SM quarks.
We introduce the index $x=u,c$ to distinguish fundamental parameters of the first and second generations, and we refer to the latter as $y_R^x$, $c_1^x$ and $M_4^x$ in the following.

We also choose to set $f=600\,$GeV for concreteness. This low scale can be in tension with EW and Higgs precision measurements in some specific CHM realizations. Nevertheless,
bounds on the fourplet states are not very sensitive to the symmetry breaking scale $f$ since the fourplet interactions with the Higgs and EW gauge bosons do not arise from EWSB. Furthermore EWSB effects  enter at $\mathcal{O}(v^2/f^2)\sim\mathcal{O}(20\%)$, so that we do not expect bounds on fourplet states to significantly change for larger $f$ scales.
 
\subsection{Exclusion limits from QCD pair production}\label{sec:pairprod}
\begin{figure}[tb!]
\centering
\includegraphics[width=0.5\textwidth]{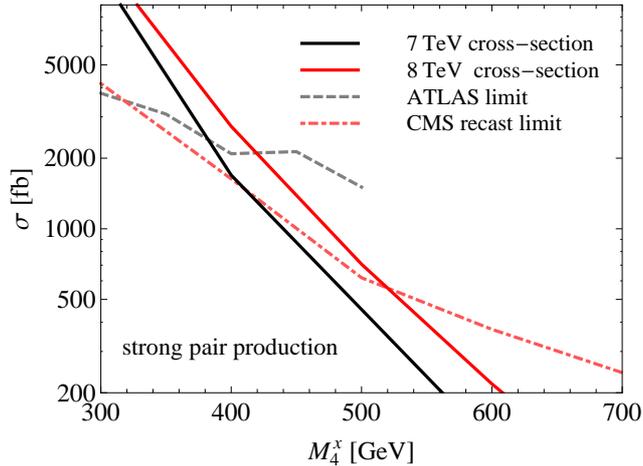}
\caption{Model Independent predictions for $WWjj$ cross sections through QCD pair production of $-1/3$ and $5/3$ charge partners of the composite right-handed up and charm quarks. The solid black (red) line stands for 
the $7\,$TeV ($8\,$TeV) cross section. They are the same for the first two generations and in both partially and fully quark scenarios.
The dashed black line represents the strongest 95$\%$ CL exclusion limit available on this channel coming from recasting the leptoquark 
CMS search or Ref.~\cite{cms8lepto} based on 8 TeV data, while the dashed red line corresponds to the 95$\%$ CL exclusion limit from
the ATLAS search of Ref.~\cite{Aad:2012bt} based on 7 TeV data.}
\label{fig:pairres}
\end{figure}

The ATLAS $WWjj$ analysis search of Ref.~\cite{Aad:2012bt} based on 7 TeV data  excludes up and charm fourplet partner masses up to $M_4^{u,c}\gtrsim390\,$GeV at 95$\%$ CL. These bounds are similar to those obtained from the Tevatron data (see Appendix~\ref{sec:app4}).
Recasting the leptoquark CMS search or Ref.~\cite{cms8lepto} based on 8 TeV data exclude fourplet partner up to 
\begin{equation}
M_4^{u,c}\gtrsim 530\,\rm GeV\,,
\end{equation}
at 95$\%$ CL. 
Note that, despite smaller efficiencies, the limit from this recast is stronger than those derived from more dedicated searches at ATLAS, as the former are based on much less luminosity. We also stress that adjusting the cuts on the $\mu^+\mu^-jj$ channel in order to optimize the sensitivity to first and second generation quark partners should result in stronger bounds. 
The model-independent bounds are shown in Fig.~\ref{fig:pairres}, assuming the resonances are only produced in pairs through QCD interactions. 

\subsection{Exclusion limits from single production}\label{sec:singleprod}
We now move to describe the exclusion limits on the fourplet partners from single production in the partially and fully composite quark cases. We assume here also that the singlet partners are decoupled.
The relevant parameters in this case are the fourplet masses and the corresponding level of right-handed quark compositeness $y_R^{u,c}$ in the partially composite case, or the coefficient of the flavor dependent $d_\mu$-term $c_1^{u,c}$ which specifies the coupling of the SM composite light quarks to the fourplet partners in  the fully composite case.
 In order to illustrate the relative impact of the searches we focus here on a benchmark point with $y_R^x=1$ and $c_1^x=1$. We discuss the implications of varying these parameters in the following subsection, in which we combine all existing bounds in order to derive the strongest available direct constraints as functions of the fundamental parameters $M_4^x$ and $y_R^x$ or $c_1^x$.

We only show in this part the strongest exclusion limits on the model parameters obtained by the ATLAS and CMS collaborations. We refer the dedicated reader to Appendix~\ref{sec:app4} for a detailed presentation of all exclusion limits obtained from the direct searches considered in Section~\ref{sec:searches}.
As the relevant analyses from ATLAS and CMS are quite different and subject to different type of systematics we summarize them separately. 
 
\subsubsection{ATLAS bounds from $7\,$TeV data}
\begin{figure}[tb!]
\centering
\begin{tabular}{cc}
\includegraphics[width=0.47\textwidth]{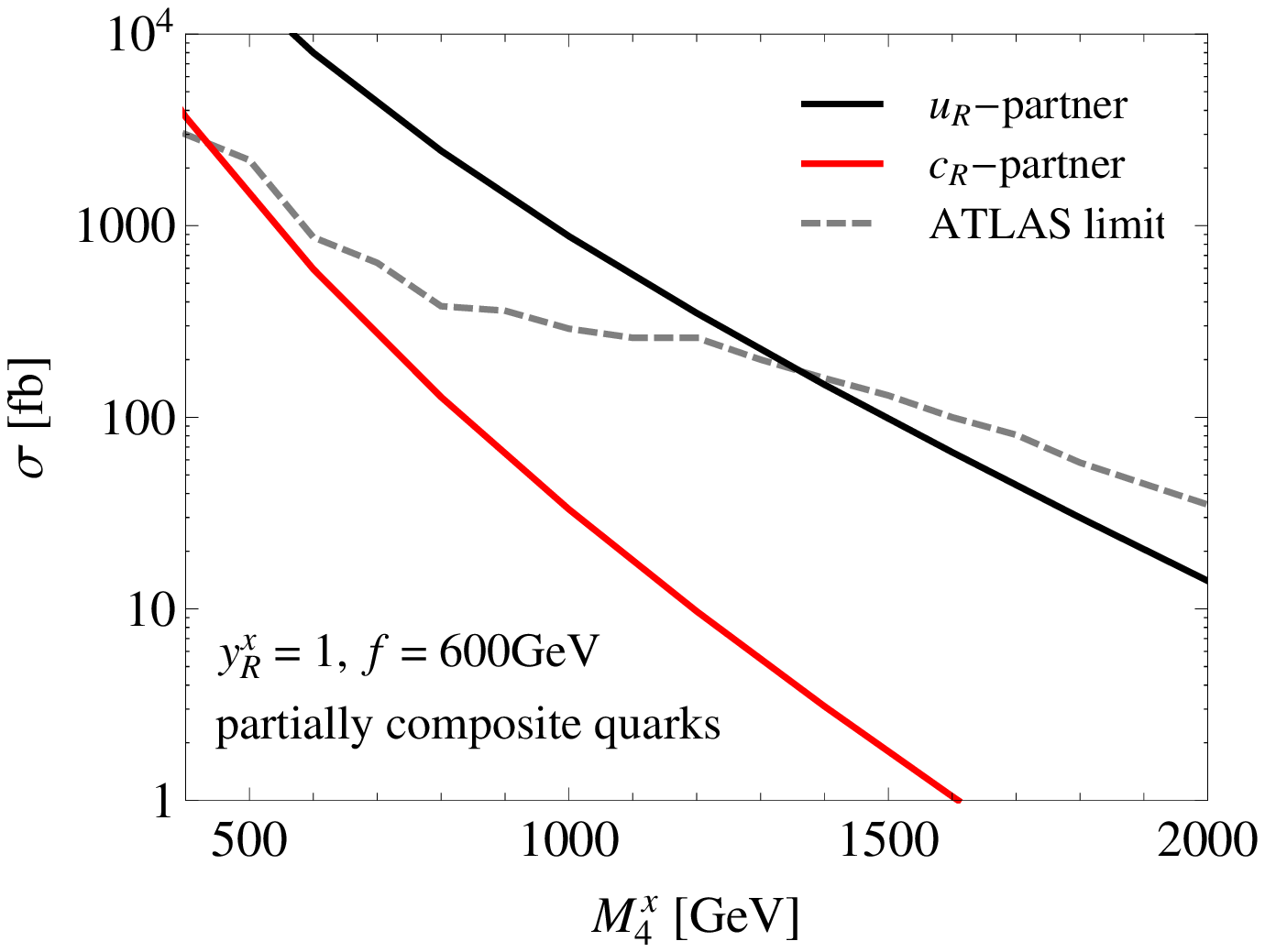}&
\includegraphics[width=0.47\textwidth]{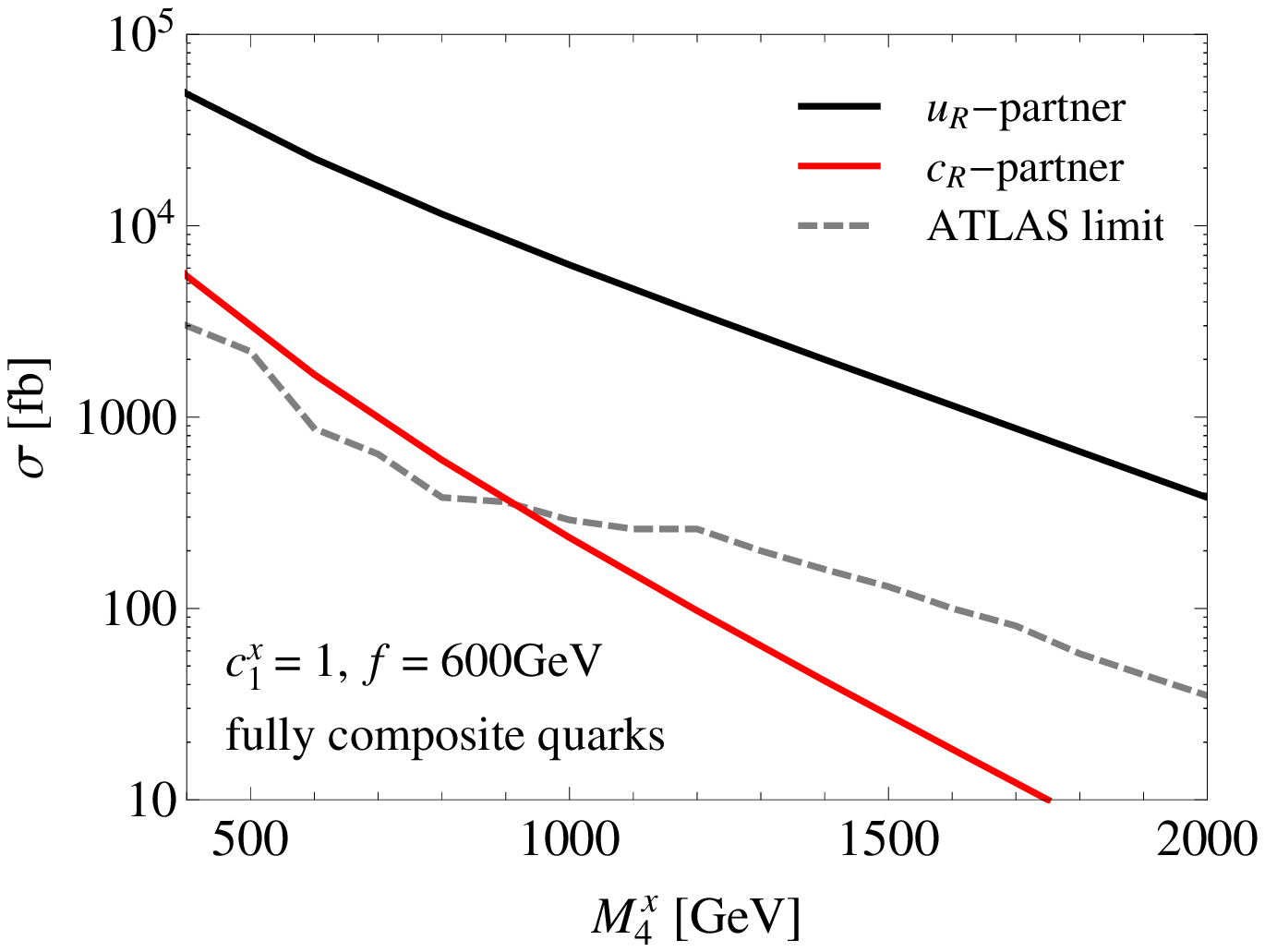}
\end{tabular}
\caption{Predictions for $Wjj$ cross sections as a function of the fourplet partner mass $M_4^x$, $x=u,c$, in the partially (left) and fully (right) composite right-handed for two generation quarks. The solid black (red) lines denote the cross section from $D$ and $X_{5/3}^u$ ($S$ and $X_{5/3}^c$) single production, while the dashed curve is the 95$\%$ CL exclusion limit from the ATLAS search of Ref.~\cite{ATLAS:pp} at the $7\,$TeV LHC run.}
\label{fig:atlassing}
\end{figure}

We first consider bounds from ATLAS analyses at the $7\,$TeV LHC~\cite{ATLAS:pp,Aad:2012bt}. The strongest bound arises from $Wjj$ final states analysed in Ref.~\cite{ATLAS:pp}. Figure~\ref{fig:atlassing}
shows that fourplet up partners are excluded up to 
\begin{equation}
M_4^u\gtrsim 1.4\,\rm TeV\,
\end{equation}
 at  95 $\%$ CL in partially composite models with $y_R^u=1$. The $Zjj$ cross section measurement also constrains the existence of up partners. However, since $Wjj$ final states receive contributions from both 
$X_{5/3}^u$ and $D$ partners, larger cross sections are expected relative to the $Zjj$ channel which receives contributions from $U_p$ production only. 
Moreover, as the current experimental limits
on $Wjj$ and $Zjj$ final states are comparable, the bound on the fourplet mass is dominated by the $Wjj$ channel.
The $Wjj$ channel is also the most sensitive probe of second generation  partners through single production of $S$ and $X_{5/3}^c$. The resulting bound on the fourplet mass is 
\begin{equation}
M_4^c\gtrsim 420\,\rm GeV\,
\end{equation} 
at 95$\%$ CL.  
The cross section for single $C_p$ production are just below present limits in the $Zjj$ channel for $y_R^c=1$ (see Fig.~\ref{fig:lhc7} in the Appendix). Besides, there is no limit from the $ZZjj$ channel sensitive to double production of $C_p$. Hence, ATLAS is most likely not directly probing the existence of this state.

For a fully composite right-handed up and charm quarks the strongest bounds on the partners also come from the  $Wjj$ channel. For $c_1^u=1$, the ATLAS limit on the $Wjj$ cross section excludes at 95$\%$ CL the presence of light first generation fourplet partners up to 
\begin{equation}
M_4^u\gtrsim 2\,\rm TeV\,,
\end{equation}
  while second generation partners as light as 
  \begin{equation}
  M_4^c\gtrsim 950\,\rm GeV\,,
  \end{equation} 
  are allowed at 95$\%$ CL.
 
\subsubsection{CMS bounds from $8\,$TeV data}
The CMS $W/Z$-tagged and dijet measurement of Ref.~\cite{CMS:wzdijet} yield the
most stringent constraint on our scenario. 
Single production of both $-1/3$ and $5/3$ charge partners modifies the $Wjj$ cross section, whereas $Zjj$ final states are produced  only through single production of  $2/3$ states, leading to weaker constraints.

\begin{figure}[tb!]
\centering
\begin{tabular}{cc}
\includegraphics[width=0.47\textwidth]{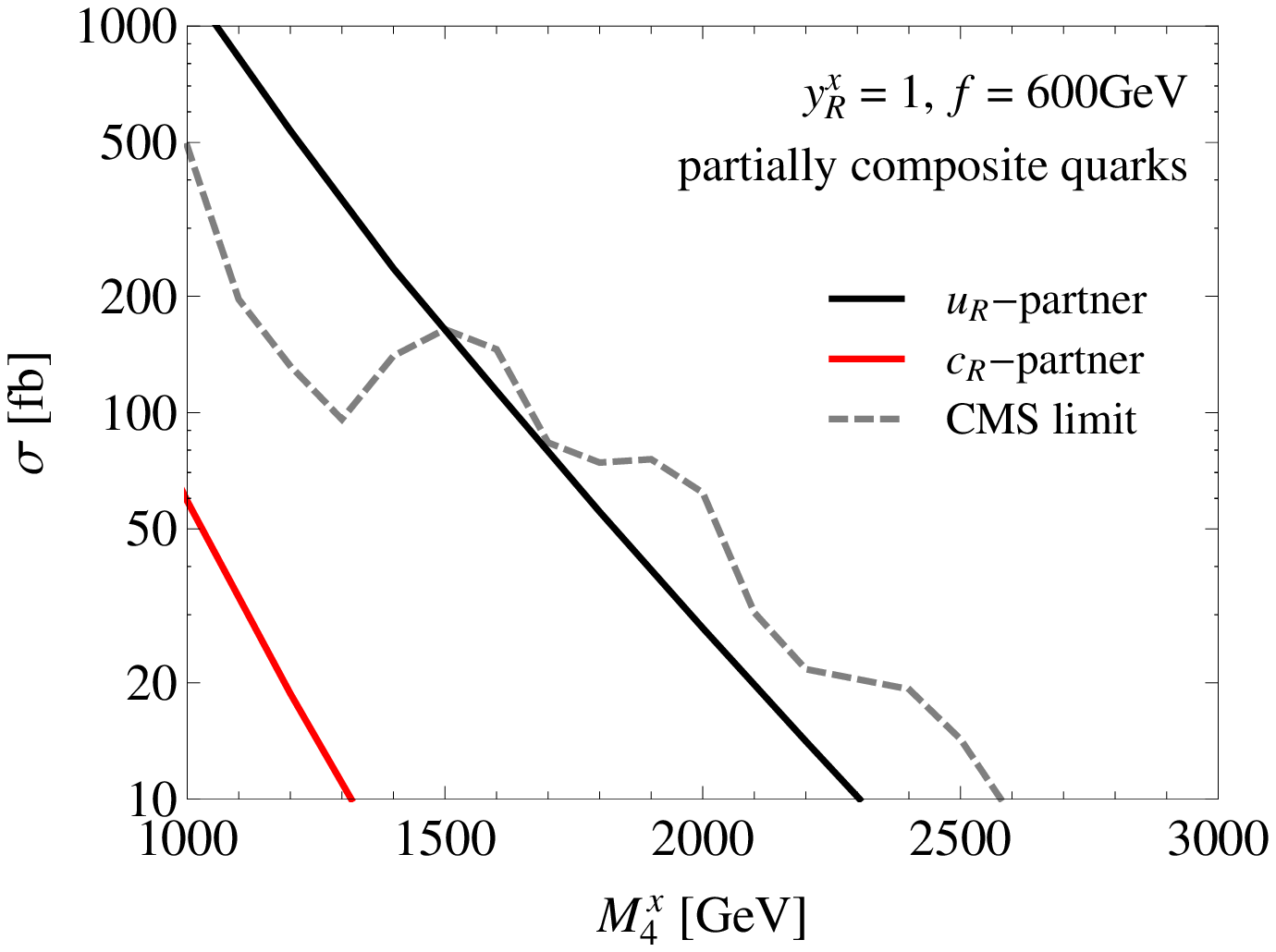}&
\includegraphics[width=0.47\textwidth]{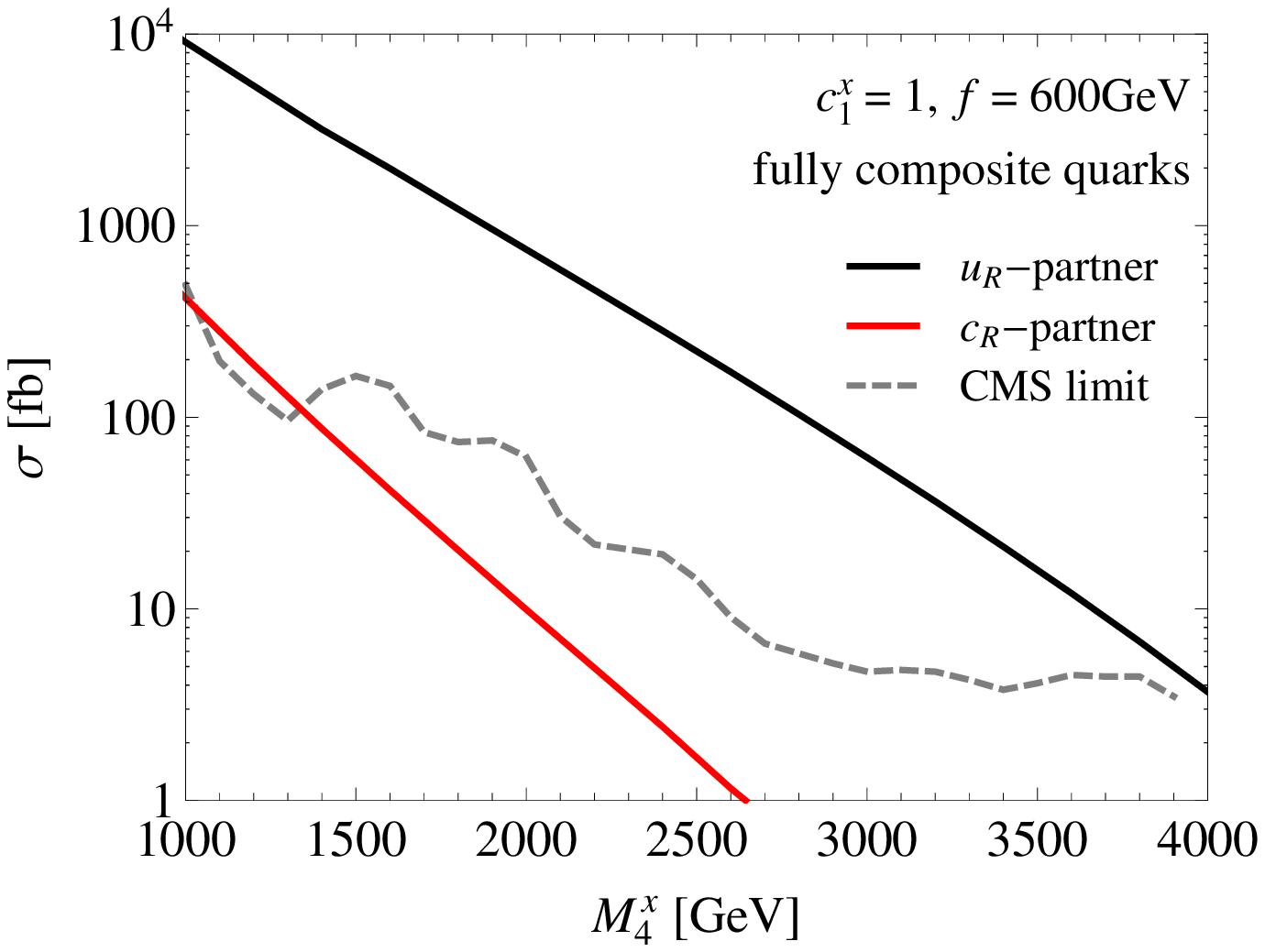}
\end{tabular}
\caption{Predictions for $Wjj$ cross sections as a function of the fourplet partner mass $M_4^x$, $x=u,c$, in the partially (left) and fully (right) composite right-handed for two generation quarks. The solid black (red) denote the cross section from $D$ and $X_{5/3}^u$ ($S$ and $X_{5/3}^c$) single production, while the dashed curve is the 95$\%$ CL exclusion limit from the CMS $W/Z$-tagged dijet search of Ref.~\cite{CMS:wzdijet} at the $8\,$TeV LHC run.}
\label{fig:cmssing}
\end{figure}

In the partially composite case, the measured $Wjj$ cross section constrains the mass of the first generation fourplet partner to 
\begin{equation}
M_4^u\gtrsim 1.7\,\rm TeV\,,
\end{equation}
 at 95$\%$ CL for $y_R^u=1$, which is the strongest
bound obtained from current existing searches in this scenario.
The corresponding bound from the $Zjj$ cross section is $M_4^u\gtrsim 1.4\,$TeV at 95$\%$ CL, which is stronger than the ATLAS bound from $7\,$TeV data in the $Zjj$ channel. For $y_R^c=1$, the existence of a light fourplet partner of a partially composite right-handed charm quark is not probed by the $W/Z$-tagged dijet analysis, due to cross sections significantly smaller  than the present experimental sensitivity. 
The $Wjj$ cross section and the corresponding experimental limits are shown in 
Fig.~\ref{fig:cmssing}.

In fully composite scenarios with $c_1^u=c_1^c=1$, the corresponding 95$\%$ CL bounds are 
\begin{equation}
M_4^u\gtrsim 3.9\,\rm TeV\,,
 \end{equation}
 and 
 \begin{equation}
 M_4^c\gtrsim 1.3\,\rm TeV\,,
 \end{equation}
  for first and second generation partners, respectively. 
Note that resonances are no longer narrow for $c_1^x=1$, with width over mass ratios exceeding $30\%$ for resonances above $2.3\,$TeV. Hence,
these bounds are to be taken with a grain of salt as the search efficiency may be significantly reduced in this case. 
They are nonetheless informative and illustrate the constraining power of the $W/Z$-tagged dijet search relative to the other final states.\\

In conclusions of this part, we find that current constraints on the fourplet partners of the first two generation quarks are
dominated by $Wjj$ searches for single production signals (though $Zjj$ searches are not far behind) and the leptoquark search in $WWjj$ final state
 for pair production signals.
Note that, despite the larger cross sections, bounds from $8\,$TeV data are only slightly more stringent than those from $7\,$TeV data. Besides the larger integrated luminosity at the $8\,$TeV LHC run, this results from the absence of forward jet requirement in the CMS $W/Z$-tagged dijet analysis. Indeed, as a forward jet is almost always radiated in single production of heavy-quark partners
the sensitivity of the $W/Z$-tagged dijet search is significantly reduced relative to $7\,$TeV searches. Note also that the presence of a light charge $2/3$ charm partner $C_p$ is not directly constrained by any existing searches for $y_R^c=1$ or $c_1^c=1$, because the experimental sensitivity to singly produced $C_p$ is currently too small and there is no available
pair production analysis of $ZZjj$ final state.

\subsection{Summary: combined limits on non-degenerate light partners}\label{sec:rescom} 

In this part we combine the bounds from all existing searches in order to derive the strongest 
limits on light fermionic partners of partially and fully composite right-handed up and charm quarks.
In the analysis presented below we also show the impact on varying the model parameters  $y_R^x$ and $c_1^x$, which were kept fixed in Section~\ref{sec:singleprod}.

\begin{figure}[tb!]
\centering
\begin{tabular}{cc}
\includegraphics[width=0.47\textwidth]{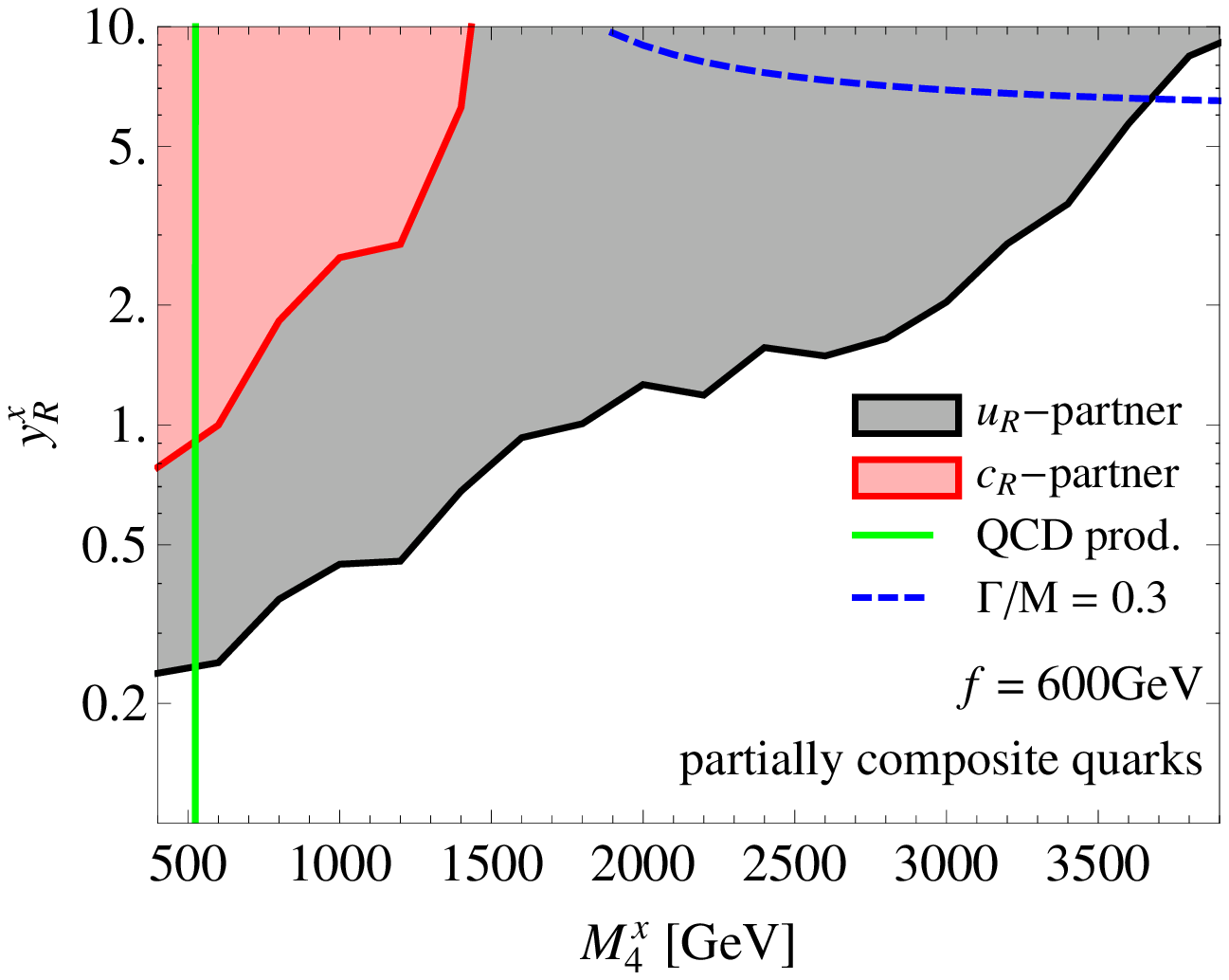}&
\includegraphics[width=0.47\textwidth]{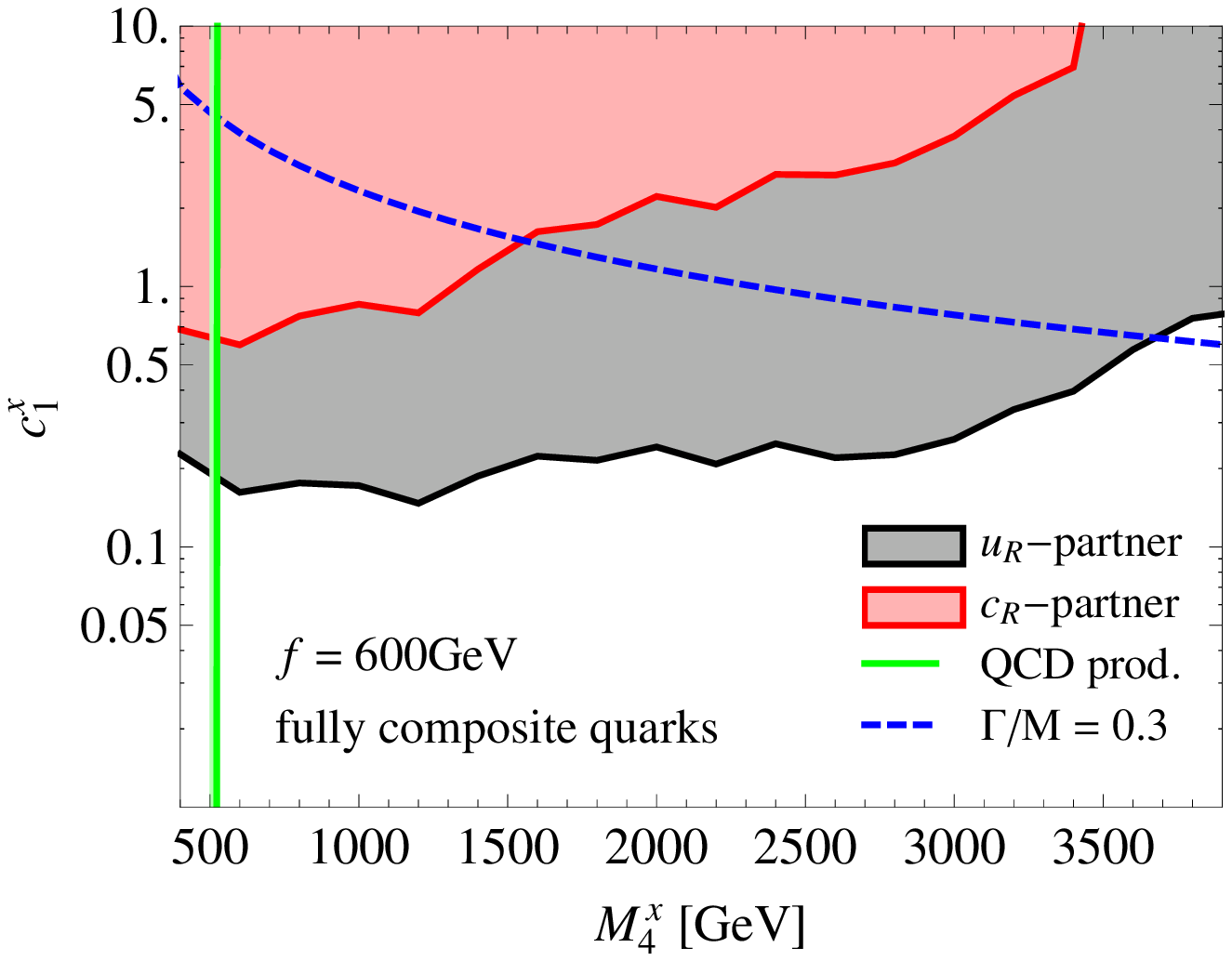}
\end{tabular}
\caption{Combined 95$\%$ CL exclusion limits in the parameter space of partially (left) and fully (right) composite up and charm quark singlets.
$M_4^x$, $x=u,c$, are the masses of the fourplet resonances mixing with the SM up and charm right-handed quarks. $y_R^x$  ($c_1^x$) are the mixing parameters in the partially (fully) composite case. Limits are derived using a $\chi^2$ analysis based on Eq.~\eqref{chi2_indiv}.
The solid black (red) line corresponds to the combined 95 $\%$ CL exclusion limit for the
up (charm) fourplet partner. The green line is the model-independent exclusion limit at 95$\%$ CL from QCD pair production. Shaded regions are excluded. The width to mass ratio of the resonances exceeds 30$\%$ above the dashed blue line.}
\label{fig:yrigc1}
\end{figure}

We first derive the combined 95$\%$ CL exclusion limit for each generation separately. In order to perform this combination  
we build a simple $\chi^2$ function as
\beq\label{chi2_indiv}
\chi^2=\sum_i \frac{\sigma(M_4^x)_i^2}{\Delta(M_4^x)_i^2}\,,
\eeq
for $x=u$ or $c$, where the $i$ index runs over the Tevatron and LHC searches listed in Section~\ref{directsearches}. $\sigma(M_4^x)_i$ is the cross section in the channel $i$ predicted from the existence of a light fourplet partner of mass $M_4^x$, while the standard deviation $\Delta(M_4^x)_i$ is obtained from  
the observed 95$\%$ CL exclusion limit $\sigma(M_4)_i^{95\%{\rm CL}}$ assuming a Gaussian error with zero mean, {\it i.e.} 
$\Delta(M_4)_i\equiv\sigma(M_4)_i^{95\%{\rm CL}}/1.96$. Figure~\ref{fig:yrigc1} shows the combined 95$\%$ CL exclusion contours in the $y_R-M_4$ and $c_1-M_4$ planes for the partially and fully composite scenarios, respectively, resulting from a $\chi^2$ analysis based on Eq.~\eqref{chi2_indiv}.
In the partially composite case the combined 95$\%$ CL bounds for $y_R^x=1$ are 
\begin{equation}
M_4^u\gtrsim 1.8\,\rm TeV\,,
\end{equation} 
and 
\begin{equation}
M_4^c\gtrsim 610\,\rm GeV\,,
\end{equation}
 for up and charm partners, respectively. Corresponding bounds in the fully composite scenario  are 
 \begin{equation} 
 M_4^u\gtrsim 3.9\,\rm TeV\,,
 \end{equation}
  and 
  \begin{equation}
  M_4^c\gtrsim 1.3\,\rm TeV\,,
  \end{equation}
 for up and charm partners, respectively.\\

Reference~\cite{Redi:2013eaa} reported stringent bounds on the right-handed charm (and top) partners in cases where both strong dynamics mass parameters and right-handed mixings are flavor universal. These strong bounds are dominantly driven by the first generation partners whose production cross sections at hadron colliders are sustained by relatively large up-quark PDFs. We derive here the bound on right-handed up and charm fourplet partners in a more general setup where the flavor universality assumption is dropped. This can be done in splitting either the mixing parameters, the strong dynamics masses or both. For simplicity we only consider below the former two cases. A careful study of the most general case where both mixings and masses are flavor non-universal, albeit interesting on its own, would require a rather involved statistical analysis which is far beyond the scope of this work. Hence, we first assume the multiplets from the strong dynamics are not degenerate, $M_4^u\neq M_4^c$, but the mixing parameters are still universal, $y_R^u=y_R^c$ or $c_1^u=c_1^c$. We then focus on the other limit where the multiplet are degenerate but the mixing parameters can differ from each other.

\begin{figure}[tb!]
\centering
\begin{tabular}{cc}
\includegraphics[width=0.47\textwidth]{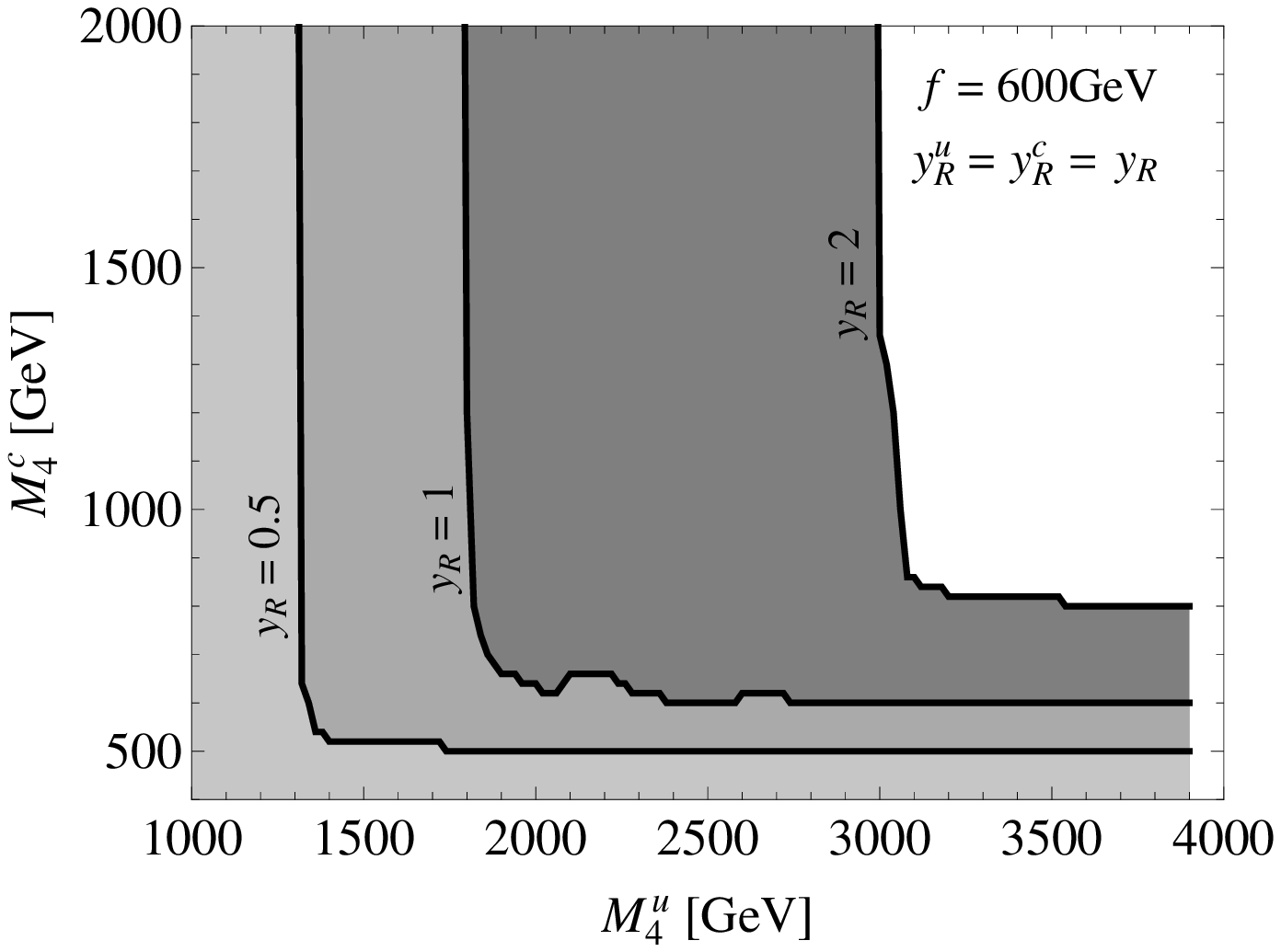}&
\includegraphics[width=0.47\textwidth]{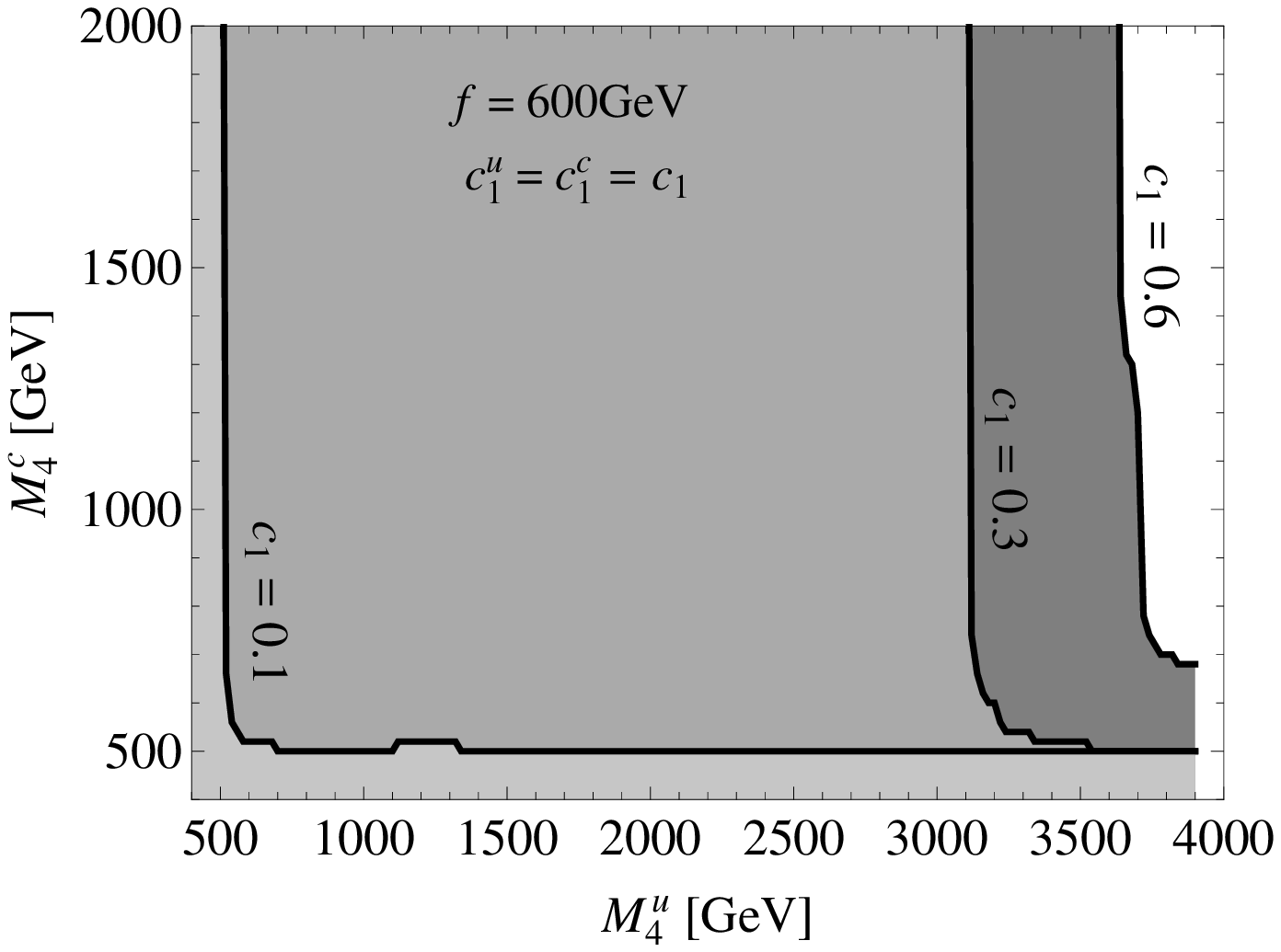}
\end{tabular}
\caption{Combined bound on fourplet partners in the $M_4^u-M_4^c$ plane assuming universal mixing parameters for the first two generation quarks. The solid black lines denote the 95$\%$ CL combined bound for various values of $y_R^u=y_R^c=y_R$ in the partially composite scenario (left) and of $c_1^u=c_1^c=c_1$ in the fully composite scenatio (right). Shaded regions on the left and below the $y_R$ or $c_1$ contours are excluded.}
\label{fig:uvscmass}
\end{figure}

In order to analyse the case where mixing parameters are degenerate, we build a $\chi^2$ function as
\beq\label{chi2_incoherent}
\chi^2= \sum_i \left[\frac{\sigma(M_4^u)_i^2}{\Delta(M_4^u)_i^2}+\frac{\sigma(M_4^c)_i^2}{\Delta(M_4^c)_i^2}\right]\,,
\eeq
where the sum goes over all experimental searches. We explicitly neglect in Eq.~\eqref{chi2_incoherent} possible correlations between the up and charm resonance contributions. We motivate this choice as follows.
Figure~\ref{fig:yrigc1} shows that, when taken individually, up partners are much more severely constrained than charm partners, assuming equal  mixing parameters. Therefore, the $\chi^2$ of Eq.~\eqref{chi2_incoherent} is minimal  generically when the up and charm partner resonances are well separated, $|M_4^u-M_4^c|\gg \Gamma$, so that their respective signals can be added incoherently. Figure~\ref{fig:uvscmass} shows the bounds resulting from a $\chi^2$ analysis based on Eq.~\eqref{chi2_incoherent} for the partially and fully composite quark scenarios. Note that, in particular, up partner masses as high as 1.3$\,$TeV, 1.8$\,$TeV and 3.0$\,$TeV are excluded at the $95\%$ CL for
$y_R=0.5$, $1$ and $2$, respectively, in the partially composite case, and so regardless of the charm partner mass.
Similarly, for fully composite quarks, up partner masses below 530$\,$GeV,  3.1$\,$TeV and 3.6$\,$TeV are excluded at the $95\%$ CL for $c_1=0.1$, $0.3$ and $0.6$, respectively, for any charm partner mass.

\begin{figure}[tb!]
\centering
\begin{tabular}{cc}
\includegraphics[width=0.47\textwidth]{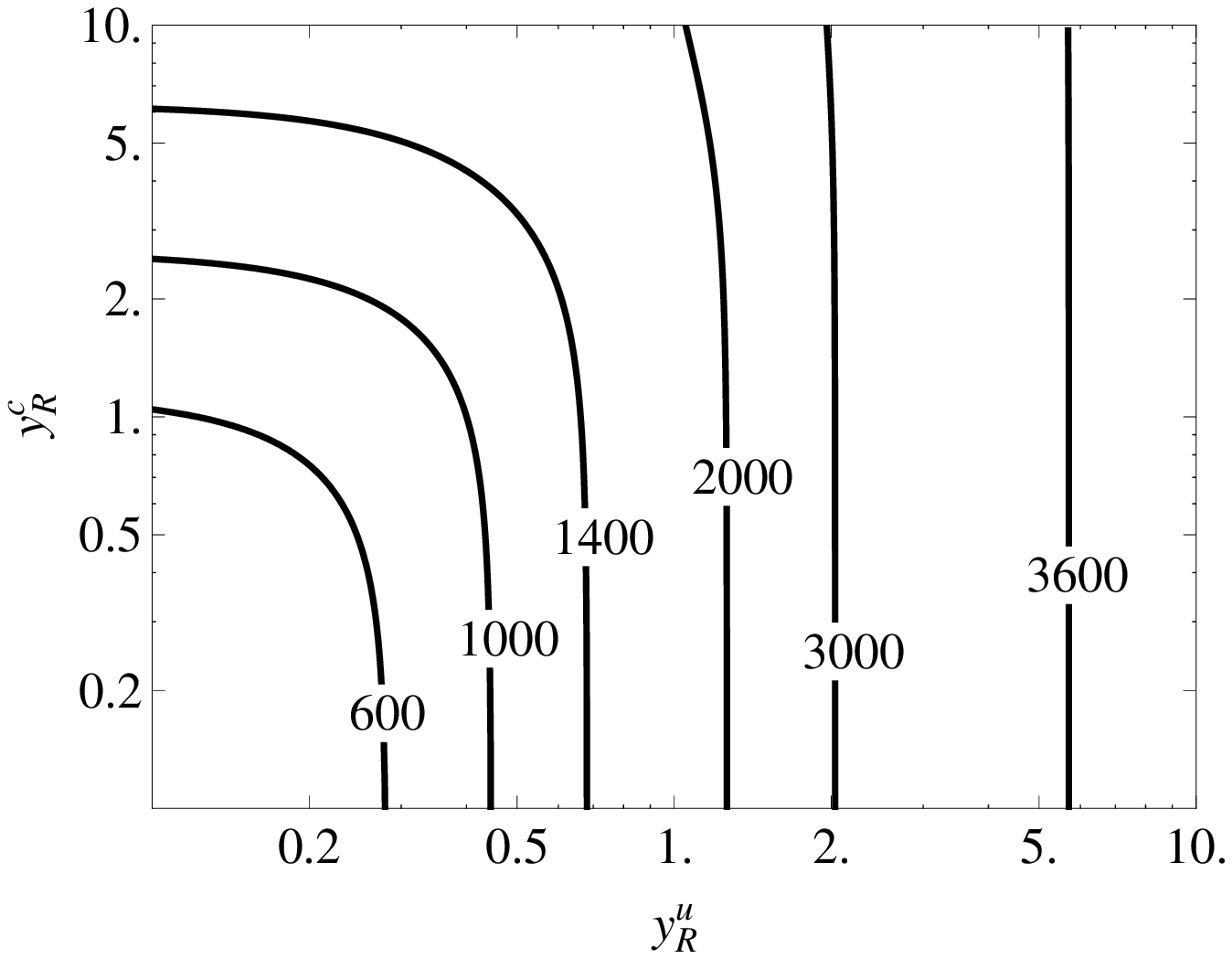}&
\includegraphics[width=0.47\textwidth]{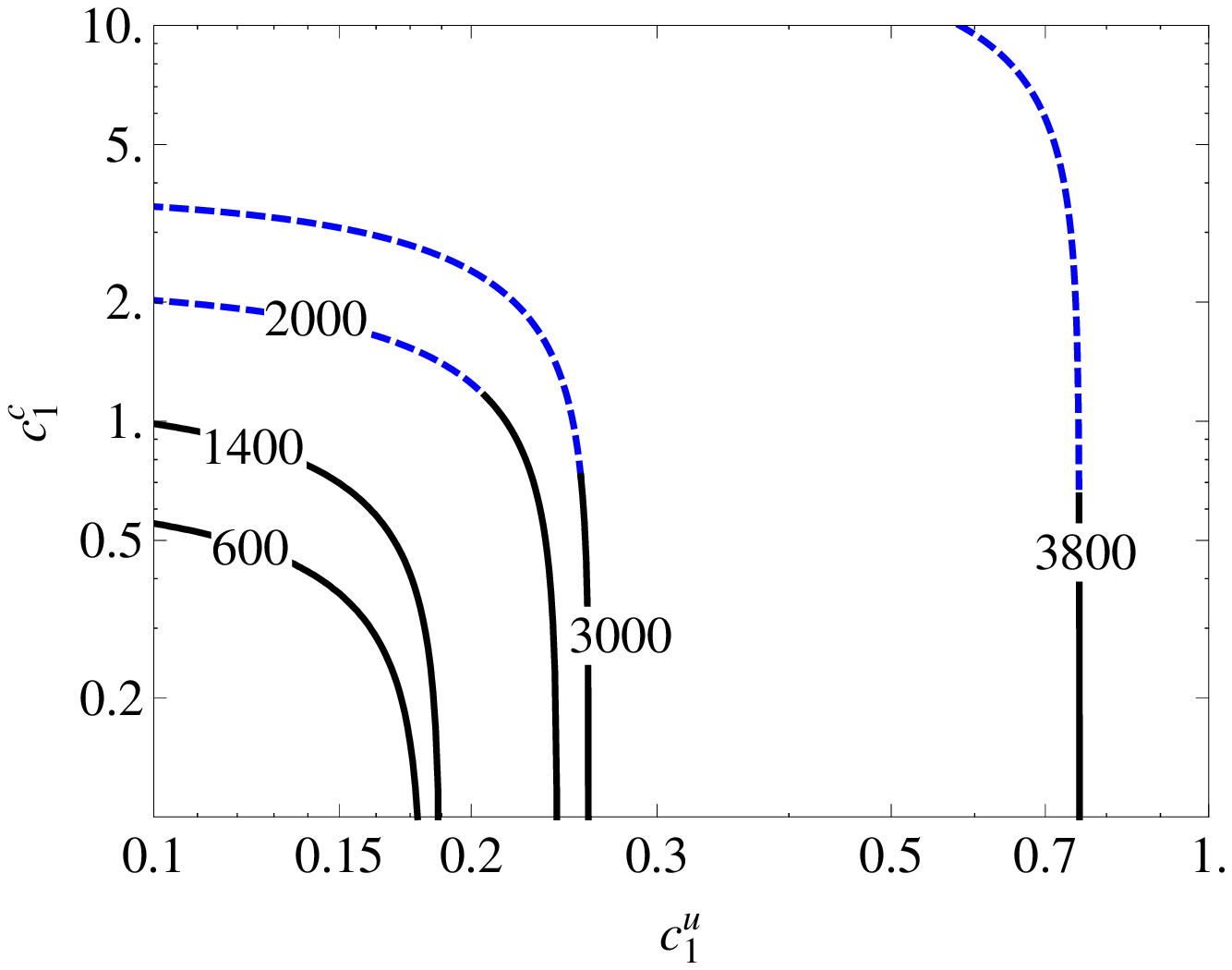}
\end{tabular}
\caption{Combined bounds on fourplet partners in the $y_R^u-y_R^c$ (left) and $c_1^u-c_1^c$ (right) planes for the partially and fully composite quark scenarios, respectively. Lines denote the 95$\%$ CL bounds for universal mass parameters, $M_4^u=M_4^c=M_4$ in units of GeV. Regions above and on the right of the lines are excluded. We denote in dashed blue the region of parameters where at least one resonance width exceeds $30\%$ of its mass.}
\label{fig:uvsccoupl}
\end{figure}

In a limit where the resonances are degenerate, $M_4^u=M_4^c\equiv M_4$ but the mixing parameters are different, we use in place of Eq.~\eqref{chi2_incoherent} the $\chi^2$ function
\beq\label{chi2_coherent}
\chi^2=\sum_i \frac{\left[\sigma(M_4^u)_i+ 
\sigma(M_4^c)_i\right]^2}{\Delta(M_4)^2_i}\,,
\eeq
in order to properly account for  correlations among the up and charm partner  signals. Figure~\ref{fig:uvsccoupl} shows the bounds in the $y_R^u-y_R^c$ and $c_1^u-c_1^c$ planes for various values of $M_4$ in partially and fully composite scenarios, respectively, resulting from a $\chi^2$ analysis based on Eq.~\eqref{chi2_coherent}.
In the partially composite case the combined 95$\%$ CL bounds for $M_4=600\,$GeV are 
\begin{equation}
y_R^u\lesssim 0.3\,,
\end{equation} 
and 
\begin{equation}
y_R^c\lesssim 1\,,
\end{equation}
 for up and charm partners, respectively. Corresponding bounds in the fully composite scenario  are 
 \begin{equation} 
 c_1^u\lesssim 0.2\,,
 \end{equation}
  and 
  \begin{equation}
  c_1^c\lesssim 0.6\,,
  \end{equation}
   for up and charm partners, respectively.

\subsection{Implications of additional light SO(4) singlet partners}\label{sec:res5plet}

We presented above constraints on the fourplet partners in the limit where singlet states were decoupled from the low-energy theory, $M_1^x\to \infty$. We comment here on the implications of having a light singlet close in mass to the fourplet states, $M_1^x\sim M_4^x$. Although these states are not currently directly probed at the LHC, their existence may still affect
production and decay of fourplet states in adequate regions of parameter space.
For illustration we only focus on describing how the existence of an additional singlet partner of the right-handed up quark modifies the fourplet bounds derived previously. Similar considerations apply to charm partners as well. In order to allow transparent comparisons with above results we set $f=600\,$GeV. We also consider for simplicity $c=0$ in the partially composite case. (A detailed study of the impact of $c\neq0$ is beyond the scope of this analysis.)\\

Fourplet bounds are modified through two main effects, which tend to reduce the EW gauge bosons plus jets signals:
\begin{itemize}
\item
$X_{5/3}$, $D$, and $U_p$ states may have reduced branching ratios into $u_R$ and an EW gauge boson. Indeed, 
for sufficiently small $M_1^x$, $X_{5/3}$, $D$, and
$U_p$ can now also decay into the singlet resonance and an EW gauge boson, with the singlet decaying further into a Higgs boson and a jet. This cascade decay leads to  different final
states which escape searches used in order to bound the fourplet parameters, thus weakening the associated constraints. This effect is common to partially and fully composite scenarios.

\item In partially composite models, single production cross sections of $X_{5/3}$, $D$ and $U_p$ are also reduced in the presence of light singlets. In this case $u_R$  mixes with a linear combination of the fourplet state $U_m$ and the singlet $\tilde{U}$. Since only $U_m$ couples to the custodial triplet $X_{5/3}$, $D$ and $U_p$, the coupling of the SM up quark to fourplet states and EW gauge boson is reduced, relative to the limit where the singlet is decoupled. This effect is absent in fully composite models as there is no large mass mixing between $u_R$ and the singlet resonance.
\end{itemize}

\begin{figure}[tb!]
\centering
\begin{tabular}{cc}
\includegraphics[width=0.47\textwidth]{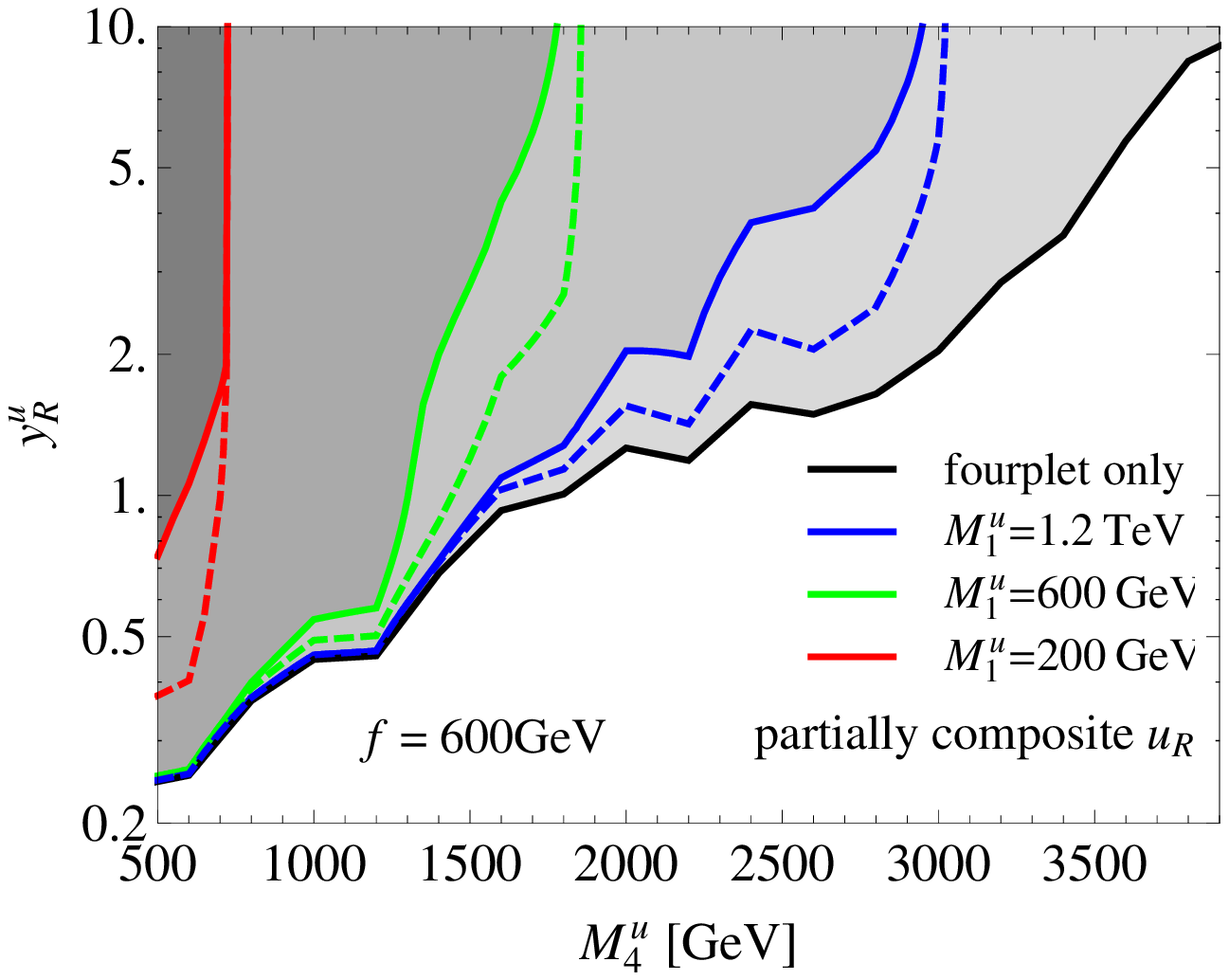}&
\includegraphics[width=0.47\textwidth]{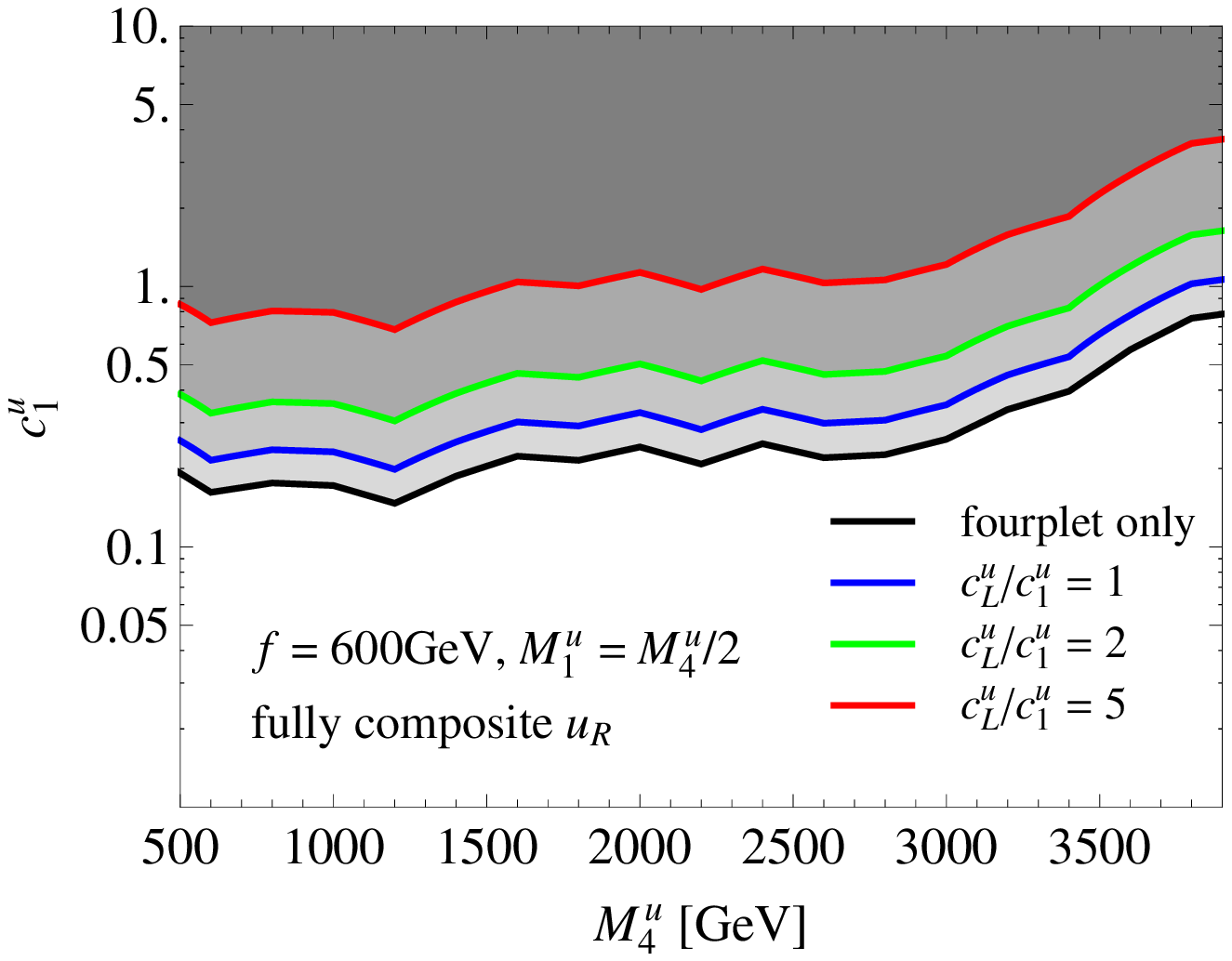}
\end{tabular}
\caption{Bounds on right-handed up quark fourplet partners in the presence of an additional light singlet partner. We set $f=600\,$GeV. (Left) 95$\%$ CL bounds in the $y_R^u-M_4^u$ plane for different values of $M_1^u$ and $c=0$ in partially composite models. Solid lines include both the reduced production cross section of fourplet states and reduced branching due to cascade decays, while dashed lines assumes a $100\%$ branching ratio of $X_{5/3}$, $D$ and $U_p$ into $W/Z+$jet. (Right) 95$\%$ CL bounds in the $c_1^u-M_4^u$ plane for $M_1^u=M_4^u/2$ and different values of $c_L^u/c_1^u$ in fully composite models. ($c_R=c_L$ was assumed for simplicity.)}
\label{fig:yrig_gen}
\end{figure}

Figure~\ref{fig:yrig_gen} shows the quantitative impact of the effects discussed above on $95\%$ CL exclusion limit in the $y^u_R-M_4^u$ and $c_1^u-M_4^u$ planes as a function of $M_1^u$, for partially and fully composite  $u_R$. 
In partially composite models the presence of a light singlet can significantly relax the bound on the fourplet state. For instance, for $y_R^u=1$, the 
$95\%$ CL bound on the fourplet mass from single production channels goes from $M_4^u\gtrsim 1.8\,$TeV for $M_1^u\to \infty$ down to $M_4^u\gtrsim 600\,$GeV for $M_1^u= 200\,$GeV.
It also appears that  the
dominant effect in this case comes from the reduced production cross sections. 
In fully composite models the presence of the extra singlet only reduces EW gauge bosons plus jets signals through eventual cascade decays. The couplings relevant for these decays are found in Eq.~\eqref{funocp}. They depend on the parameters $c_{L,R}$, while
$X_{5/3}$, $D$, and $U_p$ decays are controlled by $c_1$ (see Eq.~\eqref{funo}). The ratio of branching ratios between these two channels scales like $c_{L,R}^2/c_1^2$, so that constraints on the fourplet partner in fully composite models are substantially relaxed
when $M_1^u \lesssim M_4^u+m_{W/Z}$
and  $c_{L/R}\gg c_1$, as shown on the right panel of  Fig.~\ref{fig:yrig_gen}.\\

Several comments are in order. First of all, effects from the cascade decays are only relevant in a small region of parameter space. For  $m_{U_1}\gtrsim M_4^u+m_{W/Z}$, on-shell cascade decay is kinematically forbidden and phase-space suppressed off-shell. For $m_{U_1}\ll M_4^u$ the effects are also negligible. Indeed, in this
regime, although cascade decays would be kinematically allowed, the mass eigenstate $U_1$ almost coincide with the singlet and thus has a suppressed coupling to the custodial triplet states $X_{5/3}$, $D$ and $U_p$. Cascade decays therefore only play a role when
$m_{U_1}\lesssim M_4+m_{W/Z}$.
Note also that $c\neq0$ in partially composite models also affects production cross sections and decays of the fourplet states. In particular, $c<0$ ($c>0$) enhances (further reduces) single production of fourplet states. 
Finally, modifications due to the extra light singlet significantly depend on the value of $f$ in partially composite models. Implications of a change of the latter are however straightforward to estimate as dominant effects are controlled by the $M_1/f$ ratio.

%% file: conclusions.tex
We studied the phenomenological implications of a large degree of compositeness
for the light generation quarks in composite Higgs models. We focused in particular on
 scenarios where the right-handed SM up-type quarks  either have a sizable mixing with the
strong dynamics or are themselves pure composite states. This structure naturally arises for example in models implementing the
minimal flavor violation (MFV) hypothesis, in which the degree of compositeness of the right-handed quarks is comparable and large
for all three generations. We also investigated the more general setup in which
the strength of the elementary--composite mixing of the right-handed quarks is independent
for each family. Among this class of models, the assumption of a large compositeness only for
the right-handed charm quark component and not for the first generation quarks leads to very distinct
phenomenological features which are more challenging to probe experimentally.

For definiteness we analyzed the minimal composite Higgs realizations based on the symmetry
structure ${\rm SO}(5)/{\rm SO}(4)$. For our study we used a general low-energy parametrization
of the strong sector dynamics which only includes the lightest fermionic degrees of freedom
directly connected to the up-type quarks. In particular we considered two light multiplets
of composite resonances which transform as a fourplet and as a singlet under the unbroken
${\rm SO}(4)$ global symmetry.
For our analyses we focused on models based on the standard implementation
of partial compositeness in which each SM fermion is associated with a corresponding elementary component.
In addition we also explored the alternative setup in which the
right-handed up-type quarks are totally composite states and arise as chiral fermions from
the strong dynamics. We found that the phenomenology of these alternative models is in
qualitative agreement with the one of the standard scenarios. At the quantitative level,
however, significant differences are present. In our analysis we can distinguish two simplified frameworks in which only one composite multiplet
is present, namely the case with only a light fourplet and the one with only a light singlet.
We then derived the exclusion bounds on the masses of the resonances using the
current LHC results.\\

In the setup with only a fourplet the spectrum of the resonances for each generation
is given by two nearly degenerate $\rm{SU}(2)_L$ doublets
and contain two charge $2/3$ states, $U_{p,m}$, one state with charge $-1/3$, $D$,
and an exotic quark with charge $5/3$, $X_{5/3}$.
In cases where only the first generation quarks are composite, we found that single production
typically yields the dominant constraint.
In this case the strongest bounds come from searches of the exotic state $X_{5/3}^u$
and of the $D$ partner. The production of these two resonances contribute to the same final
state through the process $pp \rightarrow Dj + X_{5/3} j \rightarrow W jj$. 
In partially composite quarks scenarios, the combination of the $7\,{\rm TeV}$
and $8\,{\rm TeV}$ LHC results for this channel sets the tight bound $m_{X^u_{5/3}} = m_{D} \gtrsim 1.8\,{\rm TeV}$
for the benchmark configuration with a right-handed mixing $y_R^u = 1$ and $f = 600\,{\rm GeV}$.
Notice that $y_R^u \gtrsim 1$ is a necessary condition in models with three-generation universality due to the requirement
of reproducing the large top mass.
For higher values of the mixing the bounds become much stronger and reach values as high as
 $m_{X_{5/3}^u} = m_{D} \gtrsim 3\,{\rm TeV}$, for $y_R^u \gtrsim 2$.
Another interesting channel is the production of one charge
$2/3$ state which afterwards decay in
the $Z$ boson plus a jet: $pp \rightarrow U_1 j \rightarrow Zjj$. The $8\,{\rm TeV}$ LHC data
set a lower bound on the $U_p$ mass $m_{U_p} \gtrsim 1.4\,{\rm TeV}$ for the benchmark
scenario with $y_R = 1$.
Finally, if the compositeness is smaller, $y_R^u \lesssim 0.25$, the bounds from QCD pair production
become relevant. The $8\,{\rm TeV}$ LHC data sets a model-independent lower
bound $m_{X_{5/3}^u} = m_D \gtrsim 530\,{\rm GeV}$.
The phenomenology of the fully composite light quarks scenario is very similar
to that of partially composite light quarks. We have shown that for a coupling $c_1=1$
single production searches at the LHC have excluded the existence of partners for masses of almost all the range
considered, which extends up to 3.9 TeV.
The situation is fundamentally different in models where only the second generation quarks are composite. We found that 
the right-handed charm component
can be relatively composite with their partners being light. 
In this case
the single production channels are suppressed with respect to the case of first generation partners.
From the combination of the 7 and 8 TeV LHC data analyses we obtain that the strongest exclusion limits
on partially (fully) composite charm with $y_R^c=1$ ($c_1^c=1$) extend up to 610 (1300) GeV.
Bounds from QCD pair production
are also relevant. As in the previous case the model-independent lower
bound is $m_{X^c_{5/3}} = m_S \gtrsim 530\ {\rm GeV}$, irrespectively of the value of the couplings.
In order to further highlight the strong differences in the exclusion limits  in the cases where only the first or second
generation partners are present, we have also shown the exclusion bounds when partners
of both generations are present at the same time in the spectrum. We showed the exclusion limits
in the coupling plane $y_R^u-y_R^c$ ($c_1^u-c_1^c$) for
the partially (fully) composite case, assuming the same mass for both generations, as well as
the bounds in the mass plane $M_4^u-M_4^c$, assuming the same coupling
for both generations.

In the other simplified scenario with only a light singlet, the
spectrum of the resonances contains only a charge $2/3$ state.
One peculiarity of this set-up is the fact that the composite resonances are coupled with the
light fermions only through couplings involving the Higgs boson. This implies that
it can be singly produced only in association with a Higgs boson and that it almost always decays
into a Higgs boson plus a jet. QCD pair production is the dominant production mechanism and
leads to the signal $pp \rightarrow \widetilde U \widetilde U \rightarrow hhjj$, which is
very challenging at the LHC. The current experimental analyses did not focus on
this channel, thus there are currently no bounds on the mass of the resonance in this scenario.\\

The results in the simplified models with only one light composite multiplet motivated
an extension of our analysis to include a third scenario in which both a light fourplet
and a light singlet are present. We have seen that in the cases where $M_1$ and $M_4$ are
of a similar size the phenomenology of the model and thus the present exclusion limits are very similar
to the ones in the fourplet case, the main difference being a suppression of the relevant
couplings which slightly relaxes the bounds obtained in the simplified case with only a light fourplet. With large mass splitting between
the fourplet and the singlet or with new terms switched on in the Lagrangian cascade decays will be allowed for certain values
of the model parameters, leading to weaker exclusion
limits and the opening of new channels yet to be explored.

%% file: appendix1.tex
We define here notations used in the main text as well as collect some useful
expressions relative to the ${\rm SO}(5)/{\rm SO}(4)$ coset. Most of our notation follows that of Ref.~\cite{DeSimone:2012fs}.

The 10 generators of ${\rm SO}(5)$ generators in the fundamental representation are written as
\beq
(T^\alpha_{L})_{IJ} = -\frac{i}{2}\left[\frac{1}{2}\varepsilon^{\alpha\beta\gamma}
\left(\delta_I^\beta \delta_J^\gamma - \delta_J^\beta \delta_I^\gamma\right) +
\left(\delta_I^\alpha \delta_J^4 - \delta_J^\alpha \delta_I^4\right)\right]\,,\nonumber
\eeq
\beq
(T^\alpha_{R})_{IJ} = -\frac{i}{2}\left[\frac{1}{2}\varepsilon^{\alpha\beta\gamma}
\left(\delta_I^\beta \delta_J^\gamma - \delta_J^\beta \delta_I^\gamma\right) -
\left(\delta_I^\alpha \delta_J^4 - \delta_J^\alpha \delta_I^4\right)\right]\,,
\label{eq:SO4_gen}
\eeq
\begin{equation}
T^{i}_{IJ} = -\frac{i}{\sqrt{2}}\left(\delta_I^{i} \delta_J^5 - \delta_J^{i} \delta_I^5\right)\,,
\label{eq:SO5/SO4_gen}
\end{equation}
where $I,J=1,\ldots ,5$.
The above basis is convenient because it explicitly isolates the 6 unbroken generators $T^{\alpha}_{L,R}$ ($\alpha=1,2,3$) of the $\textrm{SO}(4) \simeq \textrm{SU}(2)_L \times \textrm{SU}(2)_R$ subgroup from the broken ones $T^{i}$ ($i=1,\ldots, 4$), associated with the
coset $\textrm{SO}(5)/\textrm{SO}(4)$.
The generators in eqs.~(\ref{eq:SO4_gen}) and (\ref{eq:SO5/SO4_gen}) are normalized
such that ${\rm Tr}[T^AT^B]=\delta^{AB}$. It is convenient to collectively denote $T^\alpha_{L,R}$ as $T^a$ ($a=1,\ldots,6$), where $T^{1,2,3}=T^{1,2,3}_L$ and $T^{4,5,6}=T^{1,2,3}_R$. In the basis of Eq.~\eqref{eq:SO4_gen}, $T^a$ are bock-diagonal
\begin{equation}
T^a=\left(\begin{array}{cc}t^a &0 \\ 0 &0 \end{array}\right)\,,
\end{equation}
where $t^a$ are the 6 ${\rm SO}(4)$ generators in the fundamental representation of ${\rm SO}(4)$.

The explicit form of the Goldstone matrix as a function of the Goldstone
fields $\Pi_i$ is
\beq
U_{gs}= U_{gs}(\Pi) = \exp\left[i \frac{\sqrt{2}}{f} \Pi_{i} T^{i}\right]=\left(\begin{matrix}
\displaystyle \textbf{1}_{4\times 4}-{\vec\Pi\vec\Pi^T\over \Pi^2} \left(1-\cos{\Pi\over f}\right)\hspace{1em}&
\displaystyle {\vec\Pi\over \Pi}\sin{\Pi\over f}\\
\rule{0pt}{2.em}\displaystyle -{\vec\Pi^T\over \Pi}\sin{\Pi\over f}&\displaystyle \cos{\Pi\over f}\\
\end{matrix}\right),
\label{gmatr}
\eeq
where $\vec{\Pi}\equiv (\Pi_1,\Pi_2,\Pi_3,\Pi_4)^T$ and $\Pi\equiv\sqrt{\vec{\Pi}\cdot\vec{\Pi}}$.
In unitary gauge the Goldstone multiplet reduces to
\beq
\vec{\Pi}=\left(\begin{matrix} 0 \\ 0\\ 0\\ \bar{h} \end{matrix}\right)
\label{unigauge}
\eeq
with $\bar{h}=v + h$, which yields the expression in Eq.~\eqref{gmatrU} for the $U_{gs}$ matrix.
The components of the CCWZ $d_\mu$ and $e_\mu\equiv e_\mu^at^a$ symbols are 
\begin{eqnarray}
d_\mu^{\,i}&&=\sqrt{2}\left(\frac{1}{f}-\frac{\sin{\frac{\Pi}{f}}}\Pi\right)\frac{\left(\vec{\Pi}\cdot \nabla_\mu\vec{\Pi}\right)}{\Pi^2}\Pi^{i}+\sqrt{2}\,\frac{\sin{\frac{\Pi}{f}}}\Pi\nabla_\mu\Pi^{i}\,,\nonumber\\
e_{\mu}^a&&=-A_\mu^a+4i\sin^2\left({\frac{\Pi}{2f}}\right)\ \frac{\vec{\Pi}^T t^a\nabla_\mu\vec{\Pi}}{\Pi^2}\,.
\label{dande}
\end{eqnarray}
$\nabla_\mu\Pi$ is the derivative of the Goldstone fields $\Pi$ ``covariant'' under the EW gauge group,
\begin{equation}
\nabla_\mu\Pi^{i}=\partial_\mu\Pi^{i}-i A_\mu^a\left(t^a\right)^{i}_{\  j}\Pi^{ j}\,,
\end{equation}
where $A^a_\mu$ contains the  elementary SM gauge fields written in an ${\rm SO}(5)$ notation
that is
\bea
A_\mu^a T^a &=& \frac{g}{\sqrt{2}}W^+_\mu\left(T_L^1+i T_L^2\right)+\frac{g}{\sqrt{2}}W^-_\mu\left(T_L^1-i T_L^2\right)\nonumber\\
&&+g \left(c_w Z_\mu+s_w A_\mu \right)T_L^3+g' \left(c_w A_\mu-s_w Z_\mu \right)T_R^3\,,
\label{gfd}
\eea
where $s_w$ and $c_w$ are respectively the sine and cosine of the weak mixing angle.
Note that the $d_\mu$ and $e_\mu$ symbols transform under the unbroken ${\rm SO}(4)$ symmetry as a fourplet and an adjoint, respectively. 
In unitary gauge, the $e_\mu$ symbol components reduce to
\beq\label{esymb}
e^{1,2}_\mu= -\cos^2\left(\frac{\bar{h}}{2f}\right) g W_\mu^{1,2}\,,\quad  e^3_{\mu}=-\cos^2\left(\frac{\bar{h}}{2f}\right) g W^{3}_\mu-\sin^2\left(\frac{\bar{h}}{2f}\right) g' B_\mu\,,
\eeq
\beq
e^{4,5}_\mu=-\sin^2\left(\frac{\bar{h}}{2f}\right) g W_\mu^{1,2}\,,\quad e^6_{\mu}=  -\cos^2\left(\frac{\bar{h}}{2f}\right) g' B_\mu-\sin^2\left(\frac{\bar{h}}{2f}\right) g W^{3}_\mu\,,
\eeq
with $W^1_\mu=(W^+_\mu+W^-_\mu)/\sqrt{2}$, $W_\mu^2=i(W_\mu^+-W_\mu^-)/\sqrt{2}$, $W_\mu^3=c_w Z_\mu+s_w A_\mu$ and $B_\mu=c_wA_\mu -s_w Z_\mu$,  while the $d_\mu$ components read
\beq\label{dsymb}
d^{1,2}_\mu  =  -\sin(\bar{h}/f) \frac{g W^{1,2}_{\mu}}{\sqrt{2}}\,,\quad
d^{3}_\mu  = \displaystyle \sin(\bar{h}/f) \frac{g' B_{\mu}- g W^{3}_{\mu}}{\sqrt{2}}\,,\quad
d^{4}_\mu  = \displaystyle \frac{\sqrt{2}}{f} \partial_\mu h\,.
\eeq

%% file: appendix3.tex
We derive here the couplings of the composite resonances to the SM states which are relevant
for analysing the partially composite models of Section~\ref{sec:pcmodel}. 

\subsection{Mass spectrum}

Consider the Lagrangian of Eqs.~\eqref{PCLag},\eqref{eq:pcL1},\eqref{eq:pcL2} in the $y_L=0$ limit.
Expanding the Higgs field $\bar h$ around its VEV $v$ yields the following mass term for the charge-$2/3$ states
\beq
 \mathcal{L}_{\rm mass}=-\bar{\psi}^u_L \mathcal{\hat M}_u\psi^u_R+ \mbox{h.c.}\,,\quad \psi^u=(u,U_p,U_m,\tilde{U})^T\,,
 \eeq
where
\beq
 \mathcal{\hat M}_u=\left(\begin{array}{cccc}
                0 & 0 & 0& 0 \\
		0 & M_4 & 0 & 0 \\
		 y_R f \sin\epsilon & 0 & M_4 & 0 \\
		 y_R f \cos\epsilon & 0 & 0 & M_1
               \end{array}\right)\,,\quad \epsilon=\frac{v}{f}\,.
\label{uMassmat2}
\eeq
$\mathcal{\hat M}_u$ is obtained from the mass matrix in Eq.~\eqref{uMassmat} by applying the $U_{p,m}=(U\pm X_{2/3})/\sqrt{2}$ rotation.
Note that $U_p$ does not mix the other states in $\psi^u$ as it belongs to a triplet of the custodial symmetry preserved in the $y_L=0$ limit, while $u$, $U_m$ and $\tilde U$ are singlets. Note also that $\mathcal{\hat M}_u$ has a zero eigenvalue corresponding to the SM up quark which remains massless in the $y_L=0$ limit. $\mathcal{\hat M}_u$ is further diagonalized by a bi-unitary transformation
\beq
\psi^{u\,\prime}_{L,R}=\mathcal{U}^\dagger_{L,R}\psi^u_{L,R}\,,
\label{mebdef}
\eeq
which yields a mass for $U_p$ of $M_4$, and the expressions for the masses of the $U_{l,h}$ eigenstates are given in Eq.~\eqref{U12mass}.

The $\mathcal{U}_{R}$ elements characterizing the mixing between the massless SM up quark
and its heavy partners have the simple form
\bea 
\mathcal{U}_R^{11}  =  \cos \varphi_4 \cos \tilde \varphi_1\,,\quad
\mathcal{U}_R^{31}  =  -\sin \varphi_4 \cos \tilde \varphi_1\,,\quad
\mathcal{U}_R^{41}  =- \sin \tilde \varphi_1\,, \label{ARapprox}
\eea
where the mixing angles $\varphi_4$ and $\tilde \varphi_1$ are related to the fundamental
parameters as
\beq
\tan\varphi_4= \frac{y_Rf}{M_4}\sin\epsilon\,,\quad {\rm and} \quad \tan \tilde \varphi_1 = \frac{y_R f }{M_1}\cos \epsilon \cos\varphi_4\,,
\eeq
respectively.

The $\mathcal{U}_{L}$ components can be derived analytically as well, and we used the exact
form in our simulations, but the full expressions are rather lengthy for generic values of
the parameters. Yet, simple expressions are obtained in the limits in which one SO(4)
multiplet is much lighter than the other. For instance if the fourplet is
lighter than the singlet one just finds $\mathcal{U}_L\simeq \bf{1}$.


\subsection{Higgs and EW gauge boson couplings}
In the following, we derive the EW gauge boson and Higgs interactions with one SM quark and one heavy partner quark which are relevant for the production and the decay of the partner quarks. We refer to these interactions as ``mixing" interactions. Note, that for partially composite quarks, there are no mixing interactions present in the gauge basis. These interactions are solely induced through the rotation into the mass basis as discussed above.
The couplings of the light and heavy quarks to photons and to the gluons do not induce mixing interactions thanks to the U(1)$_{\rm em}$ and to the SU(3)$_{\rm color}$ gauge invariance. Furthermore, the U(1)$_X$ charges of $u_L$, $u_R$ and $\psi$ are identical. Hence, the covariant derivative terms with respect to the U(1)$_X$ does not induce  mixing interactions when rotating into the mass basis, but only ``diagonal" couplings of the quark mass eigenstates to the $Z$ boson and the photon. Therefore, the only mixing interactions with gauge bosons arise from the $e_\mu$ and $d_\mu$ terms in the Lagrangian of Eq.~\eqref{eq:pcL1}, while the mixing interactions with the Higgs arise from the $d_\mu$ term and the Yukawa terms in Eq.~\eqref{eq:pcL2}. 
The terms relevant for mixing from the $e_\mu$-symbol interaction read
\bea
-\bar{Q}\Sla{\,e}\,Q =
\frac{g}{2}\left(\bar D\, \Sla{W}^- + \bar X_{5/3}\, \Sla{W}^+\right) U_p
+\,\frac{g}{2}\cos\epsilon\left(\bar D\, \Sla{W}^-
-\bar X_{5/3}\, \Sla{W}^+
+ \frac{1}{c_w} \bar{U}_p\,\Sla{Z}\right)U_m + \mbox{h.c.}.
\eea
Further mixing interactions are induced by the $d$-term:
\bea
i c \bar{Q}^i\Sla{\,d}^{\,i} \tilde{U} + \mbox{h.c.}&=&\left[-\frac{i\sqrt{2} c}{f}\bar{U}_{m} \gamma^\mu\left(\partial_\mu h \right)\tilde{U}\right.\nonumber\\
&&\left. \,\,\,-\frac{g}{\sqrt{2}}c \sin\frac{\bar{h}}{f}\left(\frac{1}{c_w}\bar{U}_{p}\,\Sla{Z} \tilde{U}+\bar{D}\,\Sla{W}^- \tilde{U}-\bar{X}_{5/3}\,\Sla{W}^+ \tilde{U}\right) \right]+\mbox{h.c.}. 
\eea
The leading couplings to gauge bosons directly follow by setting the Higgs field $\bar{h}$
to its VEV $v$. The derivative coupling to the Higgs can be rewritten by performing a
partial integration on the action and using the equations of motion:
\bea
i\Sla{\,\partial} U_{m,L}&=& y_R f \sin\epsilon u_R+M_4 U_{m,R}\, , \\
i\Sla{\,\partial} \tilde{U}_L&=& y_R f \cos\epsilon u_R+M_1 \tilde{U}_{R} \, ,
\label{pcdterm1}
\eea
which yields
\bea
-\frac{i \sqrt{2} c \bar{U}_{m} \gamma^\mu\left(\partial_\mu h \right)\tilde{U} }{f} &=&
\sqrt{2}c h\left[y_R\left( \cos\epsilon\, \bar{U}_{m,L} - \sin \epsilon\,  \bar{\tilde{U}}_L\right)  u_R \right.\nonumber\\
&&\left.+\frac{M_1-M_4}{f}\left(\bar{U}_{m,L}\tilde{U}_R+\bar{\tilde{U}}_L U_{m,R}\right) +\mbox{h.c.}\right].
\label{pcdterm2}
\eea
The elementary-composite mixing terms also give rise to mixing interactions involving the Higgs boson
\bea
\mathcal{L}\supset - y_R h \cos\epsilon\, \bar{U}_{m,L} u_R + y_R h \sin\epsilon\, \bar{\tilde{U}}_{L} u_R +\mbox{h.c.}. 
\label{pcyuk}
\eea

Collecting all mixing interactions from the $e$-term, $d$-term, and $y_R$ interactions,
the mixing Lagrangian in the gauge basis reads
\bea
\mathcal{L}_{\rm mix}= \sum_{\alpha=L,R}\bar{\psi}^{d}_\alpha\,\Sla{W}^-G^{D}_\alpha\psi^{u}_\alpha+\bar{X}_{5/3\,\alpha}\,\Sla{W}^+G^{X}_\alpha\psi^{u}_\alpha+\bar{\psi}^{u}_\alpha\,\Sla{Z}G^{Z}_\alpha\psi^{u}_\alpha
+\left(\bar{\psi}^{u}_{L} h G^{h}\psi^{u}_{R} +\mbox{h.c.}\right)\,,
\label{eq:Lgeb}
\eea
with $\psi^{d}_{L,R}= (d,D)^T_{L,R}$,
\bea
 G^{D}_\alpha  & = &
\frac{g}{\sqrt{2}}
\left(
\begin{array}{cccc}
\delta_\alpha^L&0&0&0\\
0&\frac{1}{\sqrt{2}}&\frac{\cos\epsilon}{\sqrt{2}}&- c \sin\epsilon
\end{array}
\right) \, , \\
G^{X}_\alpha & = &\frac{g}{\sqrt{2}}
\left(
\begin{array}{cccc}
0&\frac{1}{\sqrt{2}}&- \frac{\cos\epsilon}{\sqrt{2}}&c \sin\epsilon
\end{array}
\right) \,,\\
G^{Z}_\alpha & =& \frac{g}{2c_w}
\left(
\begin{array}{cccc}
\delta_\alpha^L&0&0&0\\
0&0&\cos\epsilon&-\sqrt{2}\,c\sin\epsilon\\
0&\cos\epsilon&0&0\\
0&-\sqrt{2}\,c\sin\epsilon&0&0
\end{array}
\right) -\frac{2g}{3}\frac{s_w^2}{c_w}\, \cdot \,\,  {\bf 1}\, ,\label{GZmat}
\eea
and 
\beq
G^{h} =
\left(
\begin{array}{cccc}
0&0&0&0\\
0&0&0&0\\
\left(\sqrt{2} c -y_R\right)\cos\epsilon&0&0&\sqrt{2} c \,\frac{M_1-M_4}{f}\\
\left(y_R-\sqrt{2} c \right)\sin\epsilon&0&\sqrt{2} c \,\frac{M_1-M_4}{f}&0
\end{array}
\right) \, .\label{Ghmat} 
\eeq
The universal part of Eq.~\eqref{GZmat} arises from the coupling to the U(1)$_X$ gauge boson and does not contribute to mixing interactions. 
The mixing couplings in the mass eigenbasis are obtained from Eq.~\eqref{eq:Lgeb} through the rotation in Eq.~\eqref{mebdef}. 
The couplings of the mixing gauge interaction involving the right-handed SM up quark are given by
\begin{equation}
g_{WuX} = - g_{WuD} = - c_w g_{ZuU_p}
= \frac{g}{2} \cos \epsilon \sin \varphi_4 \cos \tilde \varphi_1 - \frac{c}{\sqrt{2}}
\sin \epsilon \sin \tilde \varphi_1\,.
\end{equation}
The mixing interactions mediated by the Higgs take a simple form if one mupliplet
is much lighter than the other one. In the limit $\sqrt{M_4^2 + y_R^2 f^2 \sin^2 \epsilon}
\ll \sqrt{M_1^2 + y_R^2 f^2 \cos^2 \epsilon}$, one finds
\begin{equation}
\lambda_{huU_l} \approx  (y_R - \sqrt{2} c) \cos\epsilon \cos \varphi_4 \cos \tilde \varphi_1
- \sqrt{2} c \frac{M_1 - M_4}{f} \cos \tilde \varphi_1 \sin \varphi_4\,,
\end{equation}
and
\begin{equation}
\lambda_{huU_h} \approx -(y_R - \sqrt{2} c) \sin\epsilon  \cos \varphi_4 \cos \tilde \varphi_1
- \sqrt{2} c \frac{M_1 - M_4}{f} \sin \tilde \varphi_1\,.
\end{equation}
The expressions for $\lambda_{huU_l}$ and $\lambda_{huU_h}$ in the limit of a singlet
lighter than the fourplet are obtained from the above ones through a $l\leftrightarrow h$  exchange.


%% file: appendix4.tex
We present in this appendix the cross sections for the existing searches listed in Section~\ref{directsearches} as predicted in the partially and fully composite models for the first two generation quark partners. Exclusion bounds on the partner masses are also derived.
The strongest bounds from each LHC collaboration are also shown in Sec.~\ref{sec:results}, while the combination discussed in Sec.~\ref{sec:rescom} is based on all the channels considered in the following.
We only focus on fourplet partners and take a simplifying limit where singlet partners are decoupled from the low-energy effective theory. 
For illustration, we set $f=600$ GeV, as well as $y_R^x=1$ for the partially composite case and $c_1^x=1$ for the fully composite case.

\subsection{Tevatron exclusion bounds}

\begin{figure}[tb!]
\centering
\begin{tabular}{cc}
\includegraphics[width=0.45\textwidth]{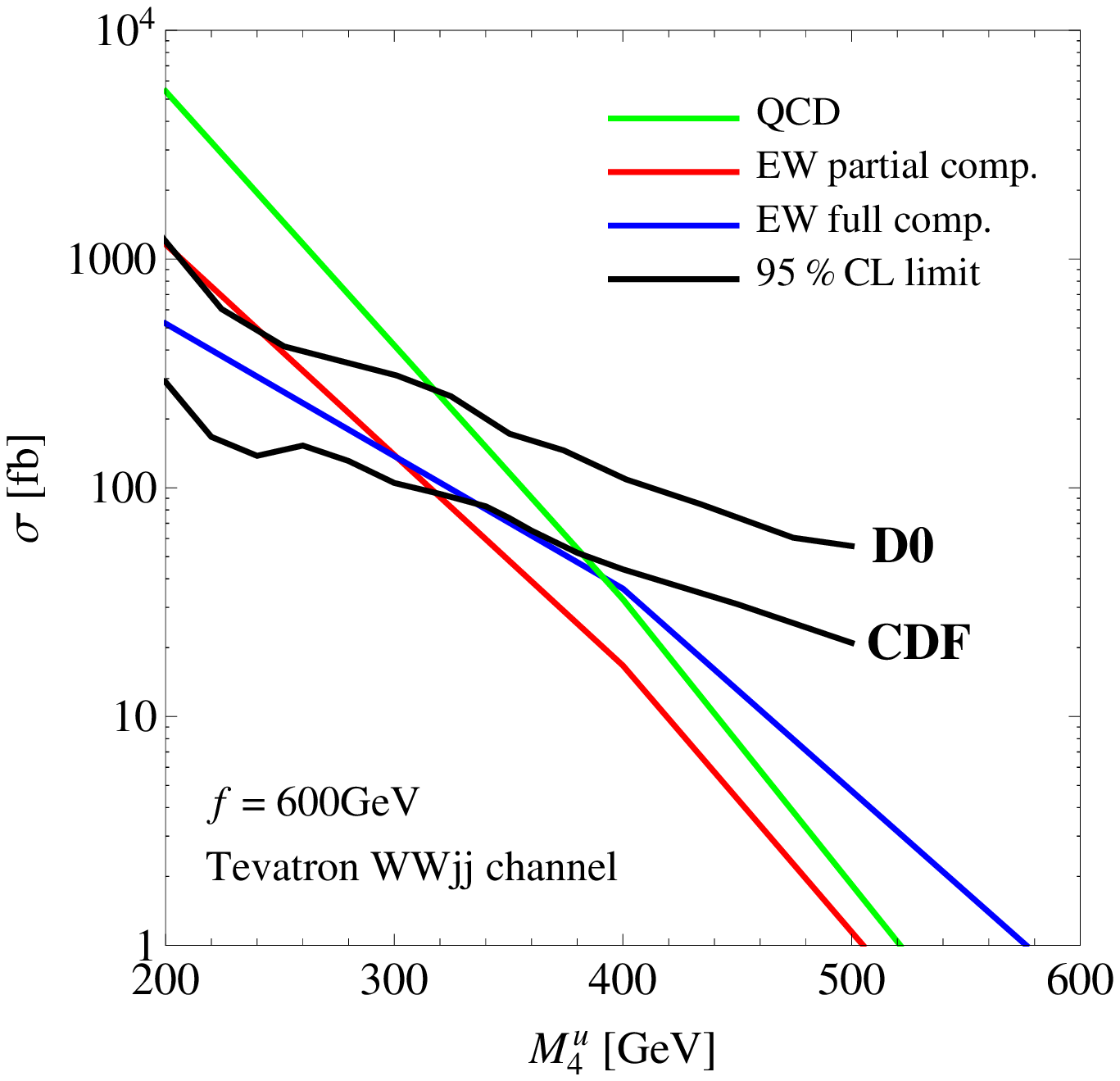}
&
\includegraphics[width=0.45\textwidth]{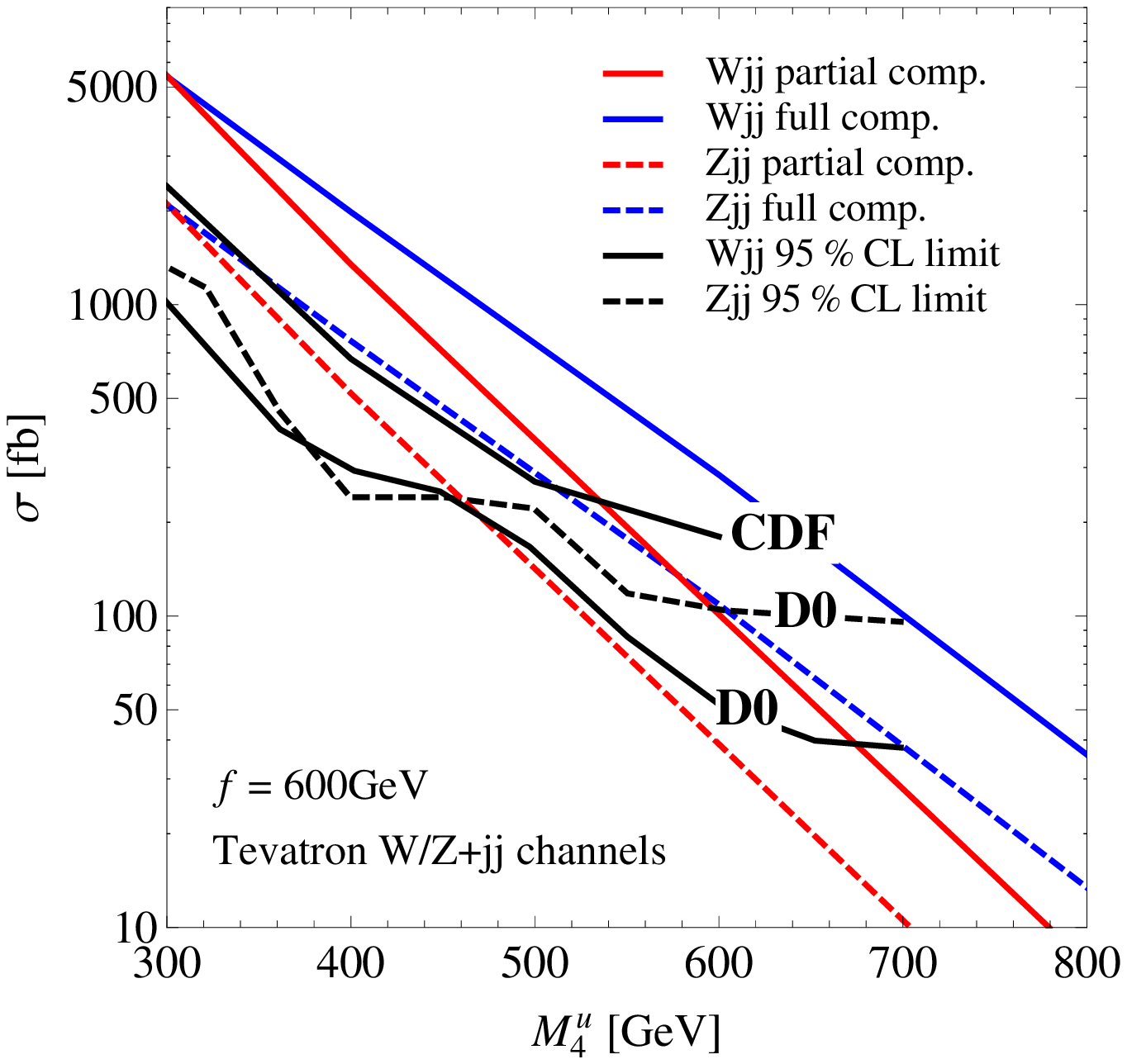}
\end{tabular}
\caption{Cross sections for pair (left) and single (right) production of the fourplet partners of $u_R^{\rm SM}$ leading to $WWjj$ and $Wjj$, $Zjj$  final states, respectively. We assumed $f=600\,$GeV and $y_R^u=1$ ($c_1^u=1$) for partially (fully) composite $u_R$. BR$(U_p\to uZ)=\,$BR$(D\to uW)=\,$BR$(X_{5/3}^u\to u W)=1$. 
95$\%$ CL exclusion limits from the Tevatron
analyses of Refs.~\cite{Abazov:2010ku,CDF:wj,Abazov:2011vy,Aaltonen:2011tq} are shown in black. Left panel:
QCD pair production cross section (green) includes both $D\bar D$ and $X_{5/3}^u\bar{X}_{5/3}^u$ contributions, while $D\bar D$, $X_{5/3}^u\bar{X}_{5/3}^u$ and $X_{5/3}^u \bar D+D \bar{X}_{5/3}^u$ states contribute to EW pair production in the partially (red) and fully (blue) composite cases. Right panel: Solid (dashed) lines denote $Wjj$ ($Zjj$) cross sections from $D$ and $X_{5/3}^u$ ($U_p$) production in partially (red) and fully (blue) composite cases.}
\label{fig:tev}
\end{figure}

We first consider Tevatron  searches~\cite{Abazov:2010ku,CDF:wj,Abazov:2011vy,Aaltonen:2011tq} described in Section~\ref{directsearches}. Tevatron experiments suffer less important QCD backgrounds than ATLAS and CMS, and thus yield interesting bounds on composite partners of the first two generations, despite a significantly smaller center of mass energy relative to the LHC.
Figure~\ref{fig:tev} shows the cross sections from right-handed up quark partners for the various final states analysed at the Tevatron. 
The cross section predicted by second generation partners are not shown as all of them, but QCD pair production, are well below Tevatron limits
for both partially and fully composite charm scenarios.

Consider first singlet production channels.
For first generation partners in the partially composite case, 
D0 analysis of  $Zjj$ final states~\cite{Abazov:2010ku} excludes a $U_p$ partner lighter than $M_4^u\simeq 460\,$GeV at 95$\%$ CL for $y_R^u=1$. 
Singly produced $D$ and $X^u_{5/3}$ contribute to the $Wjj$ cross section. Since there are two degenerate states contributing to the cross section, the D0 bound is stronger in the $Wjj$ channel. We find in this case $M_4^u\gtrsim 680\,$GeV at 95$\%$ CL. 
Assuming $c_1^u=1$, corresponding 95$\%$ CL bounds in the fully composite case are $M_4^u\gtrsim 600\,$GeV from the $Zjj$ channel and $M_4^u\gtrsim 700\,$GeV from the $Wjj$ channel, which is the high edge of the mass range covered by the experiment.
Up and charm partners can also be produced in pairs through QCD interactions with the same cross section. However, since there is no search in $ZZjj$ final states, a light $C_p$ state is not directly constrained at the Tevatron. The existence of a light fourplet partner of the second generation can nevertheless be probed through strong pair production of $S$ and $X_{5/3}^c$ states, since they contribute to the $WWjj$ cross section measured by Tevatron experiments. We find in this case $M_4^{c}\gtrsim 390\,$GeV at 95$\%$ CL from the CDF $WWjj$ analysis~\cite{Aaltonen:2011tq}. Thanks to the universality of QCD interactions, the same bound also applies to first generation partners, $M^{u}_4\gtrsim 390\,$GeV.  In contrast with single production channels, these bounds are model-independent. They are the same in both partially and fully composite models and in particular they do not depend on the values of $f$, $y_R^x$ and $c_1^x$.\\

\subsection{ATLAS exclusion bounds from $7\,$TeV data}
\begin{figure}[tb!]
\centering
\begin{tabular}{cc}
\includegraphics[width=0.45\textwidth]{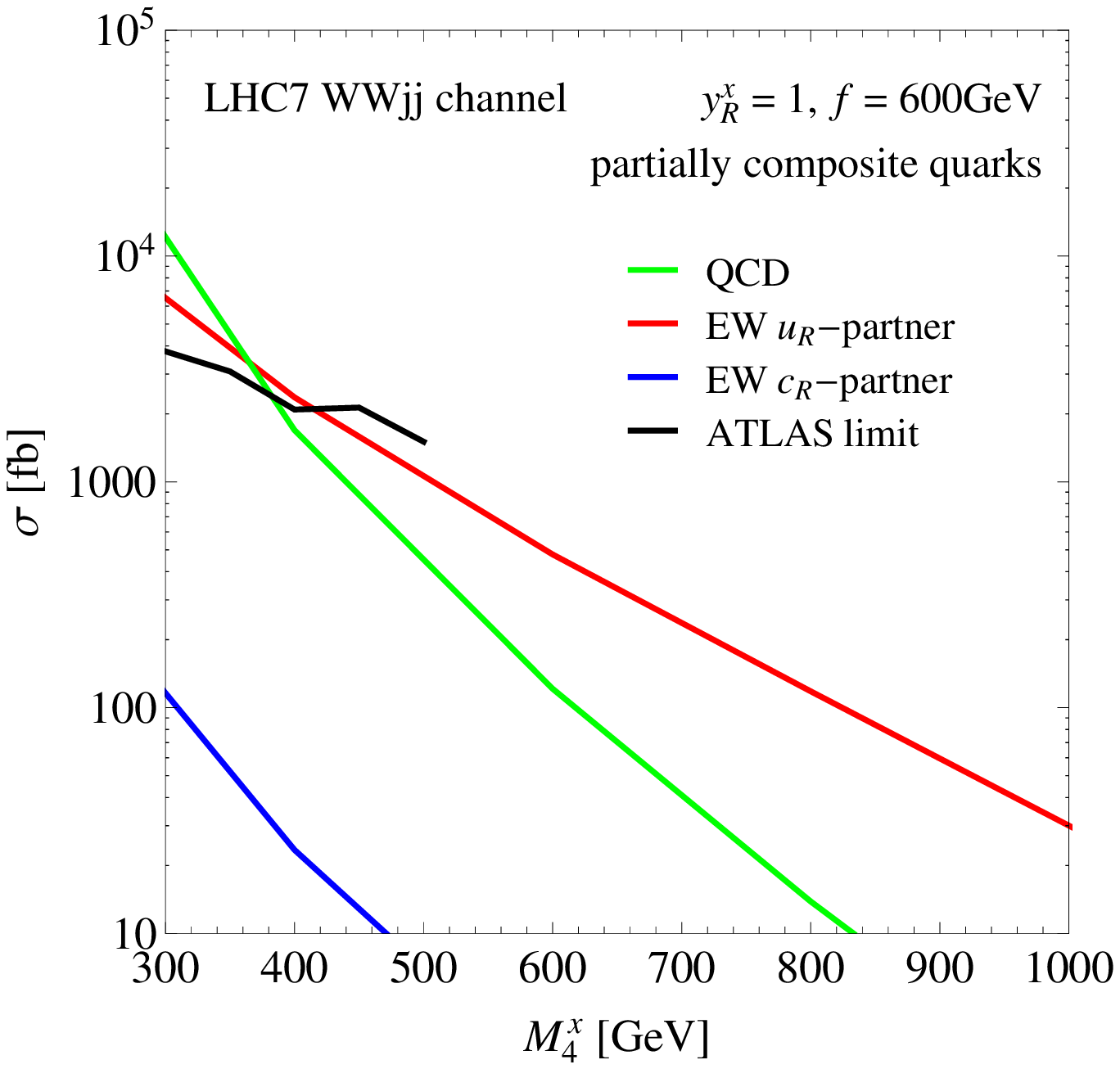}
\includegraphics[width=0.45\textwidth]{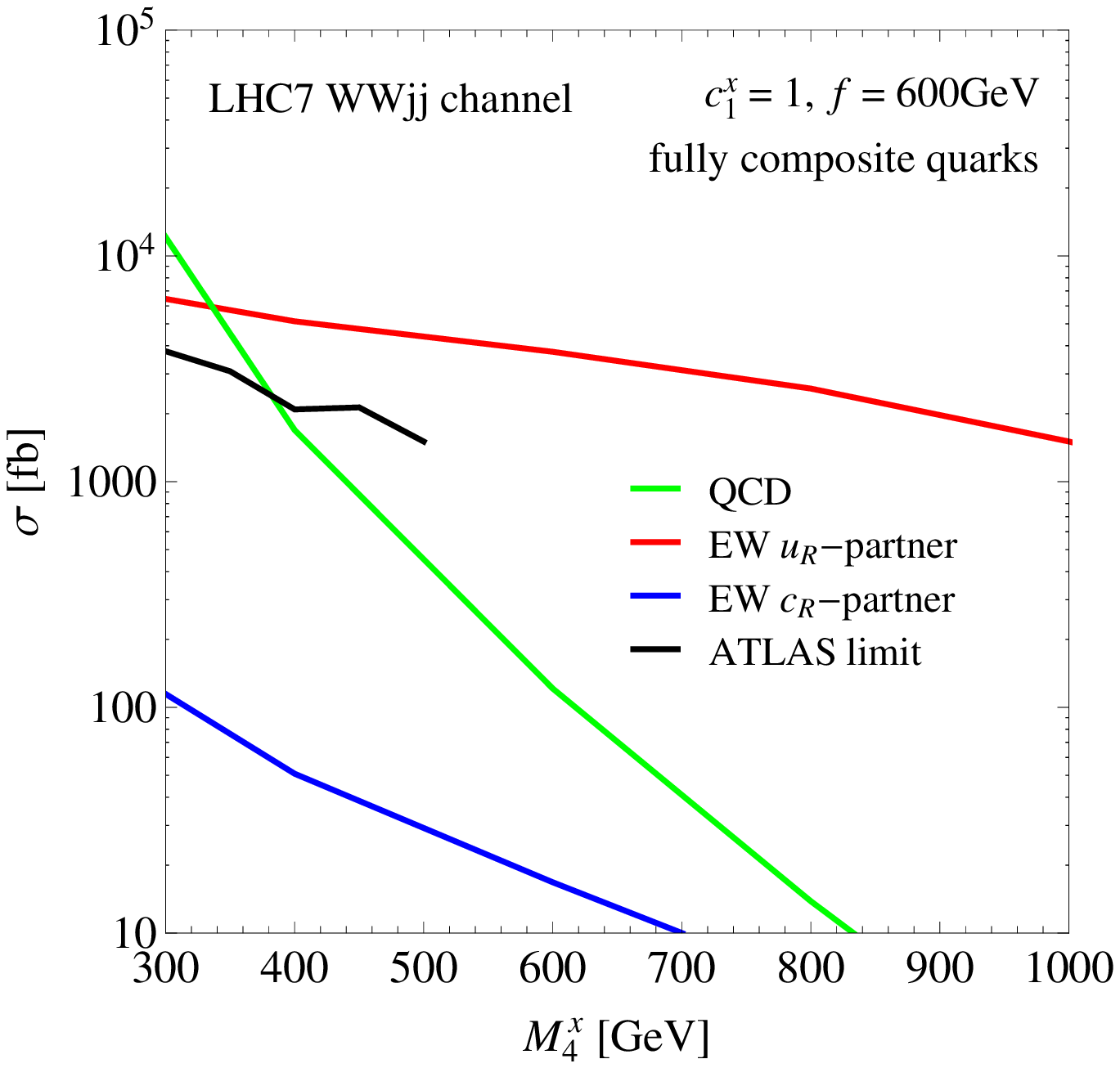}\\
\includegraphics[width=0.45\textwidth]{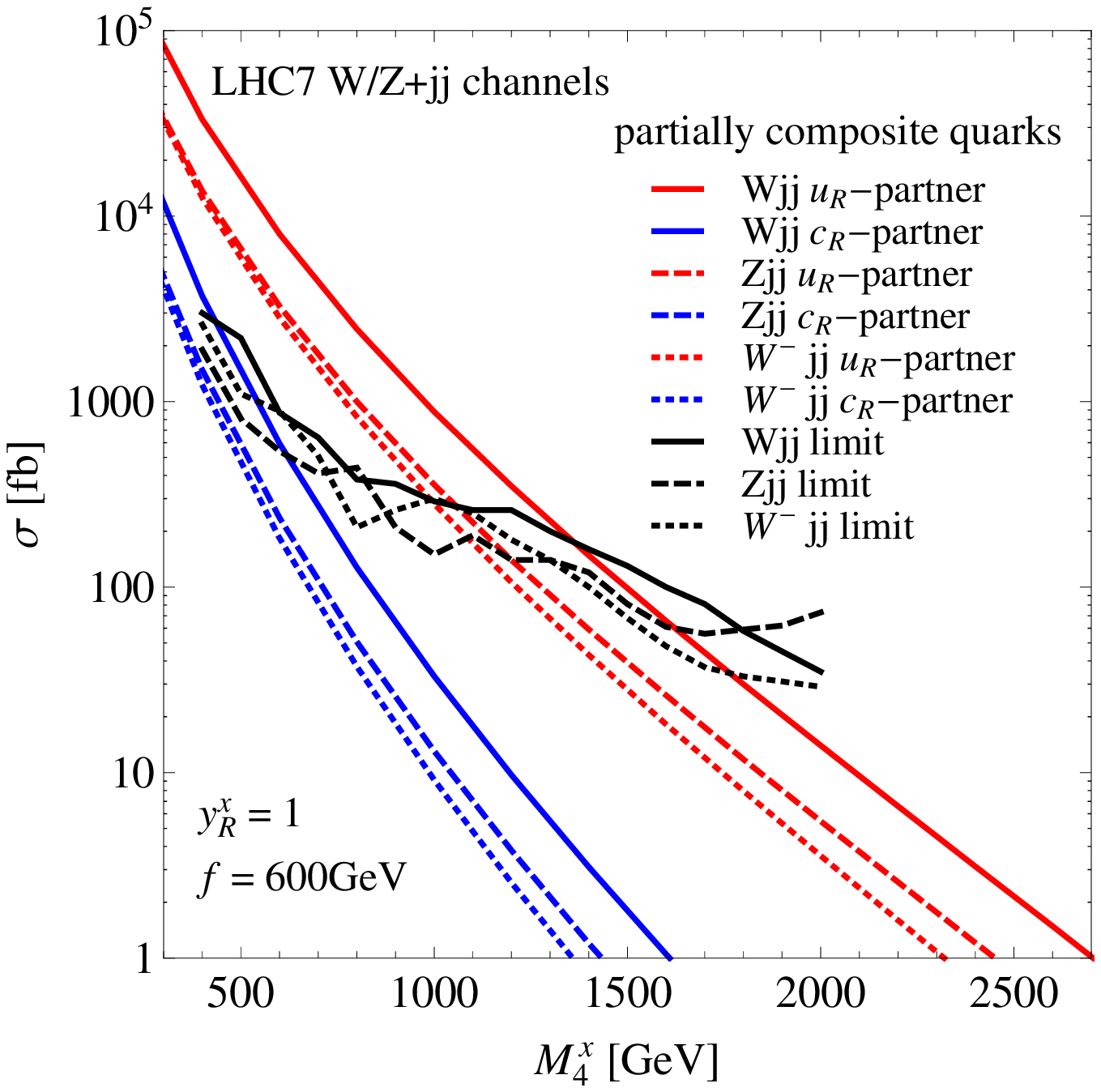}
\includegraphics[width=0.45\textwidth]{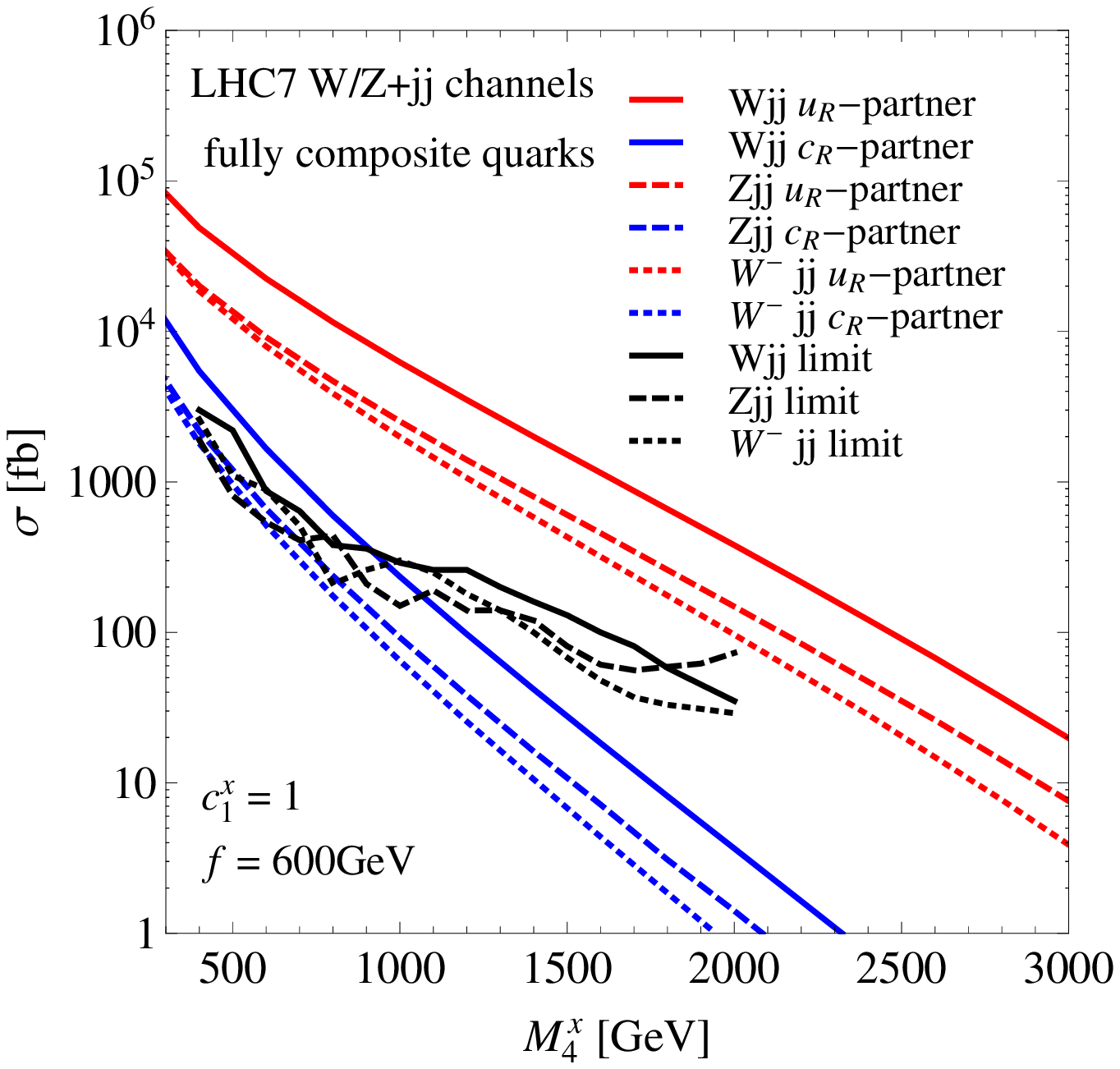}\\
\end{tabular}
\caption{Cross sections for pair (top) and single (bottom) production of the fourplet partners of partially (left) and fully (right) $u_R^{\rm SM}$ and $c_R^{\rm SM}$, leading to $WWjj$, $Wjj$ and $Zjj$  final states. We assumed $f=600\,$GeV and $y_R^u=1$ ($c_1^u=1$) for partially (fully) composite $u_R$. The $Wjj$ channel includes both $W^+$ and $W^-$ in the final state, while the $W^-jj$ channel only includes a negatively charged $W$ boson. BR$(U_p\to uZ)=\,$BR$(D\to uW)=\,$BR$(X_{5/3}^u\to u W)=1$ and BR$(C_p\to cZ)=\,$BR$(S\to cW)=\,$BR$(X_{5/3}^c\to c W)=1$. 
95$\%$ CL exclusion limits from the ATLAS
analyses of Refs.~\cite{ATLAS:pp,Aad:2012bt} are shown in black. Top panels: QCD pair production cross section (green) includes both $D\bar D$ and $X_{5/3}^u\bar{X}_{5/3}^u$ ($S\bar S$ and $X_{5/3}^c\bar{X}^c_{5/3}$) contributions for up (charm) quark  partners. $D\bar D$, $X_{5/3}^u\bar{X}_{5/3}^u$, $X_{5/3}^u \bar D+D \bar{X}_{5/3}^u$ and $S\bar S$, $X_{5/3}^c\bar{X}_{5/3}^c$, $X_{5/3}^c\bar S+S \bar{X}_{5/3}^c$ states contribute to the EW pair production cross section of up (red) and charm (blue) partners, respectively. Bottom panels: Solid (dashed) lines correspond to $Wjj$ ($Zjj$) cross sections from single production of $D$ and $X_{5/3}^u$ ($U_p$) for first generation partners, and from single production of $S$ and $X_{5/3}^c$ ($C_p$) for second generation partners. Dotted lines denote analogous cross sections in the $W^-jj$ channel.}
\label{fig:lhc7}
\end{figure}

We detail now the bounds obtained from the
ATLAS analyses~\cite{ATLAS:pp,Aad:2012bt} searching for single and pair production of first two generation partners which are described in  
Sec.~\ref{sec:searches}. 
Exclusion limits at 95$\%$ C.L. from these two analyses are shown in Fig.~\ref{fig:lhc7}, together with cross section predictions for
 partially and fully composite up and charm quarks. 

The strongest single production constraint arises from the $Wjj$ channel, which receives contributions from production and decay of $D$ and $X_{5/3}^{u}$ states in the first generation case, and $S$ and $X_{5/3}^c$ states in the second generation case. More specifically we find at 95$\%$ CL $M^u_4\gtrsim 1.4\,$TeV and  $M^c_4\gtrsim 420\,$GeV for partially composite models with $y_R^x=1$, while fully composite models with $c_1^x=1$ yields $M^u\gtrsim2.0\,$TeV, $M^c\gtrsim 950\,$GeV. Note that, in fully composite models with $c_1^x=1$ and masses larger than $2.3\,$TeV, the partner width exceeds $30\%$ of its mass,
thus breaking the narrow width approximation which the ATLAS analyses rely on.

Single production mechanisms strongly depend on the mixing parameters $y_R$ or $c_1$, but have a weak dependence on $f$. A $y_R$ and $c_1$ independent bound can be obtained from the $WWjj$ channel assuming QCD pair production of the partners. This implies a lower bound of $M_4^{u,c}\gtrsim 390$~GeV for partially and fully composite partners of up and charm quarks. $WWjj$ final states receives an additional contribution from pair produced $X_{5/3}^{u(c)}$ and $D$ $(S)$ through $t$-channel exchange of a W or Z boson (see Fig.\ref{fig:prodch}b). However, as shown in Fig.~\ref{fig:lhc7}  the $M_4^{u,c}$ bounds from EW pair production
 are much weaker than that of the $Wjj$ channel. More generally, for a fixed partner mass, the $Wjj$ channel excludes mixing parameters above a certain value. Under this constraint EW pair production is found to be subdominant to QCD production.   Thus, here we only use the $WWjj$ channel in order to determine a model independent bound on the fourplet masses $M_4^{u,c}$ through QCD pair production of the partners. Both EW and QCD pair production mechanisms leading to $WWjj$ final states are consistently added when we derive combined bounds in Section~\ref{sec:rescom}.

\subsection{CMS exclusion bounds from $8\,$TeV data}
We end with a presentation of exclusion limits and predicted cross sections from partially and fully composite models for the $8\,$TeV analyses described
in Section~\ref{sec:searches}: the CMS W/Z-tagged dijet analysis~\cite{CMS:wzdijet} and the recast of
the CMS leptoquark  search~\cite{cms8lepto}.

\begin{figure}[tb!]
\centering
\begin{tabular}{cc}
\includegraphics[width=0.45\textwidth]{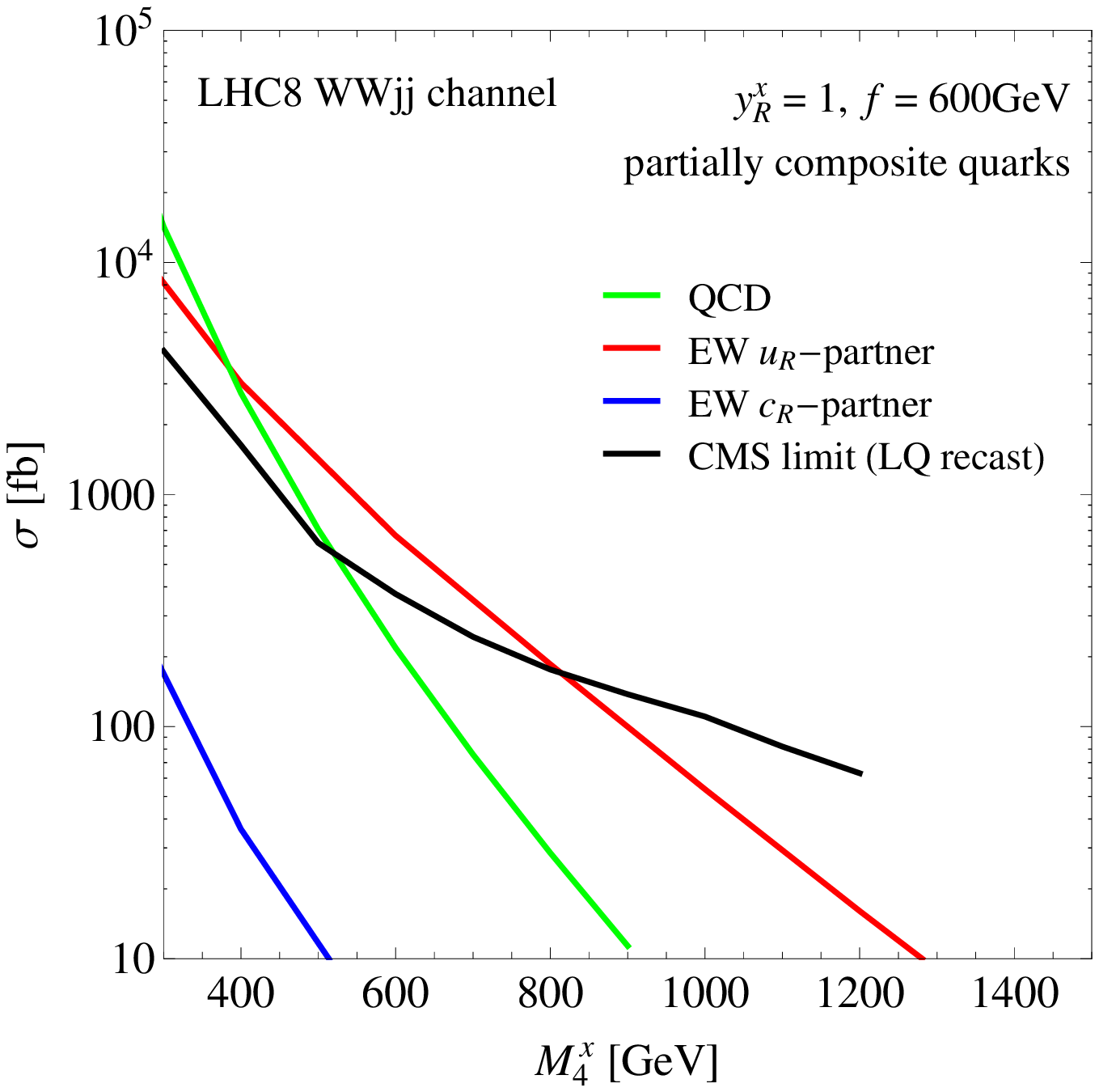}
\includegraphics[width=0.45\textwidth]{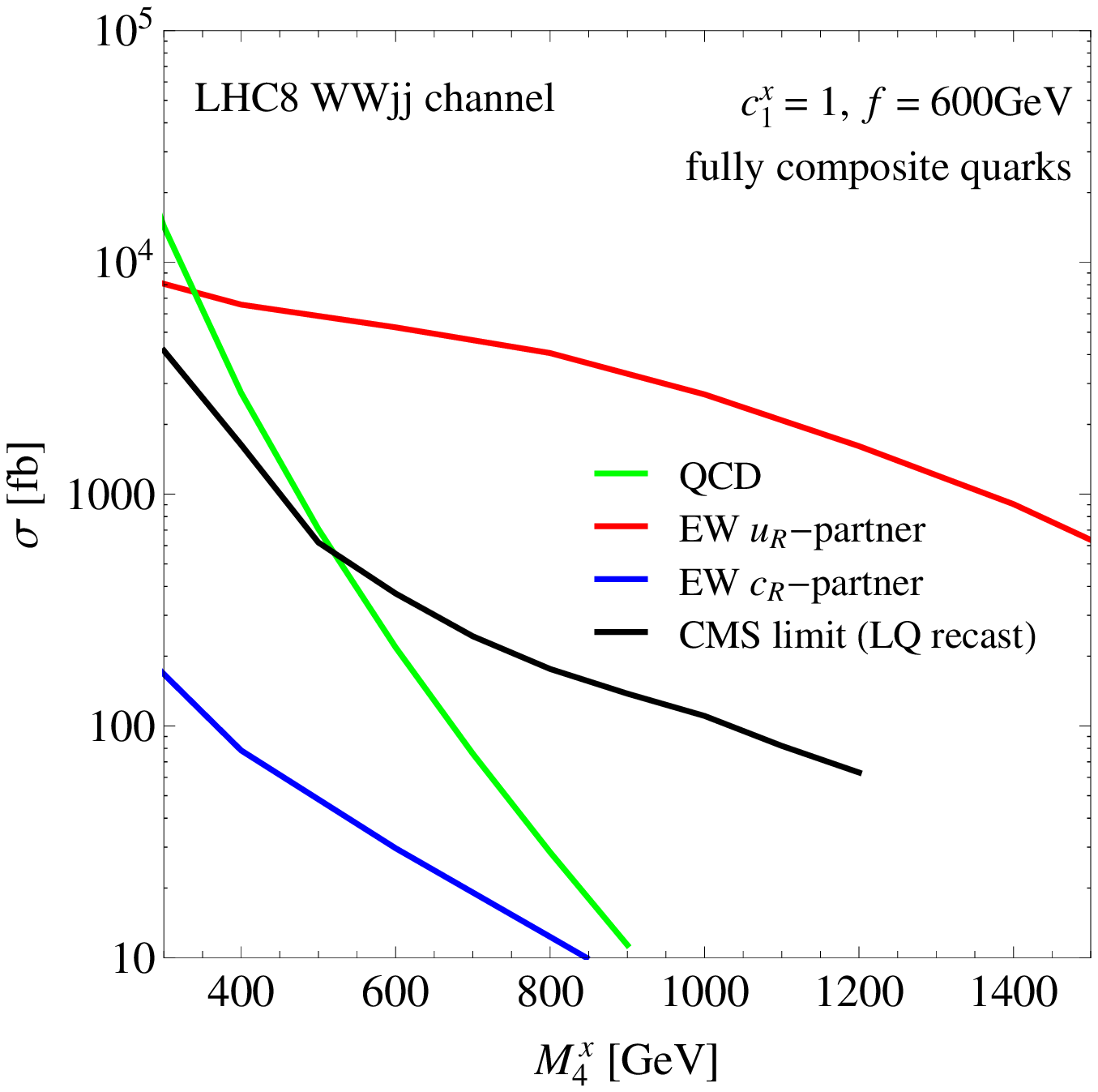}\\
\includegraphics[width=0.45\textwidth]{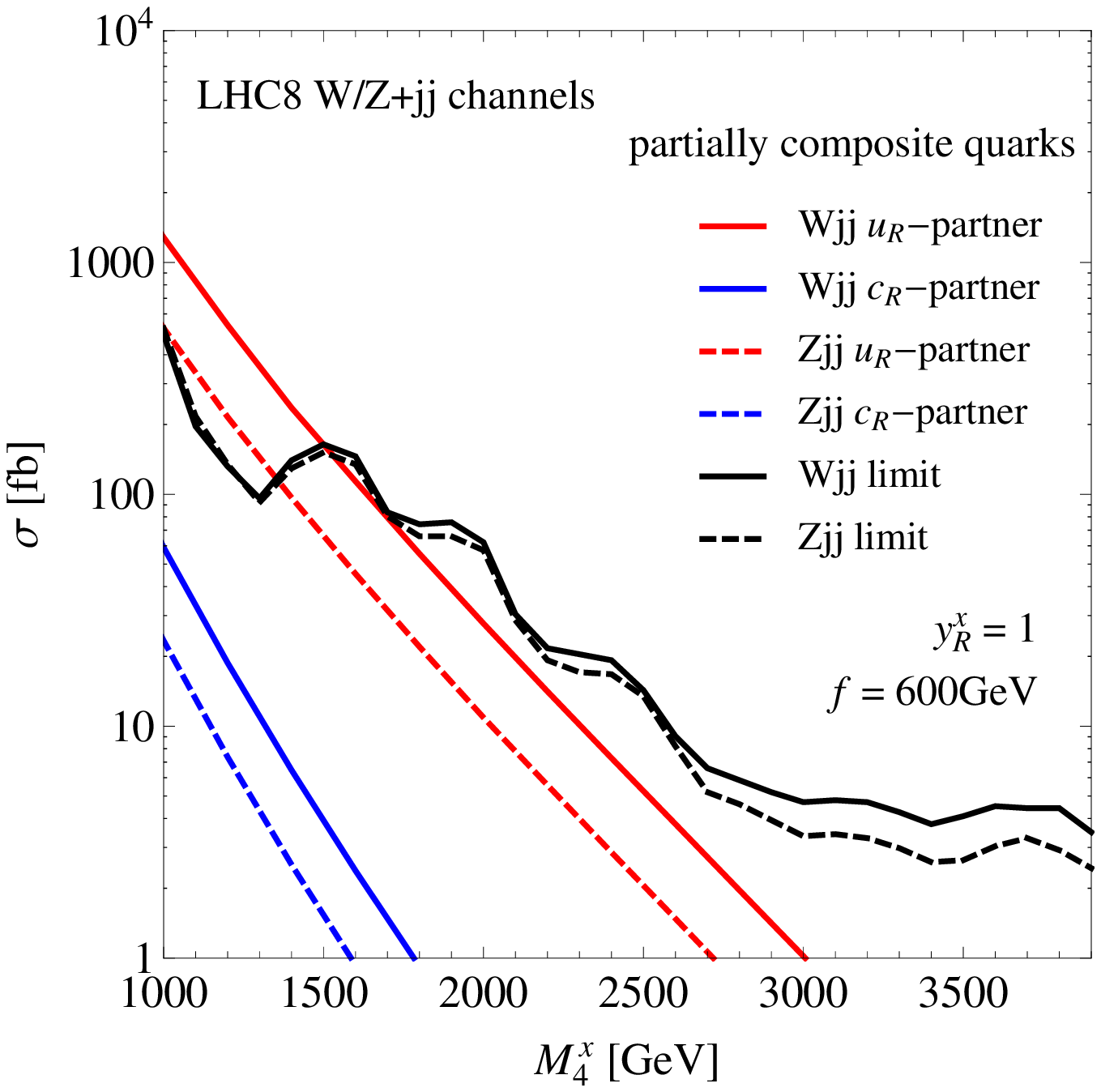}
\includegraphics[width=0.45\textwidth]{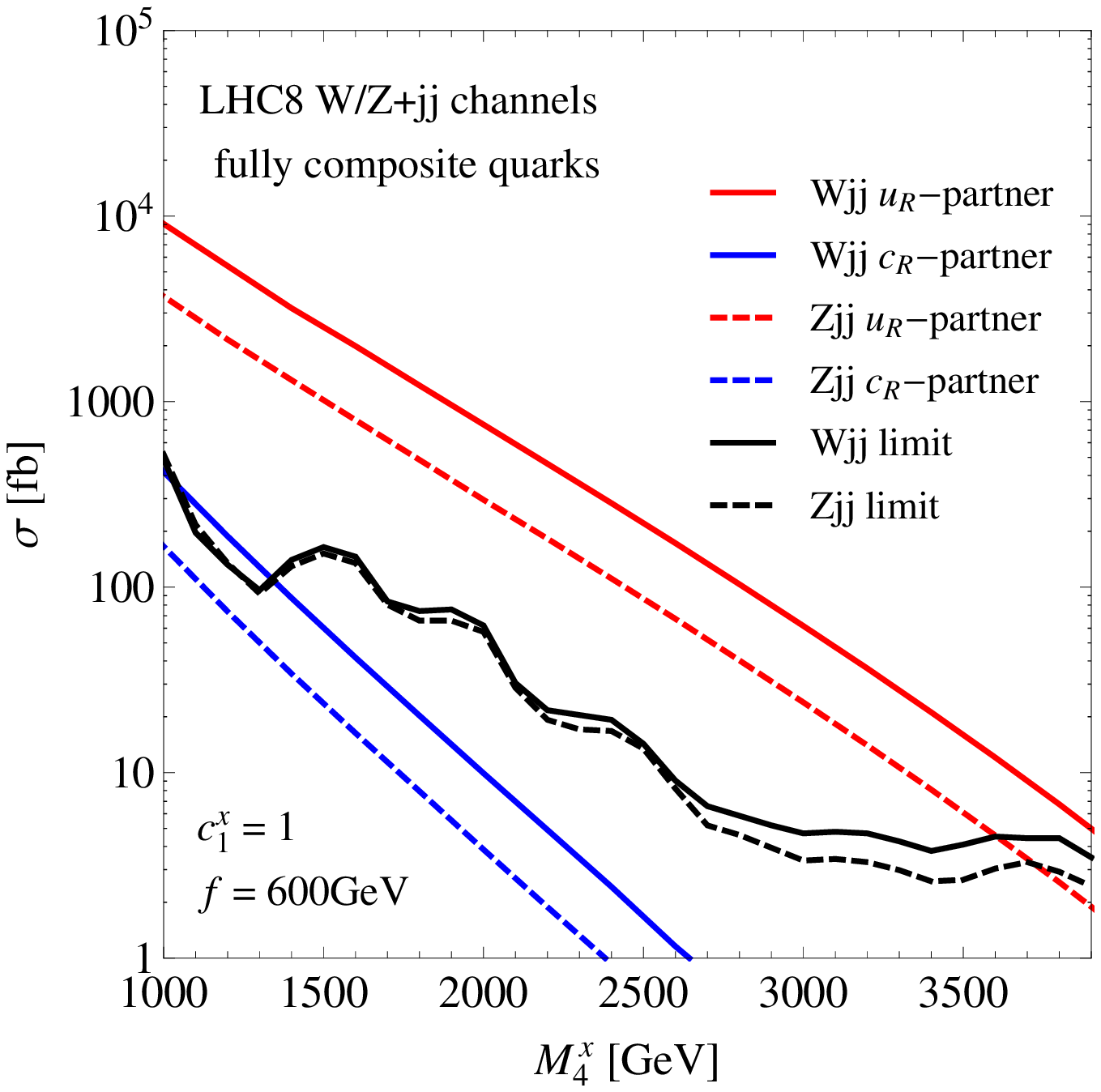}\\
\end{tabular}
\caption{Cross sections for pair (top) and single (bottom) production of the fourplet partners of partially (left) and fully (right) $u_R^{\rm SM}$ and $c_R^{\rm SM}$, leading to $WWjj$, $Wjj$ and $Zjj$  final states. We assumed $f=600\,$GeV and $y_R^u=1$ ($c_1^u=1$) for partially (fully) composite $u_R$. The $Wjj$ channel includes both $W^+$ and $W^-$ in the final state, while the $W^-jj$ channel only includes a negatively charged $W$ boson. BR$(U_p\to uZ)=\,$BR$(D\to uW)=\,$BR$(X_{5/3}^u\to u W)=1$ and BR$(C_p\to cZ)=\,$BR$(S\to cW)=\,$BR$(X_{5/3}^c\to c W)=1$. 
95$\%$ CL exclusion limits from the CMS
analyses of Refs.~\cite{CMS:wzdijet,cms8lepto} are shown in black. Top panels: QCD pair production cross section (green) includes both $D\bar D$ and $X_{5/3}^u\bar{X}_{5/3}^u$ ($S\bar S$ and $X_{5/3}^c\bar{X}^c_{5/3}$) contributions for up (charm) quark  partners. $D\bar D$, $X_{5/3}^u\bar{X}_{5/3}^u$, $X_{5/3}^u \bar D+D \bar{X}_{5/3}^u$ and $S\bar S$, $X_{5/3}^c\bar{X}_{5/3}^c$, $X_{5/3}^c\bar S+S \bar{X}_{5/3}^c$ states contribute to the EW pair production cross section of up (red) and charm (blue) partners, respectively. Bottom panels: Solid (dashed) lines correspond to $Wjj$ ($Zjj$) cross sections from single production of $D$ and $X_{5/3}^u$ ($U_p$) for first generation partners, and from single production of $S$ and $X_{5/3}^c$ ($C_p$) for second generation partners.}
\label{fig:cmslepto}
\end{figure}

The resulting 95$\%$ CL limits obtained from these analysis are shown in Fig.~\ref{fig:cmslepto}
for partially and fully composite scenarios. The constraints from the $qW$ and $qZ$ searches are  taken from Ref.~\cite{CMS:wzdijet}, while the CMS leptoquark search recast is detailed in Appendix~\ref{sec:app2}. 
As for the ATLAS searches, the dominant single production constraints arise from the $Wjj$ channel which yields, in partially composite models with $y_R^x=1$, $M^u_4\gtrsim 1.5\,$TeV, while the second generation partners are not bounded. For fully composite models with $c_1=1$, up-quark partners are excluded up to $M_4^u\gtrsim3.9\,$TeV, while the charm-quark partner mass is $M^c_4\gtrsim 1.3\,$TeV. The model independent bound obtained assuming QCD pair production is $M^{u/c}_4\gtrsim 530\,$GeV.  

%% file: appendix2.tex
We describe here the recast of the leptoquark search~\cite{cms8lepto} discussed in Section~\ref{directsearches}.
We focus on the channel with two oppositely charged muons and at least two jets in the final state. All details regarding event selection are found in the CMS report~\cite{cms8lepto}, and we limit ourselves to the criterions which are relevant to the recast.
The CMS analysis starts with the usual lepton isolation and minimum $p_T$ requirements. For muons
\beq
|\eta_\mu|<2.1\ \ \ \ \mbox{and     } p_T^\mu > 45\,\mbox{GeV}
\eeq
are imposed. The muon isolation is performed through requiring that the sum of the transverse momenta within $\Delta R<0.3$ around the muon track (excluding the muon itself)
divided by the muon transverse momenta is $<0.1$.
Jets are reconstructed using the anti-$k_T$ algorithm~\cite{Cacciari:2008gp} with a cone size of $R=0.5$.
The isolation and minimum transverse momentum cuts for the jets are
\beq
|\eta_j|<2.4\ \ \ \ \mbox{,     } p_T^{j_{\rm Lead}} > 125\,\mbox{GeV}\ \ \ \ \mbox{,     } p_T^{j_{\rm Sub}} > 45\,\mbox{GeV}
\ \ \ \ \mbox{and     } \Delta R_{\mu j} > 0.3\,,
\eeq
where $j_{\rm Lead}$ and $j_{\rm Sub}$ denote the jet of highest and next to highest $p_T$, respectively.
Finally, $S_T=p_T^{\mu_1}+p_T^{\mu_2}+p_T^{j_1}+p_T^{j_2}$ is required to be larger than 300$\,$GeV
and the invariant mass of the dimuon pair must satisfy $M_{\mu\mu}>50\,$GeV.\\

After the preselection cuts, the cuts on $S_T$, $M_{\mu\mu}$ and $M_{\mu j}^{min}$, where the latter is defined as the smallest of the
two muon-jet invariant masses which minimize the two muon-jet invariant mass difference, are optimized for the leptoquark signal.
The purpose of these cuts is to increase the signal to background ratio. See Ref.~\cite{cms8lepto} for further details.
For the recast we only consider leptoquark masses of 500 and 900$\,$GeV, which are the only cases fully described in Ref.~\cite{cms8lepto}.
Cuts corresponding to the $500\,$GeV case are
\beq
S_T>685\,\mbox{GeV,           }M_{\mu\mu}>150\,\mbox{GeV and           }M_{\mu j}^{min}>155\,\mbox{GeV}\,,
\eeq
while for the $900\,$GeV case they are
\beq
S_T>1135\,\mbox{GeV,           }M_{\mu\mu}>230\,\mbox{GeV and           }M_{\mu j}^{min}>535\,\mbox{GeV}\,.
\eeq
In addition, we also define a set
of cuts in order to suppress the dominant $Z^*\/\gamma$+jets background in the low mass region of the analysis and thus enhance the sensitivity to heavy quark partners of composite up and charm quarks. These cuts correspond to the preselection level cuts together with $M_{\mu\mu}>145$ GeV, where the value of the latter is found to maximize the sensitivity to the partners.
We refer to these cuts as ``custom preselection'' cuts.\\

We simulate the heavy quark partner signals using FeynRules~\cite{Christensen:2008py}, MadGraph5~\cite{Alwall:2011uj},
PYTHIA~\cite{Sjostrand:2006za} and PGS 4~\cite{pgs}. We also simulate the scalar leptoquark signals using the same set of tools. We then use the latter in order to tune our detector efficiencies so that they match the CMS ones for the leptoquark signals. In doing so we simulate signals for different leptoquark masses, assuming their corresponding levels of cuts. We already find a very good overall agreement between our efficiencies and those of CMS prior to any tuning of our simulation, and
we apply ``tuning factors'' ranging from 0.9 to 1.0 only.

\begin{figure}[htb!]
\centering
\includegraphics[width=0.5\textwidth]{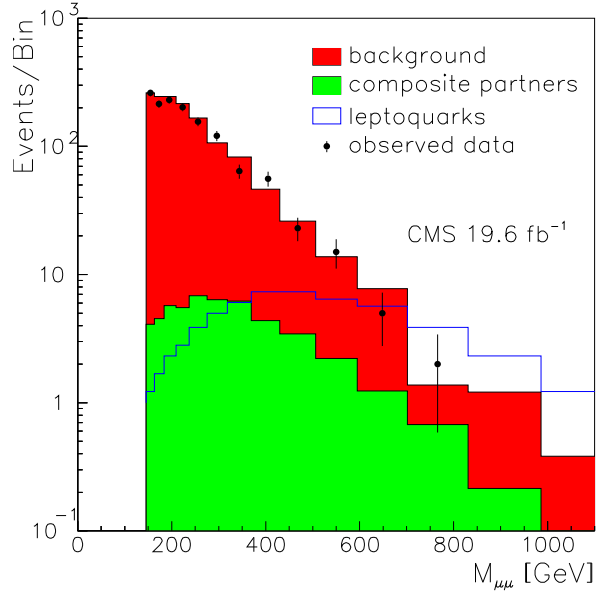}
\caption{$M_{\mu\mu}$ distributions for $\mu^+\mu^-+\geq2$jets events. The SM background as estimated by the CMS collaboration is shown in red, together with the CMS data points in black. The  signal from  QCD pair produced 500$\,$GeV  partners is shown in green, together with the CMS simulated signal of a 500$\,$ scalar leptoquark in blue. Custom preselection cuts are applied. The scalar leptoquark signal
is normalized to the total expected number of events in the composite partner case for a more transparent comparison.}
\label{fig:mumu}
\end{figure}

\begin{figure}[htb!]
\centering
\includegraphics[width=0.5\textwidth]{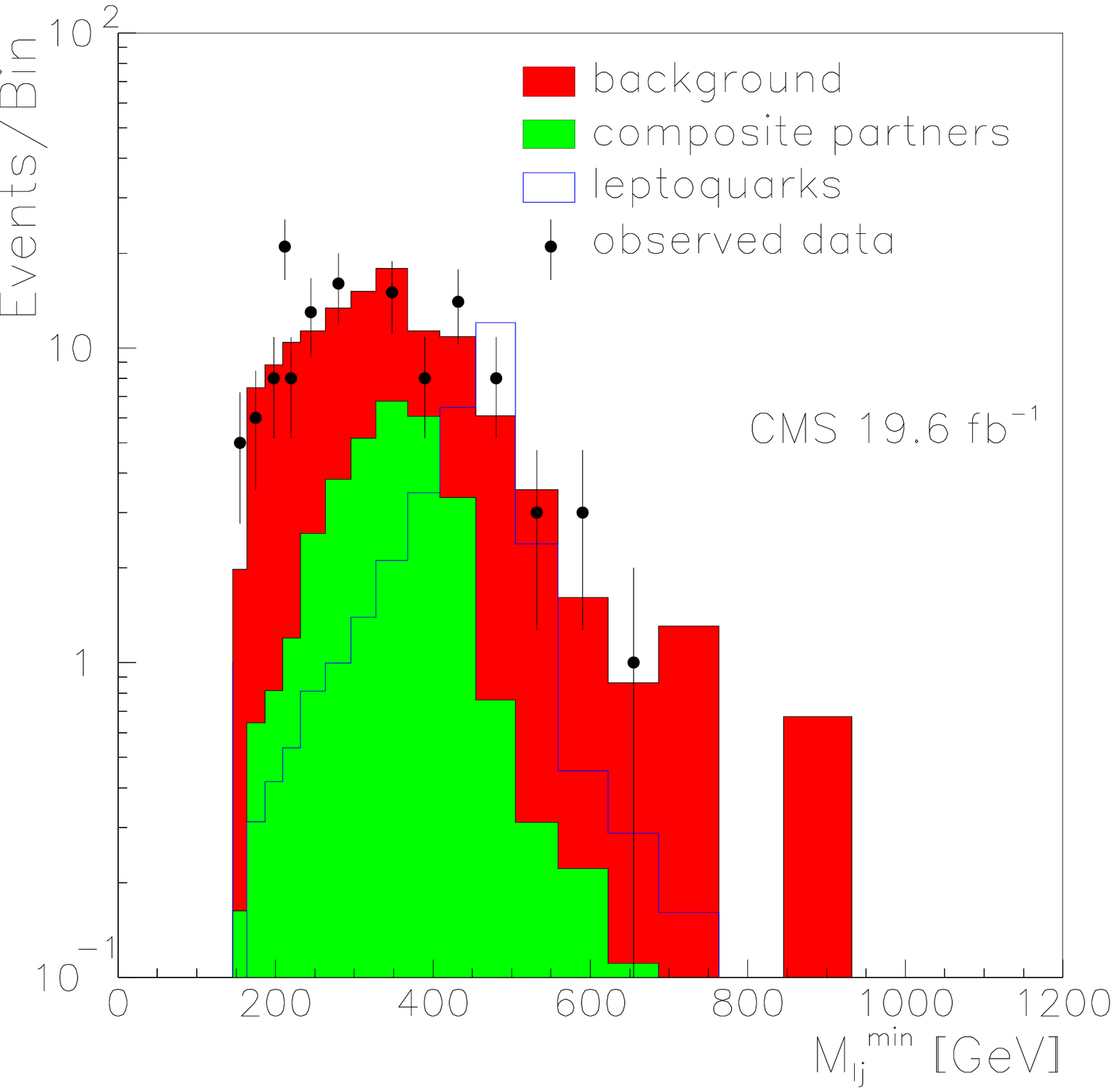}
\caption{$M_{\mu j}^{min}$ distributions for $\mu^+\mu^-+\geq2$jets events. The SM background as estimated by the CMS collaboration is shown in red, together with the CMS data points in black. The signal from  QCD pair produced 500$\,$GeV  partners is shown in green, together with the CMS simulated signal of a 500$\,$ scalar leptoquark in blue. 500$\,$GeV-like cuts are applied. The scalar leptoquark signal
is normalized to the total expected number of events in the composite partner case for a more transparent comparison.}
\label{fig:mlj}
\end{figure}

\begin{figure}[htb!]
\centering
\includegraphics[width=0.5\textwidth]{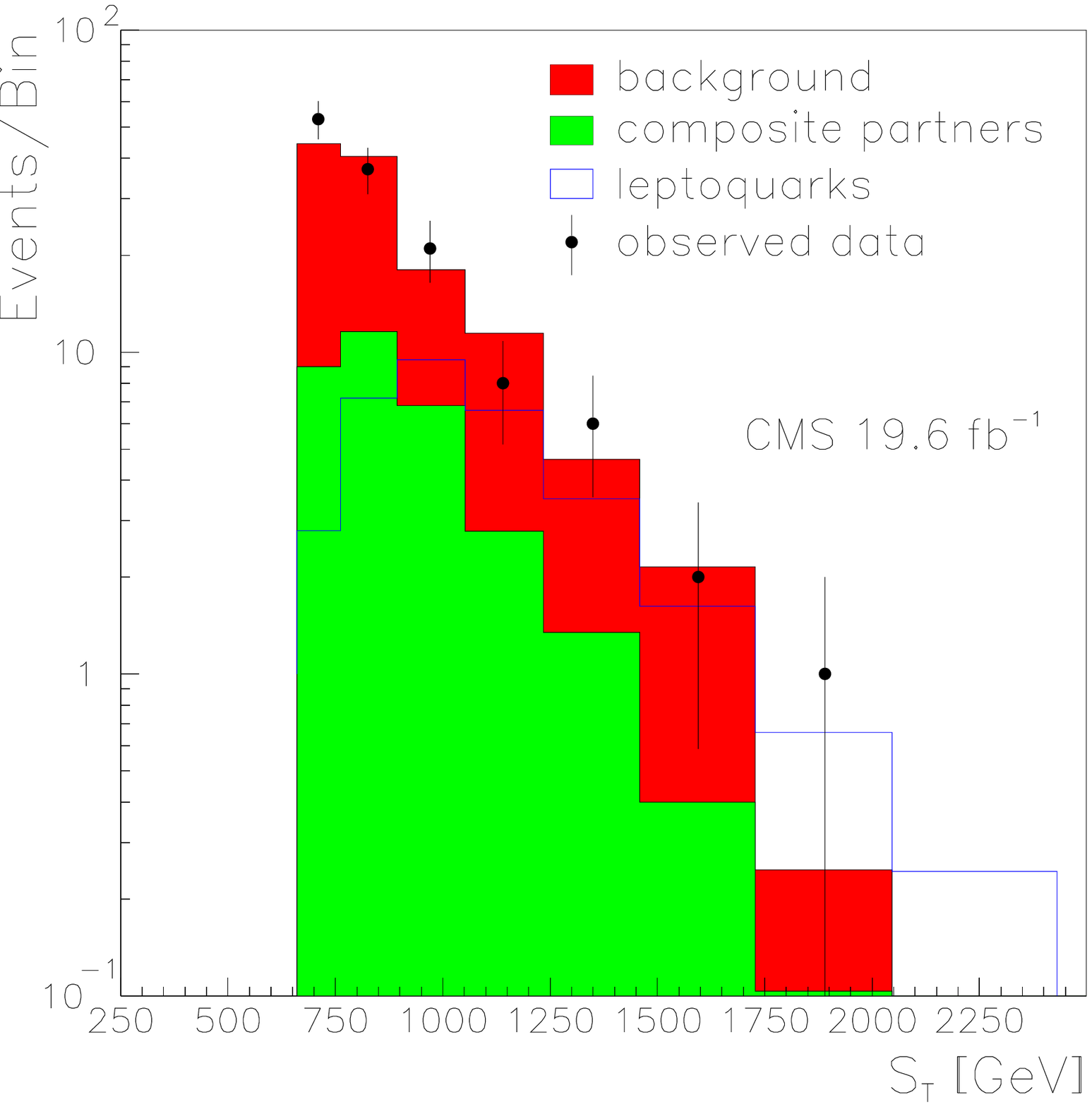}
\caption{$S_T$ distribution distributions for $\mu^+\mu^-+\geq2$jets events. The SM background as estimated by the CMS collaboration is shown in red, together with the CMS data points in black.
The  signal from QCD pair produced 500$\,$GeV  partners is shown in green, together with the CMS simulated signal of a 500$\,$ scalar leptoquark in blue. 500$\,$GeV-like cuts are applied. The scalar leptoquark signal
is normalized to the total expected number of events in the composite partner case for a more transparent comparison.}
\label{fig:st}
\end{figure}

We use the different distributions for background and signal events
that CMS made available in Ref.~\cite{cms8lepto} in order to extract the bounds on the composite partners. These are the $S_T$ and $M_{\mu j}^{min}$ distributions for cuts corresponding to a leptoquark mass of 500$\,$GeV and $900\,$GeV.
Distributions in $S_T$, $M_{\mu j}^{min}$ and $M_{\mu\mu}$ are also shown in Ref.~\cite{cms8lepto} at the preselection cut level. However, 
the $Z^*\/\gamma$+jets background is still overwhelming at this level.  We therefore use a $M_{\mu\mu}$ distribution obtained after applying
the ``custom preselection'' cut level defined above. Then, for each of the above distributions and for each heavy partner mass $M_Q$, we build a binned
log-likelihood function based on the bin content of the considered distribution. Assuming observed events in each bin are Poisson distributed and no correlation among different bins the function reads
\begin{eqnarray}
-2\ln L(M_{Q},\mu_s)\equiv
\underset{\xi_j}{\rm Min}
\left[2\sum_{i=1}^{n} \left ( N^i_B+N^i_S
-N^i_{d}+ 
N^i_{d}\log{\frac{N^i_{d}}{N^i_B+N^i_S}} \right )
+\sum_{j=s,b}\left(\frac{\xi_j}{\sigma_j}\right)^2
\right]\,,
\label{eq:chi2_atlas}
\end{eqnarray}
where $i=1,\ldots,n$ runs over the various bins of the distribution,
\begin{eqnarray}
N^i_B=N^i_b\left(1+\xi_b\right)\,, \quad N^i_S=\mu_s N^i_s\left(1+\xi_s\right)\,,
\label{eq:bin}
\end{eqnarray}
and $\mu_s$ rescales the heavy partner signal. It is used in order to derive exclusion bounds on the signal cross section for a given $M_Q$.
$N^i_d$, $N_b^i$ and $N_s^i$ are respectively the numbers of observed, expected SM background and heavy partner signal events in the bin $i$, while $N_B^i$ and $N_S^i$ are the corresponding numbers of background and signal events including systematic uncertainties.
We introduced in Eq.~\eqref{eq:chi2_atlas} the pull  parameters $\xi_{s,b}$ in order to account for systematic uncertainties in a simplified manner, as described in Ref.~\cite{pulls}. $\xi_b$ accounts for the systematic uncertainty of the background with a standard deviation $\sigma_b=0.05$~\cite{cms8lepto}, while $\xi_s$ accounts
for the systematic uncertainty originating from the signal computation with a standard deviation $\sigma_s=0.05$. 

Final exclusion bounds
are still dominated by statistics and we explicitly checked that our limits only mildly change when varying the pull values.
$\chi^2$ functions are associated the log-likelyhood functions of Eq.~\eqref{eq:chi2_atlas} through the standard relation $\chi^2(M_Q,\mu_s)=-2\log L(M_Q,\mu_s)$.

We apply the following procedure in order to extract the 95$\%$ CL limits on the heavy partner mass. For each distribution and mass $M_Q$ we first solve for the $\mu_s$ value which minimizes the associated $\chi^2$  function, or equivalently maximizes the likelihood function. We then define $\hat{\mu}_s$ as the value of $\mu_s$ which saturates the inequality
\beq
\delta\chi^2(M_Q,\mu_s)\equiv \chi^2(M_{Q},\mu_s)-\chi^2_{\rm min}(M_{Q})\geq3.84\,,
\eeq
where $\chi^2_{\rm min}$ is the minimal $\chi^2$ value. Cross sections larger than $\hat{\mu}_s$ times the assumed initial one are excluded at 95$\%$ CL.
We repeat the above minimalization procedure for each distribution described above. We  then choose for each $M_Q$ the
 strongest bound as the net 95$\%$ CL limit. The strongest bound is obtained from the
$M_{\mu\mu}$ distribution with ``custom preselection'' (see Fig.~\ref{fig:mumu}) for $M_{Q}=300$, the $M_{\mu j}^{min}$ distribution with
500$\,$GeV cuts (Fig.~\ref{fig:mlj}) for 400 and 500$\,$GeV partners, 
the $M_{\mu j}^{min}$ distribution with 900$\,$GeV cuts for $1.1$ and $1.2\,$TeV masses, and the $S_T$ distribution with 500$\,$GeV cuts (Fig.~\ref{fig:st}) for all other masses. The 95$\%$ CL exclusion limits obtained from this analysis are shown in Figure~\ref{fig:cmslepto}.
For illustration, we also plot in Figs.~\ref{fig:mumu},~\ref{fig:mlj} and ~\ref{fig:st} the 500$\,$GeV leptoquark signal
as simulated by CMS, but normalized to the total rate of the composite partner signal for $M_Q=500\,$GeV for a transparent comparison.

As a check this same binned log-likelihood analysis was applied to 500$\,$ GeV scalar leptoquarks simulated by CMS with
the 500 GeV cuts, i.e. using the information of Fig.~\ref{fig:st}. The obtained exclusion limits were compatible
with the resulting 95$\%$ CL limit obtained by CMS.

Note that the exclusion limits obtained from the recast are weaker than those obtained from a simple rescaling of the
CMS exclusion limits shown in Fig.8 of Ref.~\cite{cms8lepto} by the $W$ branching
ratio to quarks. The reason is two-fold. First of all, although the applied cuts are also suitable for
composite heavy quark partners, they are optimized to enhance scalar leptoquark signals. This results in larger acceptances for leptoquarks than for composite partners. Then, the distributions used to extract the 95$\%$ CL limits are also more suited for scalar leptoquarks than for quark partners. This is
seen in Figs.~\ref{fig:mumu},~\ref{fig:mlj} and~\ref{fig:st}, where composite quark signals are peaking slightly
more towards the background than scalar leptoquark ones, thus weakening the exclusion limits.

%% file: appendix5.tex
As we discussed in Sec.~\ref{indirectsearches}, the strongly coupled UV dynamics can give rise to four-fermion
contact interactions among the composite resonances. In addition to the dijet signals analyzed
in the main text, these high-order operators can also contribute to heavy resonances single and pair production.
Although it is easy to check that the new contributions to single
production processes are always negligible, the situation for pair production is by far less obvious.
The aim of this appendix is to clarify this issue by comparing the pair production due to four-fermion operators
with the QCD one, which is the dominant production mode for the mass window currently probed
by the experiments (see Sec.~\ref{sec:pairprod}).

Notice that the contributions coming from the contact operators could be enhanced in the channels
in which the up-quark PDF, larger than the gluonic one, can compensate the intrinsic suppression of the
higher-order operators due to the heavy scale at which they are generated and the powers of the
mixing angles originating from the mixing with the elementary states.
In spite of this effect, we find that, in the relevant mass region, the contribution from the contact operators is generally subdominant.
Only in some very specific cases, as we will point out at the end of this appendix, these new effects could become relevant.\\

As we discussed in the main text, the existing experimental searches only probe configurations in
which the resonance spectrum contains a light fourplet. For this reason in the following
we will only focus on the cases in which the fourplet $Q$ is light enough to be present in the low energy effective theory.

To start with, we consider the scenario in which the singlet $\tilde{U}$ is heavy so that it is decoupled
from the effective theory and its mixing with the elementary states can be neglected.
In this case the relevant four-fermion operator is
\beq
\frac{1}{f^2}\left[(\bar Q \gamma_\mu Q)^2\right]\,,
\eeq
which, taking into account the mixing with the right-handed
up-type quark, leads to operators of the type
\beq
c_{Qu}\left[(\bar Q_R \gamma_\mu u_R)^2\right]\,.
\label{eq:4fop}
\eeq
We only show here the case of first family partners, but it is understood that an equivalent operator appears
in the case of second family partners.
Notice that the operator in Eq.~(\ref{eq:4fop}) induces pair production processes initiated by a $uu$
state, thus its contribution is enhanced by the large up quark PDF.
Other four-fermion operators, such us
$(\bar Q_R \gamma_\mu u_R)(\bar u_R \gamma^\mu Q_R)$, which can also be present in the Lagrangian,
do not benefit from the double up quark PDF enhancement, thus they lead to subdominant contributions.

The size of the coefficient of the four-fermion operator, $c_{Qu}$, is determined by the mixing of the fourplet with the
elementary up-type quark and by the compositeness scale $f$. In the case of models with partially composite right-handed up-type quarks,
the coefficient can be estimated as $c_{Qu} \simeq \sin^2 \phi_4/f^2$, with $\sin \phi_4$ defined as in Sec.~\ref{sec:pcmodel}:
\beq
\sin\phi_4=\frac{y_R f \sin\epsilon}{\sqrt{M_4^2+y_R^2 f^2 \sin^2\epsilon}}\;.
\eeq

\begin{figure}[t!]
\centering
\begin{tabular}{cc}
\includegraphics[width=0.45\textwidth]{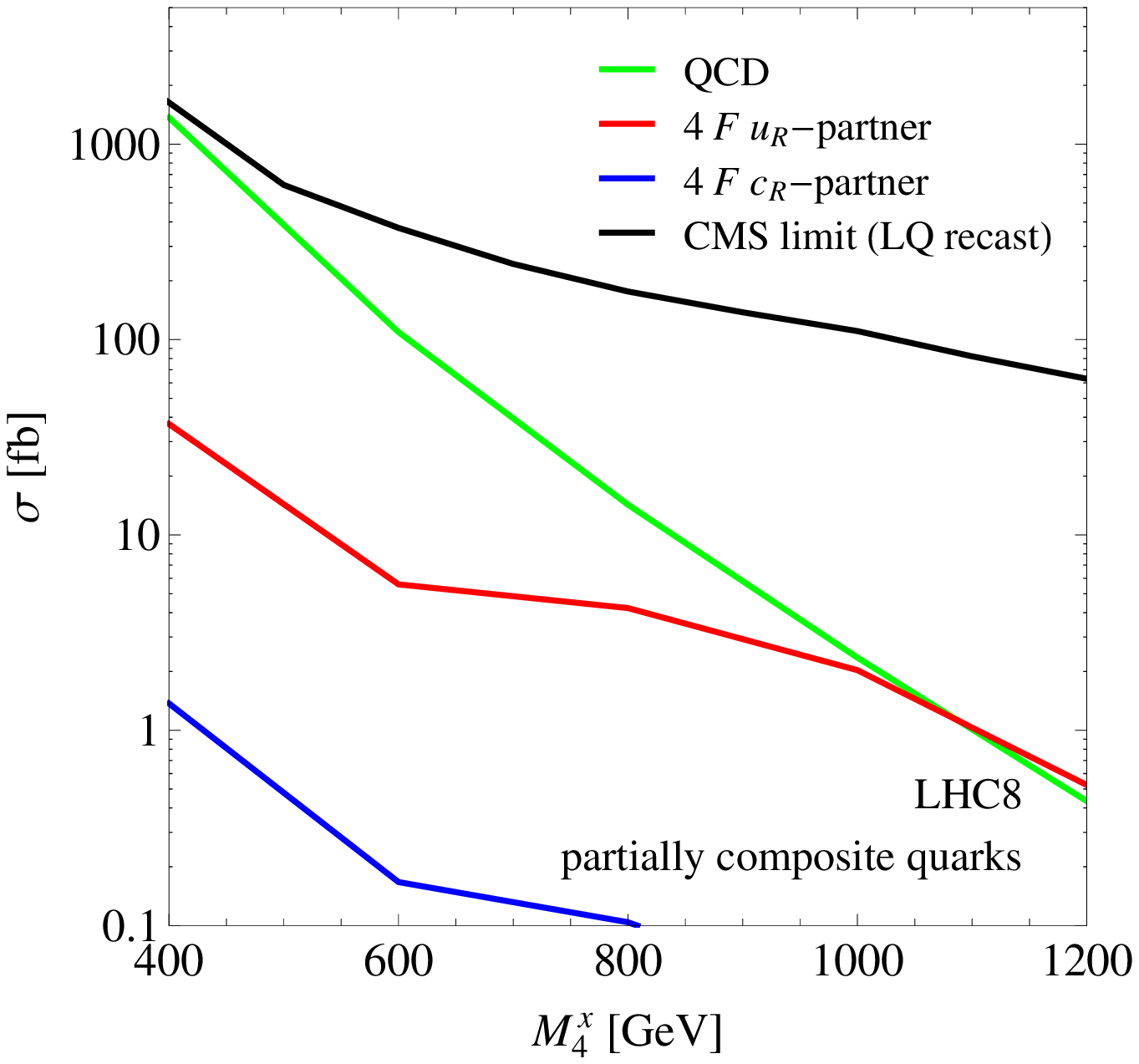}&
\includegraphics[width=0.45\textwidth]{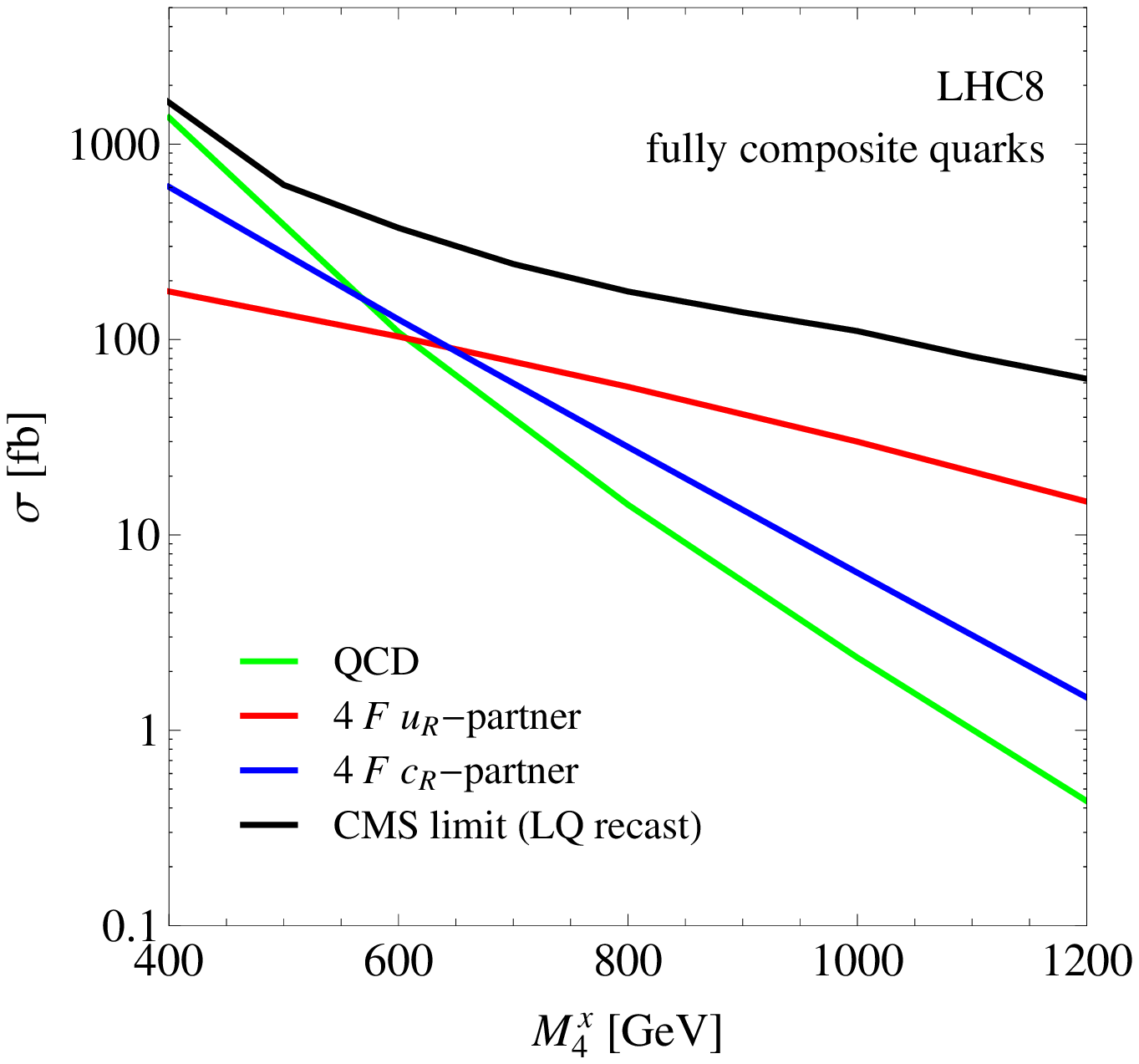}
\end{tabular}
\caption{QCD pair production of heavy partners (green line) compared with the
maximal contributions of the four-fermion
operator in Eq.~(\ref{eq:4fop}) for first family partners (red line), and second family
partners (blue line) when only a light fourplet is present. The black curve corresponds to the strongest limit on pair
production which
comes from the recast of the CMS leptoquark search~\cite{cms8lepto}. In the left (right) panel we show the
case of partially (fully) composite partners. We use $f=600$ GeV for the partially composite case, while for the fully composite case we use
$f=2.8$ TeV ($f=300$ GeV) for first (second) family partners.}
\label{fig:4f1}
\end{figure}

For a given value of $M_4$,
the size of the coefficient in front of the four-fermion operator is bounded by the experimental limits
summarized in
the left panel of Fig.~\ref{fig:yrigc1}.
The contribution of the operator in Eq.~(\ref{eq:4fop}) to pair production at the $8$ TeV LHC is shown
in the left panel of Fig.~\ref{fig:4f1} for the maximal allowed mixing and for the choice $f = 600$ GeV.
The red line corresponds to the case of first family partners while the blue curve corresponds to the case
of second family partners. For comparison we also show in the same plot the corresponding rate for QCD pair production
of heavy partners (green line), as well as the strongest limit on pair production (black curve)
which comes from the recast of the CMS leptoquark search (see Sec.~\ref{sec:pairprod}).

From the plot it can be clearly seen that the contribution coming from four-fermion operators is subleading with respect
to QCD pair production for light resonance masses, in particular below or around the current experimental limit $M_4 \lesssim 530$ GeV.
The four-fermion operator contribution becomes comparable to QCD pair production only
for heavier masses ($M_4 \sim 1$ TeV), which are well above the currently experimentally accessible region.\\

We now consider the case of models with fully composite right-handed up-type quarks.
In this scenario the coefficient of the four-fermion operators can be simply estimated as $c_{Qu}\sim 1/f^2$.
In this case the size of the contribution of four-fermion operators is limited by the experimental bound
on $f$ that we have derived from the indirect dijet bound in subsection~\ref{indirectsearches}.
For the case of first family partners
the bound is $f > 2.8$ TeV and the contribution of the four-fermion operators to pair production is shown by
the red line in the right panel of Fig.~\ref{fig:4f1}. For second family partners
the bound is weakened to $f > 300$ GeV, however the related cross section enhancement
is compensated by the reduction of the PDF of the charm quark with respect to the one of the up quark.
The contribution of second family partners is shown by the blue line in the Fig.~\ref{fig:4f1}.
As we can observe, in the case of fully composite partners
the maximal allowed value for the four-fermion contribution to pair production is still below QCD
pair production in the region of masses probed by the current searches, $M_4 \lesssim 530$ GeV.\\

We consider now the case in which a singlet partner is present in the low energy effective theory together with a
fourplet.
In this scenario, the estimate of the four-fermion contributions for fully composite partners is similar to the one
we discussed before.
However, in the case of partially composite right-handed quarks, an additional operator of the form
\beq\label{eq:4ferm_sing}
\frac{1}{f^2}\left[(\bar Q \gamma_\mu \tilde{U})^2\right]\,,
\eeq
typically leads to a bigger contribution to pair production processes with respect to the case where only a fourplet is present.
In fact the operator in Eq.~(\ref{eq:4ferm_sing})
can generate a contribution to the four-fermion interaction in Eq.~(\ref{eq:4fop}) whose size
is determined by the mixing angle of the elementary quark with the singlet state $c_{Qu}=\sin^2 \phi_1/f^2$,
which can be much larger than the mixing with the fourplet.

\begin{figure}[t!]
\centering
\includegraphics[width=0.5\textwidth]{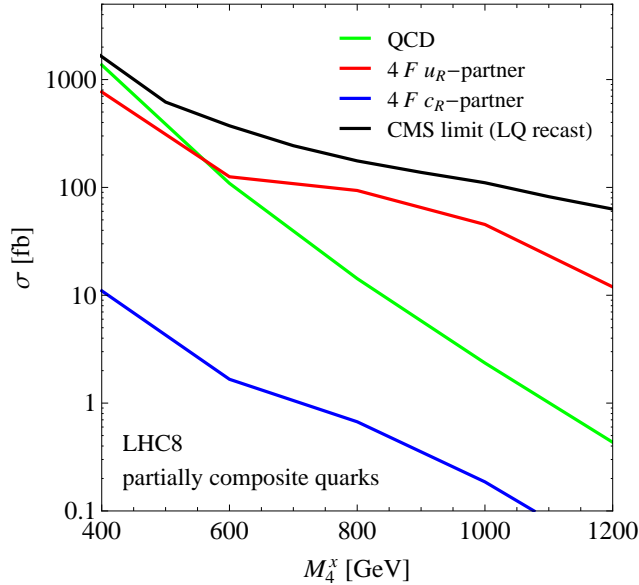}
\caption{Cross section for QCD pair production of heavy partners (green line) compared with the
maximal contributions of the four-fermion
operator in Eq.~(\ref{eq:4fop}) mixing for first family partners (red line), and second family
partners (blue line). To derive the four-fermion contributions we assumed that a fourplet and a singlet are present with
equal mass $M_1=M_4$.
The black curve corresponds to the strongest limit on pair production which
comes from the recast of the CMS leptoquark search~\cite{cms8lepto}. We use the value $f=600$ GeV.}
\label{fig:4f2}
\end{figure}

By using the results of Sec.~\ref{sec:pcmodel}, and restricting the analysis to the mass range $M_1 \gtrsim M_4$ on which
we mainly focused in this paper, we find that
\beq
\sin\phi_1 \sim \frac{y_R f \cos\epsilon}{\sqrt{M_1^2+y_R^2 f^2 \cos^2\epsilon}}\;.
\eeq
If we set $y_R^{u/c}$ to the maximal allowed value (see the left panel
of Fig.~\ref{fig:yrigc1}), the estimate of the contribution
of the four-fermion operators to pair-production of fourplet states is shown in Figure~\ref{fig:4f2} by the red (blue)
line for first (second) family partners.

It can be seen that the estimate of the four-fermion contribution is similar to the one we found for the
fully composite framework in the scenario with first family partners (right panel
of Fig.~\ref{fig:4f1}). Also in the present case for first generation partners the new contribution is subdominant with
respect to QCD pair production for the mass range probed by the current experimental data. For second generation
resonances the four-fermion processes are always negligible.

We conclude this appendix noting that in the scenarios in which the singlet is lighter than the fourplet a plethora of
possibilities opens. In this case the bound on the mixing with the singlet could be much weaker for a combination
of reasons: first of all the experimental analysis leads to weaker bounds on $y_R^{u/c}$,
moreover the smaller value of $M_1$ allows for a larger amount of compositeness at fixed value of $y_R^{u/c}$.
The enhanced mixing can of course lead to a stronger impact of the four-fermion operators in pair
production processes and can make this production mechanism dominant with respect to the usual QCD one.
Notice moreover that the presence of a singlet lighter than the fourplet can also lead to new cascade decay
channels making the extraction of the bounds on the resonance masses much more involved. The detailed analysis of this
generic case however is out of the scope of the present paper.

%% file: non-degenerate_v1.bbl
\begin{thebibliography}{}

\bibitem{Aad:2012tfa} 
  G.~Aad {\it et al.}  [ATLAS Collaboration],
  Phys.\ Lett.\ B {\bf 716}, 1 (2012)
  [arXiv:1207.7214 [hep-ex]].

\bibitem{Chatrchyan:2012ufa} 
  S.~Chatrchyan {\it et al.}  [CMS Collaboration],
  Phys.\ Lett.\ B {\bf 716}, 30 (2012)
  [arXiv:1207.7235 [hep-ex]].

\bibitem{thooft}
G.~'t Hooft, in Proc. of 1979 Carg\`ese Institute on {\it Recent Developments in Gauge Theories}, p.~135, Plenum Press, New York 1980;
 K.~G.~Wilson,
  Phys.\ Rev.\ D {\bf 3} (1971) 1818;
   E.~Gildener,
  Phys.\ Rev.\ D {\bf 14}, 1667 (1976);
S.~Weinberg,
  Phys.\ Lett.\ B {\bf 82}, 387 (1979).

\bibitem{Farina:2013mla} 
  M.~Farina, D.~Pappadopulo and A.~Strumia,
  JHEP {\bf 1308}, 022 (2013)
  [arXiv:1303.7244 [hep-ph]].

\bibitem{Giudice:2013yca} 
  G.~F.~Giudice,
  arXiv:1307.7879 [hep-ph].

\bibitem{Weinberg:1987dv} 
  S.~Weinberg,
  Phys.\ Rev.\ Lett.\  {\bf 59}, 2607 (1987).
  
\bibitem{Agrawal:1997gf} 
  V.~Agrawal, S.~M.~Barr, J.~F.~Donoghue and D.~Seckel,
  Phys.\ Rev.\ D {\bf 57}, 5480 (1998)
  [hep-ph/9707380].
  
  

\bibitem{Harnik:2006vj} 
  R.~Harnik, G.~D.~Kribs and G.~Perez,
  Phys.\ Rev.\ D {\bf 74}, 035006 (2006)
  [hep-ph/0604027].


\bibitem{Gedalia:2010iy} 
  O.~Gedalia, A.~Jenkins and G.~Perez,
  Phys.\ Rev.\ D {\bf 83}, 115020 (2011)
  [arXiv:1010.2626 [hep-ph]].

\bibitem{Blanke:2013uia}
  M.~Blanke, G.~F.~Giudice, P.~Paradisi, G.~Perez and J.~Zupan,
  JHEP {\bf 1306}, 022 (2013) 
  [arXiv:1302.7232 [hep-ph]].
  

\bibitem{Gedalia-Kamenik-Ligeti-Perez}
  O.~Gedalia, J.~F.~Kamenik, Z.~Ligeti and G.~Perez,
  Phys.\ Lett.\ B {\bf 714}, 55  (2012) 
  [arXiv:1202.5038 [hep-ph]].


\bibitem{Nir-Seiberg}
  Y.~Nir and N.~Seiberg,
  Phys.\ Lett.\ B {\bf 309}, 337  (1993)  
  [hep-ph/9304307].

\bibitem{Galon-Perez-Shadmi}
  I.~Galon, G.~Perez and Y.~Shadmi,
  arXiv:1306.6631 [hep-ph].
 
\bibitem{Fitzpatrick-Perez-Randall}
  A.~L.~Fitzpatrick, G.~Perez and L.~Randall,
  Phys.\ Rev.\ Lett.\  {\bf 100}, 171604  (2008) 
  [arXiv:0710.1869 [hep-ph]].
 
 
\bibitem{Csaki-Perez-Surujon-Weiler}
  C.~Csaki, G.~Perez, Z.~Surujon and A.~Weiler,
  Phys.\ Rev.\ D {\bf 81}, 075025  (2010) 
  [arXiv:0907.0474 [hep-ph]].
  
\bibitem{Mahbubani:2012qq}
  R.~Mahbubani, M.~Papucci, G.~Perez, J.~T.~Ruderman and A.~Weiler,
  Phys.\ Rev.\ Lett.\  {\bf 110}, 151804  (2010)
  [arXiv:1212.3328 [hep-ph]].


\bibitem{HPGB}
D.~B.~Kaplan, H.~Georgi,
  Phys.\ Lett.\  {\bf B136 } (1984)  183.
  D.~B.~Kaplan, H.~Georgi, S.~Dimopoulos,
  Phys.\ Lett.\  {\bf B136 } (1984)  187.
  H.~Georgi, D.~B.~Kaplan, P.~Galison,
  Phys.\ Lett.\  {\bf B143 } (1984)  152.
  T.~Banks,
  Nucl.\ Phys.\  {\bf B243 } (1984)  125.
  H.~Georgi, D.~B.~Kaplan,
  Phys.\ Lett.\  {\bf B145 } (1984)  216.
  M.~J.~Dugan, H.~Georgi, D.~B.~Kaplan,
  Nucl.\ Phys.\  {\bf B254 } (1985)  299.
N.~Arkani-Hamed, A.~G.~Cohen, E.~Katz and A.~E.~Nelson,
  JHEP {\bf 0207}, 034 (2002)
  [hep-ph/0206021];
R.~Contino, Y.~Nomura and A.~Pomarol,
  Nucl.\ Phys.\ B {\bf 671}, 148 (2003)
  [hep-ph/0306259].


\bibitem{Agashe:2004rs}
  K.~Agashe, R.~Contino and A.~Pomarol,
  Nucl.\ Phys.\ B {\bf 719}, 165 (2005)
  [hep-ph/0412089].
  
\bibitem{DP}
S.~Dimopoulos, J.~Preskill,
  Nucl.\ Phys.\  {\bf B199 } (1982)  206.
  
  
\bibitem{flavor-triviality1}
  C.~Delaunay, O.~Gedalia, S.~J.~Lee, G.~Perez and E.~Ponton,
  Phys.\ Rev.\ D {\bf 83}, 115003  (2011) 
  [arXiv:1007.0243 [hep-ph]].
  
  
\bibitem{Redi-Weiler}
  M.~Redi and A.~Weiler,
  JHEP {\bf 1111}, 108  (2011)
  [arXiv:1106.6357 [hep-ph]].
  
\bibitem{Zbb} 
  K.~Agashe, R.~Contino, L.~Da Rold and A.~Pomarol,
  Phys.\ Lett.\ B {\bf 641}, 62 (2006)
  [hep-ph/0605341].

 
\bibitem{flavor-triviality2}
  C.~Delaunay, O.~Gedalia, S.~J.~Lee, G.~Perez and E.~Ponton,
  Phys.\ Lett.\ B {\bf 703}, 486  (2011) 
  [arXiv:1101.2902 [hep-ph]].
  
\bibitem{Da-Rold-Delaunay-Grojean-Perez}
  L.~Da Rold, C.~Delaunay, C.~Grojean and G.~Perez,
  JHEP {\bf 1302}, 149  (2013) 
  [arXiv:1208.1499 [hep-ph]].
 
\bibitem{Delaunay-Grojean-Perez}
  C.~Delaunay, C.~Grojean and G.~Perez,
  arXiv:1303.5701 [hep-ph].
  
  
  
  
  
\bibitem{Redi:2013eaa}
  M.~Redi, V.~Sanz, M.~de Vries and A.~Weiler,
  JHEP {\bf 1308}, 008  (2013) 
  [arXiv:1305.3818 [hep-ph]].
  
  
   
\bibitem{Matsedonskyi:2012ym} 
  O.~Matsedonskyi, G.~Panico and A.~Wulzer,
  JHEP {\bf 1301}, 164 (2013) 
  [arXiv:1204.6333 [hep-ph]].
  
\bibitem{Redi:2012ha}
  M.~Redi and A.~Tesi,
  JHEP {\bf 1210}, 166  (2012) 
  [arXiv:1205.0232 [hep-ph]].
   
\bibitem{Marzocca:2012zn}
  D.~Marzocca, M.~Serone and J.~Shu,
  JHEP {\bf 1208}, 013  (2012) 
  [arXiv:1205.0770 [hep-ph]].
  
\bibitem{Pomarol:2012qf}
  A.~Pomarol and F.~Riva,
  JHEP {\bf 1208}, 135  (2012) 
  [arXiv:1205.6434 [hep-ph]].
  
\bibitem{Panico:2012uw}
  G.~Panico, M.~Redi, A.~Tesi and A.~Wulzer,
  JHEP {\bf 1303}, 051  (2013) 
  [arXiv:1210.7114 [hep-ph]].


\bibitem{Contino:2006qr}
  R.~Contino, L.~Da Rold and A.~Pomarol,
  Phys.\ Rev.\ D {\bf 75}, 055014  (2007) 
  [hep-ph/0612048].
   
  \bibitem{ccwz}
  S.~R.~Coleman, J.~Wess and B.~Zumino,
  Phys.\ Rev.\  {\bf 177}, 2239 (1969); 
  C.~G.~Callan, Jr., S.~R.~Coleman, J.~Wess and B.~Zumino,
  Phys.\ Rev.\  {\bf 177}, 2247  (1969).
  
   
\bibitem{DeSimone:2012fs}
  A.~De Simone, O.~Matsedonskyi, R.~Rattazzi and A.~Wulzer,
  arXiv:1211.5663 [hep-ph].

  
\bibitem{Grojean:2013qca}
  C.~Grojean, O.~Matsedonskyi and G.~Panico,
  arXiv:1306.4655 [hep-ph].
  
\bibitem{Kaplan:1991dc}
  D.~B.~Kaplan,
  Nucl.\ Phys.\ B {\bf 365}, 259  (1991); 
  R.~Contino, T.~Kramer, M.~Son and R.~Sundrum,
  JHEP {\bf 0705}, 074  (1991) 
  [hep-ph/0612180].

  
\bibitem{Panico:2011pw} 
  G.~Panico and A.~Wulzer,
  JHEP {\bf 1109}, 135 (2011) 
  [arXiv:1106.2719 [hep-ph]].


  \bibitem{Carmona:2012jk}
  A.~Carmona, M.~Chala and J.~Santiago,
  JHEP {\bf 1207}, 049  (2012) 
  [arXiv:1205.2378 [hep-ph]].
  

   
\bibitem{ContinoServant}
R.~Contino and G.~Servant,
  JHEP {\bf 0806}, 026 (2008) 
  [arXiv:0801.1679 [hep-ph]].

\bibitem{MrazekWulzer}
J.~Mrazek and A.~Wulzer,
  Phys.\ Rev.\ D {\bf 81}, 075006 (2010) 
  [arXiv:0909.3977 [hep-ph]].



  \bibitem{Atre:2013ap} 
  A.~Atre, M.~Chala and J.~Santiago,
  JHEP {\bf 1305}, 099 (2013) 
  [arXiv:1302.0270 [hep-ph]].
  


\bibitem{ATLAS:topsearches}
G.~Aad {\it et al.} [ATLAS Collaboration], 	ATLAS-CONF-2013-060; ATLAS-CONF-2013-056; ATLAS-CONF-2013-018;
  Phys.\ Lett.\ B {\bf 718}, 1284 (2013)
  [arXiv:1210.5468 [hep-ex]];
  Phys.\ Rev.\ Lett.\  {\bf 109}, 032001 (2012)
  [arXiv:1202.6540 [hep-ex]];
  JHEP {\bf 1204}, 069 (2012)
  [arXiv:1202.5520 [hep-ex]];
  Phys.\ Rev.\ Lett.\  {\bf 108}, 261802 (2012)
  [arXiv:1202.3076 [hep-ex]].

\bibitem{CMS:topsearches}
S.~Chatrchyan {\it et al.} [CMS Collaboration], CMS-PAS-B2G-12-021; CMS-PAS-B2G-12-019; CMS-PAS-B2G-12-015;
  JHEP {\bf 1301}, 154 (2013)
  [arXiv:1210.7471 [hep-ex]];
  Phys.\ Lett.\ B {\bf 718}, 307 (2012)
  [arXiv:1209.0471 [hep-ex]];
  JHEP {\bf 1205}, 123 (2012)
  [arXiv:1204.1088 [hep-ex]];
  Phys.\ Lett.\ B {\bf 716}, 103 (2012)
  [arXiv:1203.5410 [hep-ex]];
  Phys.\ Rev.\ Lett.\  {\bf 107}, 271802 (2011)
  [arXiv:1109.4985 [hep-ex]].

\bibitem{ATLAS:53searches}
G.~Aad {\it et al.} [ATLAS Collaboration],  ATLAS-CONF-2013-051; ATLAS-CONF-2012-130.

\bibitem{CMS:53searches}
S.~Chatrchyan {\it et al.} [CMS Collaboration], CMS-PAS-B2G-12-012; CMS-PAS-B2G-12-003


  \bibitem{CDF:wj}
  [CDF Collaboration], CDF/PUB/EXOTIC/PUBLIC/1026.
  \begin{verbatim}
   http://www-cdf.fnal.gov/physics/new/top/2010/tprop/HQ_public/HQpub.pdf
  \end{verbatim}


  \bibitem{Abazov:2010ku} 
  V.~M.~Abazov {\it et al.}  [D0 Collaboration],
  Phys.\ Rev.\ Lett.\  {\bf 106}, 081801 (2011)
  [arXiv:1010.1466 [hep-ex]


  \bibitem{Aaltonen:2011tq} 
  T.~Aaltonen {\it et al.}  [CDF Collaboration],
  Phys.\ Rev.\ Lett.\  {\bf 107}, 261801 (2011)
  [arXiv:1107.3875 [hep-ex]].

  
  \bibitem{Abazov:2011vy} 
  V.~M.~Abazov {\it et al.}  [D0 Collaboration],
  Phys.\ Rev.\ Lett.\  {\bf 107}, 082001 (2011)
  [arXiv:1104.4522 [hep-ex]].
    

    \bibitem{ATLAS:pp}
G.~Aad {\it et al.} [ATLAS Collaboration], ATLAS-CONF-2012-137.

  \bibitem{Aad:2012bt} 
  G.~Aad {\it et al.}  [ATLAS Collaboration],
  Phys.\ Rev.\ D {\bf 86}, 012007 (2012)
  [arXiv:1202.3389 [hep-ex]].



   \bibitem{CMS:wzdijet}
S.~Chatrchyan {\it et al.}  [CMS Collaboration], 
  CMS-PAS-EXO-12-024.
  
\bibitem{Christensen:2008py}
  N.~D. Christensen and C.~Duhr,
  \newblock Comput.Phys.Commun. {\bf 180}, 1614 (2009) [arXiv:0806.4194].
  

  \bibitem{Alwall:2011uj} 
  J.~Alwall, M.~Herquet, F.~Maltoni, O.~Mattelaer and T.~Stelzer,
  JHEP {\bf 1106}, 128 (2011)
  [arXiv:1106.0522 [hep-ph]].
  
   
  \bibitem{Chatrchyan:2012uxa} 
  S.~Chatrchyan {\it et al.}  [CMS Collaboration],
  Phys.\ Lett.\ B {\bf 718}, 329 (2012)
  [arXiv:1208.2931 [hep-ex]].

  \bibitem{CMS:2012ab} 
  S.~Chatrchyan {\it et al.}  [CMS Collaboration],
  Phys.\ Lett.\ B {\bf 716}, 103 (2012)
  [arXiv:1203.5410 [hep-ex]].  
  
  \bibitem{ATLAS:2012qe} 
  G.~Aad {\it et al.}  [ATLAS Collaboration],
  Phys.\ Lett.\ B {\bf 718}, 1284 (2013)
  [arXiv:1210.5468 [hep-ex]].
  
  \bibitem{Aad:2012xc} 
  G.~Aad {\it et al.}  [ATLAS Collaboration],
  Phys.\ Rev.\ Lett.\  {\bf 108}, 261802 (2012)
  [arXiv:1202.3076 [hep-ex]].
  
  \bibitem{Chatrchyan:2012vu} 
  S.~Chatrchyan {\it et al.}  [CMS Collaboration],
  Phys.\ Lett.\ B {\bf 718}, 307 (2012)
  [arXiv:1209.0471 [hep-ex]].
  
  
   \bibitem{CMS:2012jwa}  
  S.~Chatrchyan {\it et al.}   [CMS Collaboration],
  CMS-PAS-EXO-11-066.
 
   
  \bibitem{cms8lepto}
  S.~Chatrchyan {\it et al.}   [CMS Collaboration],
 CMS-PAS-EXO-12-042.
   
  
  \bibitem{Sjostrand:2006za} 
  T.~Sjostrand, S.~Mrenna and P.~Z.~Skands,
  JHEP {\bf 0605}, 026 (2006)
  [hep-ph/0603175].
  
  \bibitem{pgs}
  John Conway, PGS4
  \begin{verbatim}
   http://www.physics.ucdavis.edu/~conway/research/software/pgs/pgs4-general.htm
  \end{verbatim}
   

\bibitem{Eichten} 
  E.~Eichten, K.~D.~Lane and M.~E.~Peskin,
  Phys.\ Rev.\ Lett.\  {\bf 50}, 811 (1983).


 

\bibitem{NDA} 
  A.~Manohar and H.~Georgi,
  Nucl.\ Phys.\ B {\bf 234}, 189 (1984).

 
\bibitem{CMSdijet} 
  S.~Chatrchyan {\it et al.}  [CMS Collaboration],
  JHEP {\bf 1205}, 055 (2012)
  [arXiv:1202.5535 [hep-ex]].

\bibitem{ATLASdijet} 
  G.~Aad {\it et al.}  [ATLAS Collaboration],
  JHEP {\bf 1301}, 029 (2013)
  [arXiv:1210.1718 [hep-ex]].



\bibitem{Pomarol-dijets} 
  O.~Domenech, A.~Pomarol and J.~Serra,
  Phys.\ Rev.\ D {\bf 85}, 074030 (2012)
  [arXiv:1201.6510 [hep-ph]].

  
  \bibitem{Cacciari:2008gp} 
  M.~Cacciari, G.~P.~Salam and G.~Soyez,
  JHEP {\bf 0804}, 063 (2008)
  [arXiv:0802.1189 [hep-ph]].
  
      
  \bibitem{pulls}
  G.~L.~Fogli, E.~Lisi, A.~Marrone, D.~Montanino, A.~Palazzo,
  Phys.\ Rev.\  {\bf D66}, 053010 (2002) 
  [hep-ph/0206162];
 M.~C.~Gonzalez-Garcia, M.~Maltoni,
  Phys.\ Rept.\  {\bf 460}, 1-129 (2008) 
  [arXiv:0704.1800 [hep-ph]].
  

      

\end{thebibliography}
